\documentclass[a4paper,fleqn,usenatbib]{mnras}

\usepackage[T1]{fontenc}
\usepackage{ae,aecompl}
\usepackage{fixltx2e}

\usepackage{graphicx}	
\usepackage{amsmath}	
\usepackage{amssymb}	
\usepackage{comment}
\usepackage{ulem}
\usepackage{subcaption}%

\usepackage{xcolor}		
\usepackage{mathrsfs} 
\usepackage{breakcites} 
\usepackage{caption} 
\usepackage{multicol}
\usepackage{rotating} 
\usepackage{threeparttable}
\usepackage{float}

\usepackage{dblfloatfix}

\usepackage{letltxmacro} 
\usepackage{xparse}

\usepackage{newtxtext,newtxmath}

\newcommand{\angstrom}{\mbox{\normalfont\AA}}

\newcommand{\todo}{\ifmmode {\color{red}\text{\Huge{\(\bullet\)}}} \else {\color{red} \Huge$\bullet$}\fi}
\newcommand{\tido}{\ifmmode {\bullet} \else $\bullet$\fi}

\newcommand{\E        }[1]{\ifmmode 10^{#1} \else $10^{#1}$\fi}
\newcommand{\tE        }[1]{\ifmmode \times10^{#1} \else $\times10^{#1}$\fi}
\newcommand{\til}{\ifmmode \sim \else $\sim$\fi}
\renewcommand{\~} {\ifmmode \sim \else $\sim$\fi}

\newcommand{\pc}	{\ifmmode {\rm pc} \else pc\fi}
\newcommand{\ld}	{\ifmmode {\rm l.d.} \else l.d.\fi}
\newcommand{\kms}	{\ifmmode {\rm km\,s}^{-1} \else km\,s$^{-1}$\fi}
\newcommand{\Jykms}	{\ifmmode {\rm Jy\,km\,s}^{-1} \else Jy\,km\,s$^{-1}$\fi}
\newcommand{\cc}	{\ifmmode {\rm cm}^{-3}    \else cm$^{-3}$\fi}
\newcommand{\cmii}	{\ifmmode {\rm cm}^{-2}    \else cm$^{-2}$\fi}
\newcommand{\ergs}	{\ifmmode {\rm erg\,s}^{-1} \else erg s$^{-1}$\fi}
\newcommand{\ergcms}	{\ifmmode {\rm erg\,cm}^{-2}\,{\rm s}^{-1} \else erg\,cm$^{-2}$\,s$^{-1}$\fi}
\newcommand{\ergcmsA}	{\ifmmode {\rm erg\,cm}^{-2}\,{\rm s}^{-1}\,{\rm\AA}^{-1}
\else erg\,cm$^{-2}$\,s$^{-1}$\,\AA$^{-1}$\fi}
\newcommand{  \ergcmsHz  }{\ifmmode{\rm erg\,cm}^{-2}\,{\rm s}^{-1}\,{\rm Hz}^{-1}
                       \else ergs\,cm$^{-2}$\,s$^{-1}$\,Hz$^{-1}$\fi}
\newcommand{\kev}	{\ifmmode {\rm keV} \else keV\fi}

\newcommand{\mic}	{\ifmmode {\rm \mu m} \else $\mu$m\fi}
\newcommand{\vFWHM}	{\ifmmode v_{\mbox{\tiny FWHM}} \else $v_{\mbox{\tiny FWHM}}$\fi}
\newcommand{\vBLR}	{\ifmmode v_{\mbox{\tiny BLR}} \else $v_{\mbox{\tiny BLR}}$\fi}
\newcommand{\sigBLR}	{\ifmmode \sigma_{\mbox{\tiny BLR}} \else $\sigma_{\mbox{\tiny BLR}}$\fi}
\newcommand{\vNLR}	{\ifmmode v_{\mbox{\tiny NLR}} \else $v_{\mbox{\tiny NLR}}$\fi}
\newcommand{\tauBLR}	{\ifmmode \tau_{\mbox{\tiny BLR}} \else $\tau_{\mbox{\tiny BLR}}$\fi}

\newcommand{\Hubble}	{\ifmmode {\rm km\,s}^{-1}\,{\rm Mpc}^{-1} \else km\,s$^{-1}$\,Mpc$^{-1}$\fi}
\newcommand{\NDunit}	{\ifmmode {\rm Mpc}^{-3} \else Mpc$^{-3}$\fi}
\newcommand{\LFunit}	{\ifmmode {\rm Mpc}^{-3}\,{\rm mag}^{-1} \else Mpc$^{-3}$\,mag$^{-1}$\fi}
\newcommand{\MFunit}	{\ifmmode {\rm Mpc}^{-3}\,{\rm dex}^{-1} \else Mpc$^{-3}$\,dex$^{-1}$\fi}

\newcommand{\Msun}{\ifmmode M_{\odot} \else $M_{\odot}$\fi}
\newcommand{\Lsun}{\ifmmode L_{\odot} \else $L_{\odot}$\fi}
\newcommand{\Zsun}{\ifmmode Z_{\odot} \else $Z_{\odot}$\fi}
\newcommand{\mpyr}{\ifmmode \Msun\,{\rm yr}^{-1} \else $\Msun\,{\rm yr}^{-1}$\fi}

\newcommand{\qnote}{\ifmmode q_{0} \else $q_{0}$\fi}
\newcommand{\Hnote}{\ifmmode H_{0} \else $H_{0}$\fi}
\newcommand{\hnote}{\ifmmode h_{0} \else $h_{0}$\fi}
\newcommand{\anote}{\ifmmode a_{0} \else $a_{0}$\fi}

\def\gsim{\;\rlap{\lower 2.5pt \hbox{$\sim$}}\raise 1.5pt\hbox{$>$}\;}
\def\lsim{\;\rlap{\lower 2.5pt \hbox{$\sim$}}\raise 1.5pt\hbox{$<$}\;}

\newcommand{  \Halpha   }{\ifmmode {\rm H}\alpha \else H$\alpha$\fi}

\newcommand{  \ha       }{\Halpha}
\newcommand{  \Hbeta    }{\ifmmode {\rm H}\beta \else H$\beta$\fi}

\newcommand{  \hb       }{\Hbeta}
\newcommand{  \Hgamma   }{\ifmmode {\rm H}\gamma \else H$\gamma$\fi}
\newcommand{  \Hdelta   }{\ifmmode {\rm H}\delta \else H$\delta$\fi}
\newcommand{  \Lya      }{\ifmmode {\rm Ly}\alpha \else Ly$\alpha$\fi}
\newcommand{  \Lyb      }{\ifmmode {\rm Ly}\beta \else Ly$\beta$\fi}
\newcommand{  \Pa       }{\ifmmode {\rm P}\alpha \else P$\alpha$\fi}
\newcommand{  \Pb       }{\ifmmode {\rm P}\beta \else P$\beta$\fi}
\newcommand{  \Bra      }{\ifmmode {\rm Br}\alpha \else Br$\alpha$\fi}
\newcommand{  \Brg      }{\ifmmode {\rm Br}\gamma \else Br$\gamma$\fi}
\newcommand{  \hi       }{\ifmmode {\rm H}\,\textsc{i} \else H\,\textsc{i}\fi}
\newcommand{  \HI       }{\ifmmode {\rm H}\,\textsc{i} \else H\,\textsc{i}\fi}
\newcommand{  \hii      }{\ifmmode {\rm H}\,\textsc{ii} \else H\,\textsc{ii}\fi}
\newcommand{  \hei      }{\ifmmode {\rm He}\,\textsc{i} \else He\,\textsc{i}\fi}
\newcommand{  \heii     }{\ifmmode {\rm He}\,\textsc{ii} \else He\,\textsc{ii}\fi}
\newcommand{  \HeIIuv   }{\ifmmode {\rm He}\,\textsc{ii}\,\lambda1640 \else He\,\textsc{ii}\,$\lambda1640$\fi}
\newcommand{  \HeIIop   }{\ifmmode {\rm He}\,\textsc{ii}\,\lambda4686 \else He\,\textsc{ii}\,$\lambda4686$\fi}
\newcommand{  \CII	}{\ifmmode \left[{\rm C}\,\textsc{ii}\right]\,\lambda157.74\,\mu{\rm m} \else [C\,{\sc ii}]\ $\lambda157.74\,\mu{\rm m}$\fi}
\newcommand{  \cii	}{\ifmmode \left[{\rm C}\,\textsc{ii}\right] \else [C\,{\sc ii}]\fi}

\newcommand{  \ciii     }{\ifmmode {\rm C}\,\textsc{iii}\right] \else C\,\textsc{iii}]\fi}
\newcommand{  \CIII     }{\ifmmode {\rm C}\,\textsc{iii}\right]\,\lambda1909 \else C\,\textsc{iii}]\,$\lambda1909$\fi}
\newcommand{  \civ      }{\ifmmode {\rm C}\,\textsc{iv}  \else C\,\textsc{iv}\fi}
\newcommand{  \CIV      }{\ifmmode {\rm C}\,\textsc{iv}\,\lambda1549 \else C\,\textsc{iv}\,$\lambda1549$\fi}
\newcommand{  \nii      }{\ifmmode {\rm N}\,\textsc{ii}  \else N\,\textsc{ii}\fi}
\newcommand{  \niii     }{\ifmmode {\rm N}\,\textsc{iii} \else N\,\textsc{iii}\fi}
\newcommand{  \niv      }{\ifmmode {\rm N}\,\textsc{iv}  \else N\,\textsc{iv}\fi}
\newcommand{  \NIVuv    }{\ifmmode {\rm N}\,\textsc{iv}\,\lambda1486 \else N\,\textsc{iv}\,$\lambda1486$\fi}
\newcommand{  \nv       }{\ifmmode {\rm N}\,\textsc{v}   \else N\,\textsc{v}\fi}
\newcommand{\oi}{\ifmmode \left[{\rm O}\,\textsc{i}\right] \else [O\,{\sc i}]\fi}
\newcommand{\OI}{\ifmmode \left[{\rm O}\,\textsc{i}\right]\,\lambda6300 \else [O\,{\sc i}]$\,\lambda6300$\fi}
\newcommand{\oii}{\ifmmode \left[{\rm O}\,\textsc{ii}\right] \else [O\,{\sc ii}]\fi}
\newcommand{\OII}{\ifmmode \left[{\rm O}\,\textsc{ii}\right]\,\lambda3727 \else [O\,{\sc ii}]\,$\lambda3727$\fi}
\newcommand{\oiii}{\ifmmode \left[{\rm O}\,\textsc{iii}\right] \else [O\,{\sc iii}]\fi}
\newcommand{\OIII}{\ifmmode \left[{\rm O}\,\textsc{iii}\right]\,\lambda5007 \else [O\,{\sc iii}]\,$\lambda5007$\fi}
\newcommand{  \OIIIuv   }{\ifmmode {\rm O}\,\textsc{iii}\,\lambda1663 \else O\,\textsc{iii}\,$\lambda1663$\fi}
\newcommand{  \oiv      }{\ifmmode {\rm O}\,\textsc{iv}  \else O\,\textsc{iv}\fi}
\newcommand{  \OIVuv    }{\ifmmode {\rm O}\,\textsc{iv}\,\lambda1402  \else O\,\textsc{iv}\,$\lambda1402$\fi}
\newcommand{  \OIVIR    }{\ifmmode {\rm O}\,\textsc{iv}\,25.9\,\mu {\rm m} \else O\,\textsc{iv}\,$25.9\,\mu$m\fi}
\newcommand{  \ovi      }{\ifmmode {\rm O}\,\textsc{vi}   \else O\,\textsc{vi}\fi}
\newcommand{  \Ovi      }{\ifmmode {\rm O}\,\textsc{vi}\,\lambda1035 \else O\,\textsc{vi}\,$\lambda1035$\fi}
\newcommand{  \nei      }{\ifmmode {\rm Ne}\,\textsc{i}   \else Ne\,\textsc{i}\fi}
\newcommand{  \neii     }{\ifmmode {\rm Ne}\,\textsc{ii}  \else Ne\,\textsc{ii}\fi}
\newcommand{  \NeiiIR   }{\ifmmode {\rm Ne}\,\textsc{ii}\,12.8\,\mu {\rm m} \else Ne\,\textsc{ii}\,$12.8\,\mu$m\fi}
\newcommand{  \neiii    }{\ifmmode {\rm Ne}\,\textsc{iii} \else Ne\,\textsc{iii}\fi}
\newcommand{  \neiv     }{\ifmmode {\rm Ne}\,\textsc{iv}  \else Ne\,\textsc{iv}\fi}
\newcommand{  \nev      }{\ifmmode {\rm Ne}\,\textsc{v}   \else Ne\,\textsc{v}\fi}
\newcommand{  \NevIR    }{\ifmmode {\rm Ne}\,\textsc{v}\,24.3\,\mu {\rm m} \else Ne\,\textsc{v}\,$24.3\,\mu$m\fi}
\newcommand{  \nevi     }{\ifmmode {\rm Ne}\,\textsc{vi}  \else Ne\,\textsc{vi}\fi}
\newcommand{  \mgi      }{\ifmmode {\rm Mg}\,\textsc{i} \else Mg\,\textsc{i}\fi}
\newcommand{  \mgii     }{\ifmmode {\rm Mg}\,\textsc{ii} \else Mg\,\textsc{ii}\fi}
\newcommand{  \MgII     }{\ifmmode {\rm Mg}\,\textsc{ii}\,\lambda2798 \else Mg\,\textsc{ii}\,$\lambda2798$\fi}
\newcommand{  \sii      }{\ifmmode {\rm S}\,\textsc{ii} \else S\,\textsc{ii}\fi}
\newcommand{  \siii     }{\ifmmode {\rm S}\,\textsc{iii} \else S\,\textsc{iii}\fi}
\newcommand{  \siv      }{\ifmmode {\rm S}\,\textsc{iv} \else S\,\textsc{iv}\fi}
\newcommand{  \sili     }{\ifmmode {\rm Si}\,\textsc{i}   \else Si\,\textsc{i}\fi}
\newcommand{  \silii    }{\ifmmode {\rm Si}\,\textsc{ii}  \else Si\,\textsc{ii}\fi}
\newcommand{  \Siliv    }{\ifmmode {\rm Si}\,\textsc{iv}  \else Si\,\textsc{iv}\fi}
\newcommand{  \SilIVuv  }{\ifmmode {\rm Si}\,\textsc{iv}\,\lambda1400  \else Si\,\textsc{iv}\,$\lambda1400$\fi}
\newcommand{  \AlIII   }{\ifmmode {\rm Al}\,\textsc{iii}\,\lambda1857 \else Al\,\textsc{iii}\,$\lambda1857$\fi}
\newcommand{  \Aliii   }{\ifmmode {\rm Al}\,\textsc{iii} \else Al\,\textsc{iii}\fi}
\newcommand{  \caii     }{\ifmmode {\rm Ca}\,\textsc{ii} \else Ca\,\textsc{ii}\fi}
\newcommand{  \feii     }{\ifmmode {\rm Fe}\,\textsc{ii} \else Fe\,\textsc{ii}\fi}
\newcommand{  \feiii    }{\ifmmode {\rm Fe}\,\textsc{iii} \else Fe\,\textsc{iii}\fi}
\newcommand{  \Kalpha   }{\ifmmode {\rm K}\alpha \else K$\alpha$\fi}

\newcommand{ \Lhb   }{\ifmmode L_{\hb} \else $L_{\hb}$\fi}
\newcommand{ \Lha   }{\ifmmode L_{\ha} \else $L_{\ha}$\fi}
\newcommand{ \fwhb  }{\ifmmode {\rm FWHM}\left(\hb\right) \else FWHM(\hb)\fi}
\newcommand{\sighb  }{\ifmmode \sigma\left(\hb\right) \else $\sigma\left(\hb\right)$\fi}
\newcommand{ \ewhb  }{\ifmmode {\rm EW}\left(\hb\right) \else EW(\hb)\fi}
\newcommand{ \fwha  }{\ifmmode {\rm FWHM}\left(\ha\right) \else FWHM(\ha)\fi}
\newcommand{ \ewha  }{\ifmmode {\rm EW}\left(\ha\right) \else EW(\ha)\fi}
\newcommand{ \Lmg   }{\ifmmode L\left(\mgii\right) \else $L\left(\mgii\right)$\fi}
\newcommand{ \fwmg  }{\ifmmode {\rm FWHM}\left(\mgii\right) \else FWHM(\mgii)\fi}
\newcommand{ \Lciv  }{\ifmmode L\left(\civ\right) \else $L\left(\civ\right)$\fi}
\newcommand{ \fwciv }{\ifmmode {\rm FWHM}\left(\civ\right) \else FWHM(\civ)\fi}
\newcommand{ \fwhm  }{\ifmmode {\rm FWHM} \else FWHM\fi} 
\newcommand{ \voff  }{\ifmmode v_{\rm off} \else $v_{\rm off}$\fi} 
\newcommand{ \vmax  }{\ifmmode v_{\rm max} \else $v_{\rm max}$\fi} 

\newcommand{ \mumg  }{\ifmmode \mu\left(\mgii\right) \else $\mu\left(\mgii\right)$\fi}
\newcommand{ \fmg   }{\ifmmode f\left(\mgii\right) \else $f\left(\mgii\right)$\fi}
\newcommand{ \muciv }{\ifmmode \mu\left(\civ\right) \else $\mu\left(\civ\right)$\fi}
\newcommand{ \fciv  }{\ifmmode f\left(\civ\right) \else $f\left(\civ\right)$\fi}

\newcommand{  \auvo     }{\ifmmode \alpha_{\nu,{\rm UVO}} \else $\alpha_{\nu,{\rm UVO}}$\fi}
\newcommand{  \Ledd     }{\ifmmode L_{\rm Edd} \else $L_{\rm Edd}$\fi}
\newcommand{  \lamLlam  }{\ifmmode \lambda L_{\lambda} \else $\lambda L_{\lambda}$\fi}
\newcommand{  \lLl      }{\ifmmode \lambda L_{\lambda} \else $\lambda L_{\lambda}$\fi}
\newcommand{  \nuLnu    }{\ifmmode \nu L_{\nu} \else $\nu L_{\nu}$\fi}
\newcommand{  \nLn      }{\ifmmode \nu L_{\nu} \else $\nu L_{\nu}$\fi}
\newcommand{  \Luv      }{\ifmmode L_{1450} \else $L_{1450}$\fi}
\newcommand{  \Lop      }{\ifmmode L_{5100} \else $L_{5100}$\fi}
\newcommand{  \lLop     }{\ifmmode \log\left(\Lop/\ergs\right) \else $\log\left(\Lop/\ergs\right)$\fi}
\newcommand{  \Lthree   }{\ifmmode L_{3000} \else $L_{3000}$\fi}
\newcommand{  \lLthree  }{\ifmmode \log\left(\Lthree/\ergs\right) \else $\log\left(\Lthree/\ergs\right)$\fi}
\newcommand{  \Lsix      }{\ifmmode L_{6200} \else $L_{6200}$\fi}
\newcommand{  \lLisx     }{\ifmmode \log\left(\Lop/\ergs\right) \else $\log\left(\Lop/\ergs\right)$\fi}
\newcommand{  \Lxray    }{\ifmmode L_{\rm X} \else $L_{\rm X}$\fi}
\newcommand{  \Lhard    }{\ifmmode L_{\rm 2-10} \else $L_{\rm 2-10}$\fi}
\newcommand{  \Lsoft    }{\ifmmode L_{\rm 0.5-2} \else $L_{\rm 0.5-2}$\fi}

\newcommand{\Fthree}{\ifmmode F_{3000} \else $F_{3000}$\fi}
\newcommand{\fuv}{\ifmmode f_{\lambda}\left(1450{\rm \AA}\right) \else $f_{\lambda}\left(1450 {\rm \AA}\right)$\fi}
\newcommand{\fthree}{\ifmmode f_{\lambda}\left(3000{\rm \AA}\right) \else $f_{\lambda}\left(3000{\rm \AA}\right)$\fi}
\newcommand{\fH}{\ifmmode f_{\lambda}\left(1.65\micron\right) \else
$f_{\lambda}\left(1.65\micron\right)$\fi}

\newcommand{\fbol}{\ifmmode f_{\rm bol} \else $f_{\rm bol}$\fi}
\newcommand{\fbolwv}{\ifmmode f_{\rm bol}\left(\lambda\right) \else $f_{\rm bol}\left(\lambda\right)$\fi}
\newcommand{\fbolopt}{\ifmmode f_{\rm bol}\left(5100{\rm \AA}\right) \else $f_{\rm bol}\left(5100{\rm \AA}\right)$\fi}
\newcommand{\fbolthree}{\ifmmode f_{\rm bol}\left(3000{\rm \AA}\right) \else $f_{\rm bol}\left(3000{\rm \AA}\right)$\fi}
\newcommand{\fboluv}{\ifmmode f_{\rm bol}\left(1450{\rm \AA}\right) \else $f_{\rm bol}\left(1450{\rm \AA}\right)$\fi}

\newcommand{\fobs}{\ifmmode f_{\rm obs} \else $f_{\rm obs}$\fi}

\newcommand{  \mbh      }{\ifmmode M_{\rm BH} \else $M_{\rm BH}$\fi}
\newcommand{  \lmbh     }{\ifmmode \log\left(\mbh/\Msun\right) \else $\log\left(\mbh/\Msun\right)$\fi} 
\newcommand{  \lledd    }{\ifmmode L/L_{\rm Edd} \else $L/L_{\rm Edd}$\fi}
\newcommand{  \mmedd    }{\ifmmode \dot{m}/\dot{m}_{\rm \,Edd} \else $\dot{m}/\dot{m}_{\rm \,Edd}$\fi}
\newcommand{  \Lbol     }{\ifmmode L_{\rm bol} \else $L_{\rm bol}$\fi}
\newcommand{  \lbol     }{\ifmmode L_{\rm bol} \else $L_{\rm bol}$\fi}
\newcommand{  \lLbol    }{\ifmmode \log\left(\Lbol/\ergs\right) \else $\log\left(\Lbol/\ergs\right)$\fi} 
\newcommand{  \Lagn     }{\ifmmode L_{\rm AGN} \else $L_{\rm AGN}$\fi}
\newcommand{  \lagn     }{\ifmmode L_{\rm AGN} \else $L_{\rm AGN}$\fi}

\newcommand{  \tgrow     }{\ifmmode t_{\rm growth} \else $t_{\rm growth}$\fi}
\newcommand{  \tUni      }{\ifmmode t_{\rm Universe} \else $t_{\rm Universe}$\fi}

\newcommand{  \Mindot	}{\ifmmode \dot{M}_{\rm infall} \else $\dot{M}_{\rm infall}$\fi}
\newcommand{  \Mbhdot	}{\ifmmode \dot{M}_{\rm BH} \else $\dot{M}_{\rm BH}$\fi}

\newcommand{  \Maddot	}{\ifmmode \dot{M}_{\rm AD} \else $\dot{M}_{\rm AD}$\fi}

\newcommand{  \Mdot	}{\ifmmode \dot{M} \else $\dot{M}$\fi}

\newcommand{  \as	}{\ifmmode a_{\rm *} \else $a_{\rm *}$\fi}
\newcommand{  \avec	}{\ifmmode \vec{a}_{\rm *} \else $\vec{a}_{\rm *}$\fi}
\newcommand{  \re	}{\ifmmode \eta      	 \else $\eta$\fi}
\newcommand{  \RISCO	}{\ifmmode R_{\rm ISCO}  \else $R_{\rm ISCO}$\fi}
\newcommand{  \rg	}{\ifmmode r_{\rm g}  \else $r_{\rm g}$\fi}
\newcommand{  \rS	}{\ifmmode r_{\rm S}  \else $r_{\rm S}$\fi}

\newcommand{  \mseed    }{\ifmmode M_{\rm seed} \else $M_{\rm seed}$\fi}
\newcommand{  \mbul     }{\ifmmode M_{\rm Bulge} \else $M_{\rm Bulge}$\fi} 
\newcommand{  \mstar    }{\ifmmode M_{*} \else $M_{*}$\fi} 
\newcommand{  \mgal     }{\ifmmode M_{*} \else $M_{*}$\fi} 
\newcommand{  \mhost    }{\ifmmode M_{\rm Host} \else $M_{\rm Host}$\fi}
\newcommand{  \mmsmall  }{\ifmmode M_{\rm BH}/M_{*} \else $M_{\rm BH}/M_{*}$\fi}
\newcommand{  \mmlarge  }{\ifmmode M_{*}/M_{\rm BH} \else $M_{*}/M_{\rm BH}$\fi}

\newcommand{  \mmwp     }{\ifmmode \left(M_{*}/M_{\rm BH}\right) \else $\left(M_{*}/M_{\rm BH}\right)$\fi}
\newcommand{  \ml       }{\ifmmode M_{*}/L_{*} \else $M_{*}/L_{*}$\fi}
\newcommand{  \mlwp     }{\ifmmode \left(M_{*}/L\right) \else $\left(M_{*}/L\right)$\fi}
\newcommand{  \mlk      }{\ifmmode \left(M_{*}/L_{K}\right) \else $\left(M_{*}/L_{K}\right)$\fi}
\newcommand{  \sigs     }{\ifmmode \sigma_{*} \else $\sigma_{*}$\fi}
\newcommand{  \Reff     }{\ifmmode R_{\rm e} \else $R_{\rm e}$\fi}
\newcommand{  \nser     }{\ifmmode n_{\rm s} \else $n_{\rm s}$\fi}
\newcommand{  \LFIR     }{\ifmmode L_{\rm FIR} \else $L_{\rm FIR}$\fi}
\newcommand{  \Lfir     }{\ifmmode L_{\rm FIR} \else $L_{\rm FIR}$\fi}

\newcommand{  \mdyn     }{\ifmmode M_{\rm dyn} \else $M_{\rm dyn}$\fi} 
\newcommand{  \mgas     }{\ifmmode M_{\rm gas} \else $M_{\rm gas}$\fi} 
\newcommand{  \mh       }{\ifmmode M_{\rm h} \else $M_{\rm h}$\fi}
\newcommand{  \mhalo    }{\ifmmode M_{\rm halo} \else $M_{\rm halo}$\fi}

\newcommand{  \Lcii     }{\ifmmode L_{\cii} \else $L_{\cii}$\fi}

\newcommand{\bj}{\ifmmode b_{\rm J} \else $b_{\rm J}$\fi}

\newcommand{\iab}{\ifmmode i_{\rm AB} \else $i_{\rm AB}$\fi}

\newcommand{\jab}{\ifmmode J_{\rm AB} \else $J_{\rm AB}$\fi}
\newcommand{\hab}{\ifmmode H_{\rm AB} \else $H_{\rm AB}$\fi}
\newcommand{\kab}{\ifmmode K_{\rm AB} \else $K_{\rm AB}$\fi}

\newcommand{\jveg}{\ifmmode J_{\rm Vega} \else $J_{\rm Vega}$\fi}
\newcommand{\hveg}{\ifmmode H_{\rm Vega} \else $H_{\rm Vega}$\fi}
\newcommand{\kveg}{\ifmmode K_{\rm Vega} \else $K_{\rm Vega}$\fi}

\newcommand{  \Chisq    }{\ifmmode \chi^{2} \else $\chi^{2}$}
\newcommand{  \nelec    }{\ifmmode n_{e} \else $n_{e}$\fi}     
\newcommand{  \nh       }{\ifmmode n_{H} \else $n_{H}$\fi}     
\newcommand{  \Ncol     }{\ifmmode N_{col} \else $N_{col}$\fi} 
\newcommand{  \NH       }{\ifmmode N_{H} \else $N_{H}$\fi}     

\newcommand{\MgasCO}{\ifmmode M_{\rm gas, \, CO} \else $M_{\rm gas, \, CO}$\fi}
\newcommand{\Mgasdust}{\ifmmode M_{\rm gas, \, dust} \else $M_{\rm gas, \, dust}$\fi}
\newcommand{\lMgasCO}{\relax\ifmmode \log \left( M_{\rm gas, \, CO} \right) \else $ \log \left( M_{\rm gas, \, CO}  \right)$\fi}
\newcommand{\lMgasdust}{\ifmmode  \log \left( M_{\rm gas, \, dust} \right) \else $ \log \left( M_{\rm gas, \, dust} \right)$\fi}
\newcommand{\Mdust}{\ifmmode M_{\rm dust} \else $M_{\rm dust}$\fi}
\newcommand{\lMdust}{\ifmmode  \log \left( M_{\rm dust} \right) \else $ \log \left( M_{\rm dust} \right)$\fi}

\newcommand{\Hmol }{\ifmmode {\rm H_2} \else ${\rm H_2}$\fi}

\def\deg{\hbox{$^\circ$}}

\def\micron{\hbox{$\mu$m}}
\def\ion#1#2{#1$\;${\small\rm\@Roman{#2}}\relax}

\LetLtxMacro\oldcitep\citep 
\RenewDocumentCommand{\citep}{O{} O{} m}{\oldcitep[#1][#2]{#3}}
\NewDocumentCommand{\citex}{O{} O{} m}{\oldcitep{#3}}

\LetLtxMacro\oldcitet\citet 
\RenewDocumentCommand{\citet}{O{} O{} m}{\oldcitet[#1][#2]{#3}}

\newcommand\blfootnote[1]{%
  \begingroup
  \renewcommand\thefootnote{}\footnote{#1}%
  \addtocounter{footnote}{-1}%
  \endgroup
}

\title[Dust attenuation and star-dust geometry of SFGs]{Dust attenuation, dust content and geometry of star-forming galaxies}

\author[Zhang et al.]{
Junkai Zhang$^{1,{\color{blue} \star}}$,
Stijn Wuyts$^{1}$,
Sam E. Cutler$^{2}$,
Lamiya A. Mowla$^{3}$,
\newauthor
Gabriel B. Brammer$^{4,5}$,
Ivelina G. Momcheva$^{6}$,
Katherine E. Whitaker$^{7,4}$,
\newauthor
Pieter van Dokkum$^{8}$,
Natascha M. F\"{o}rster Schreiber$^{9}$,
Erica J. Nelson$^{10}$,
\newauthor
Patricia Schady$^{1}$,
Carolin Villforth$^{1}$,
David Wake$^{11}$,
Arjen van der Wel$^{12}$
\\
$^{1}$ Department of Physics, University of Bath, Claverton Down, Bath, BA2 7AY, UK\\
$^{2}$ Department of Astronomy, University of Massachusetts, Amherst, MA 01003, USA\\
$^{3}$ Dunlap Institute for Astronomy and Astrophysics, University of Toronto, 50 St. George Street, Toronto, ON M5S 3H4, Canada\\
$^{4}$ Cosmic Dawn Center (DAWN), Copenhagen, Denmark\\
$^{5}$ Niels Bohr Institute, University of Copenhagen, Lyngbyvej 2, DK-2100 Copenhagen, Denmark\\
$^{6}$ Max-Planck-Institut f\"{u}r Astronomie, K\"{o}nigstuhl 17, D-69117 Heidelberg, Germany\\
$^{7}$ Department of Astronomy, University of Massachusetts, Amherst, MA 01003, USA\\
$^{8}$ Astronomy Department, Yale University, New Haven, CT 06511, USA\\
$^{9}$ Max-Planck-Institut f\"{u}r extraterrestrische Physik, Giessenbachstrasse 1, D-85748 Garching, Germany\\
$^{10}$ Department for Astrophysical and Planetary Science, University of Colorado, Boulder, CO 80309, USA\\
$^{11}$ Department of Physics and Astronomy, University of North Carolina Asheville, 1 University Heights, Asheville, NC 28804, USA\\
$^{12}$ Sterrenkundig Observatorium, Universiteit Gent, Krijgslaan 281 S9, B-9000 Gent, Belgium
}

\date{Accepted 2023 June 09; in original form 2023 March 08}

\pubyear{2023}

\begin{document}
\label{firstpage}
\pagerange{\pageref{firstpage}--\pageref{lastpage}}
\maketitle
\raggedbottom  

\begin{abstract}
We analyse the joint distribution of dust attenuation and projected axis ratios, together with galaxy size and surface brightness profile information, to infer lessons on the dust content and star/dust geometry within star-forming galaxies at $0<z<2.5$.  To do so, we make use of large observational datasets from KiDS+VIKING+HSC-SSP and extend the analysis out to redshift $z = 2.5$ using the HST surveys CANDELS and 3D-DASH.  We construct suites of SKIRT radiative transfer models for idealized galaxies observed under random viewing angles with the aim of reproducing the aforementioned distributions, including the level and inclination dependence of dust attenuation. We find that attenuation-based dust mass estimates are at odds with constraints from far-infrared observations, especially at higher redshifts, when assuming smooth star and dust geometries of equal extent. We demonstrate that UV-to-near-IR and far-infrared constraints can be reconciled by invoking clumpier dust geometries for galaxies at higher redshifts and/or very compact dust cores. We discuss implications for the significant wavelength- and redshift-dependent differences between half-light and half-mass radii that result from spatially varying dust columns within -especially massive- star-forming galaxies.

\end{abstract}


\section{Introduction}
\label{sec:Intro}

\blfootnote{ {\color{blue} $\star$} {E-mail: jz2192@bath.ac.uk}}

The primary source of astronomical analysis, starlight, is unavoidably influenced by attenuation due to dust in the interstellar medium (ISM). In order to characterise physical properties of galaxies, such as their stellar mass and star formation rate (SFR), astronomers therefore need to consider dust attenuation to correct the observed luminosities and spectral energy distributions (SEDs).  Dust is also an important component to galaxies in its own right.  Besides reprocessing radiation, it provides shielding and acts as a catalyst for the formation of molecules.  A number of comprehensive review articles, with focuses ranging from the characteristics of dust grains to the bulk dust properties of galaxies and the complexities of radiative transfer, have captured progress in our understanding of dust-related phenomena over the past two decades \citep{Draine2003, Steinacker2013, Galliano2018, Salim2020}.

Broadly speaking, inferences on dust in galaxies are either derived from the UV-to-near-IR part of the spectral energy distribution, or from observations in the far-IR.\footnote{For convenience, we will frequently use the shorthand far-IR throughout this paper to refer to the full far-infrared to (sub)millimetre wavelength range where diagnostics of the cold ISM can be observed.}  The former is where the net dimming and reddening effects of absorption and scattering are imprinted, themselves a function of grain properties as well as geometry.  The latter is where the emission reprocessed by dust emerges and molecular gas tracers provide an additional probe of the ISM.  Approaches that combine the panchromatic information under the condition that the energy absorbed matches the energy re-emitted have been developed as well \citep[e.g.,]{daCunha2008,Boquien2019}.  The latter inherently adopt an angle-averaged approach, effectively assuming the emission to be isotropic at all wavelengths, an ansatz that ultimately may not hold at short wavelengths \citep[see, e.g.,]{Qin2022} and for dusty starbursts has even been questioned for longer wavelengths \citep{Lovell2022}.  The consistency (or apparent inconsistencies) between attenuation-based and far-IR based inferences on dust have also been considered by \citet{Shapley2022}, and will be a key focus of our analysis. 

More pragmatically though, the desired rich panchromatic sampling of SEDs from UV to far-IR wavelengths is not generally available for the average object in mass-selected samples of galaxies spanning appreciable lookback times.  For this reason, significant effort has been devoted to casting far-IR/sub-mm observational datasets in the form of scaling relations that allow the dust and molecular gas mass of galaxies to be estimated as a function of more readily available parameters, such as their redshift, stellar mass and star formation rate \citep[e.g.,]{Magdis2012, Santini2014, Bethermin2015, Berta2016, Tacconi2018, Tacconi2020}.\footnote{While parameterised for convenience as a function of the more readily available parameters, this does not imply the latter are necessarily to be considered the independent variables in the context of causality.  We refer the reader to \citet{Baker2022} for a discussion on which scaling relations are more fundamental.} 

On the shorter wavelength end, on the other hand, we do have the advantage that basic estimates of the visual attenuation ($A_V$) extracted from UV-to-near-IR SEDs can be obtained for tens of thousands of individual objects, as a by-product from stellar population modelling, albeit under relatively simplified assumptions \citep[see, e.g.,][for an overview of the intricacies of stellar population synthesis]{Conroy2013}.  When paired with basic structural measurements (galaxy sizes and projected axis ratios) these diagnostics can be modelled at the population level, following the ansatz that as observers we are studying ensembles of galaxies observed under random viewing angles.  This is the approach of our paper.

Without the consideration of dust, applications of such population modelling, tackling the inversion problem of projected axial ratio (and in some cases size) distributions in order to characterise intrinsic 3D shapes have been presented for quiescent galaxies \citep[e.g.,]{van2009,Holden2012,Chang2013,Hill2019,Satoh2019,Zhang2022}, as well as star-forming galaxies (SFGs).  The latter class are not necessarily flattened axisymmetric disks at all masses and redshifts.  Specifically, there is an emerging notion that prolate shapes contribute significantly among low-mass, high-redshift SFGs \citep{van2014a, Zhang2019}.  This process of structural disk settling finds its counterpart in observational trends of dynamical disk settling, with lower mass SFGs gaining a rotation-dominated status at later cosmic times \citep{Kassin2012, Simons2017, Wisnioski2019}.

Once folding in dust, its presence can result in a relation between the SED reddening on the one hand and galaxy projected shapes and sizes as observed in a particular waveband on the other hand.  This inter-relation can take two forms.  

First, more inclined disk galaxies can suffer enhanced dust reddening due to the thicker projected dust columns.  \citet{Wild2011} exploit this effect for an analysis of attenuation laws in nearby galaxies.  \citet{Patel2012} illustrate that the effect is seen over a range in lookback times, and \citet{Zuckerman2021} compose a simple model to argue that inclination, rather than bulk dust mass, is largely responsible for explaining the spread of similar-mass SFGs across the rest-frame $UVJ$ colour-colour space.  Assumptions made in the latter analysis include an approximation of SFG shapes as axisymmetric, and an inclination-dependent scaling of $A_V$ that is proportional to the length of the sightline through the galaxy (i.e., neglecting emission from stars that are mixed with the dust).

A second connection between dust reddening and observed structural parameters can arise in the presence of spatially varying dust columns.  Owing to larger central dust columns, the net effect in such a scenario would typically result in half-light radii that exceed the half-mass radius of the stellar distribution.  Empirical evidence for such size differences, mass-to-light ratio gradients and significant dust contributions to them have been revealed on the basis of multi-band HST imaging \citep[e.g.,]{Wuyts2012, Liu2017, Suess2019}, resolved grism spectroscopic measurements of the Balmer decrement \citep{Nelson2016}, and have been augmented recently by early JWST analyses \citep{Suess2022, Miller2022}.

In this paper, we leverage the number statistics provided by a wedding cake of wide-area and deep surveys with high imaging quality yielding size and axial ratio ($q$) measurements.  Their associated multi-band coverage allows the estimation of redshifts, stellar masses, star formation rates and attenuation levels. We describe the datasets and sample in Section\ \ref{data.sec}, and provide more details on inferred galaxy properties in Section\ \ref{physprop.sec}.  Overall, our sample comprises 439,965 SFGs spanning a range in stellar mass from $10^9$ to $10^{11.5}\ M_{\odot}$ and a redshift range $0 < z < 2.5$.  In Section\ \ref{methodology.sec}, we lay out the methodology.  We describe how intrinsic 3D shapes are inferred from projected axial ratio distributions.  We introduce the SKIRT code used to compute radiative transfer on toy model galaxies, and illustrate lessons learned from analysing such toy models. Next, we lay out the Bayesian framework to combine both aspects in a joint modelling of the distribution of structural and attenuation properties of SFGs of a particular mass and redshift.  Section\ \ref{results.sec} presents our results in the following steps: Section\ \ref{Avobs.sec} considers the observational trends in $A_V$ -- $q$ space as a function of mass and redshift.  In Section\ \ref{smoothidenticalmodel.sec}, we apply a first round of population modelling, adopting simplified star and dust geometries (smooth profiles with identical spatial distributions for stars and dust), arriving at the conclusion that dust masses inferred from such an approach are at odds with those inferred from far-IR scaling relations.  We next impose the far-IR constraints on dust content as priors and relax the assumptions regarding identical and smooth star-dust geometries (Section\ \ref{altgeometry.sec}).  Specifically, we consider the effect of potential differences between dust and stellar scale-heights/lengths, and demonstrate how clumpier dust geometries at higher redshifts can reconcile the attenuation and far-IR results.  Implications for the size measurements of galaxies, with emphasis on the relation between half-light and half-stellar-mass radii, are discussed in Section\ \ref{size.sec}.  We summarize our results in Section\ \ref{summary.sec}.

Throughout the paper, we adopt a \citet{Chabrier2003} stellar initial mass function (IMF) and a flat $\Lambda$CDM cosmology with $\Omega_{\Lambda} = 0.7$, $\Omega_m = 0.3$ and $H_0 = 70\ \rm{km}\ \rm{s}^{-1}\ \rm{Mpc}^{-1}$.


\section{Data and sample selection}
\label{data.sec}

Our galaxy sample contains 439,965 star-forming galaxies in a wide range of redshift ($0.0<z<2.5$) and stellar mass ($9.0<\log(M_*/M_\odot)<11.5$). We take advantage of the complementary qualities provided by different lookback surveys in terms of area and depth by combining them. Table \ref{tab:combinedsurveys} indicates which surveys we use for a particular redshift and stellar mass bin. Briefly speaking, for the low-redshift and high-mass galaxies, we exploit the overlap between the wide-area HSC-SSP survey and the KiDS+VIKING multi-wavelength catalogs. For high-redshift and low-mass galaxies, sources come from deep fields with near-infrared Hubble Space Telescope (HST) imaging and rich ancillary multi-wavelength coverage: CANDELS+3D-HST and 3D-DASH+COSMOS2020.  The subset of galaxies at $z > 1$ counts 27,685 SFGs.

As in \citet{Zhang2022}, star-forming galaxies are separated from quiescent galaxies (QGs) using a redshift-dependent cut in specific star formation rate: ${\rm SFR}/{M_*} > 1/({3 t_{\rm H}(z)})$, where $t_{\rm H}(z)$ represents the Hubble time at the redshift of the galaxy under consideration.  Details on the SED modelling are provided in Section\ \ref{SEDmodelling.sec}.

\begin{table} 
\centering 
\resizebox{\linewidth}{!}{%
\begin{tabular}{c|ccccc} 
\hline \hline
Stellar mass  & \multicolumn{5}{c}{Redshift} \\
$\log(M_*/M_\odot)$ & $0.0-0.5$ & $0.5-1.0$ & $1.0-1.5$ & $1.5-2.0$ & $2.0-2.5$ \\
\hline 
$9.0-9.5$ & C+3D & C & C & C & C \\
 & 3311 & 3351 & 3945 & 5264 & 3056 \\
$9.5-10.0$ & C+3D & C+3D & C & C & C \\
 & 1961 & 8401 & 2152 & 2537 & 2179 \\
$10.0-10.5$ & K & K+C+3D & C & C & C \\
 & 176021 & 108178 & 1243 & 1141 & 1035 \\
$10.5-11.0$ & K & K+C+3D & C+3D & C & C \\
 & 56668 & 46796 & 2691 & 566 & 449 \\
$11.0-11.5$ & K & K+C+3D & C+3D & C+3D & C+3D \\
 & 3719 & 3874 & 501 & 538 & 388 \\

\hline \hline  
\end{tabular}%
}
\caption{Surveys used for different stellar mass and redshift regimes. The abbreviations K, C and 3D stand for KiDS+VIKING+HSC-SSP, CANDELS+3D-HST and 3D-DASH+COSMOS2020, respectively. Numbers of SFGs are quoted on the second line for each $(z, M_*)$ bin.}
\label{tab:combinedsurveys}
\end{table}

\subsection{KiDS, VIKING and HSC-SSP}
\label{KiDS_VIKING_HSC.sec}

The fourth data release of the Kilo-Degree Survey (KiDS, \citealt{Kuijken2019}) provides optical wide-field imaging with OmegaCAM on the ESO VLT Survey Telescope (\citealt{Capaccioli2011}, \citealt{Capaccioli2012}) in four filters ($ugri$). The VISTA Kilo-degree Infrared Galaxy Survey (VIKING, \citealt{Edge2013}) is highly overlapping with KiDS, and provides complementary near-infrared ($zYJHK_s$) coverage. The joint $ugrizYJHK_s$ nine band photometry from KiDS+VIKING serves as input to our redshift estimation and SED modelling. On the other hand, the Hyper Suprime-Cam Subaru Strategic Program (HSC-SSP, \citealt{Aihara2019}) is a wide-field optical imaging survey using the 8.2 m Subaru Telescope. The depth of the $0\farcs 6$ resolution $i$-band imaging in the HSC-SSP Wide layer is 25.9 magnitude ($5\sigma$ limits within two arcsec diameter apertures). The high signal-to-noise images provide us with measurements of the projected axial ratio $q$ and semi-major axis half-light radius $a$. The cross-matched KiDS+VIKING+HSC-SSP catalogue covers a few disjoint regions positioned along a long strip in the sky between $129^{\circ}$ and $227^{\circ}$ in longitude and $-2.3^{\circ}$ to $3.0^{\circ}$ in latitude, making for a total of 257 square degrees with 9-band photometric coverage.

\subsection{CANDELS and 3D-HST}
\label{CANDELS.sec}
The Cosmic Assembly Near-infrared Deep Extragalactic Legacy Survey (CANDELS, \citealt{Grogin2011}) is a deep extragalactic Treasury programme carried out with the Hubble Space Telescope (HST). It contains five disjoint fields: GOODS-N, GOODS-S, COSMOS, EGS and UDS, covering the combined 0.25 square degrees area to a $5\sigma$ point source limit of $H = 27.0 $ mag and imaging over 250,000 distant galaxies.  Specifically, we make use of the $U$-to-$8\mu \rm{m}$ multi-wavelength photometric catalogs and grism+photometric redshifts compiled for these fields by the 3D-HST team \citep{Skelton2014, Momcheva2016}.

\subsection{3D-DASH and COSMOS2020}
\label{3D-DASH_COSMOS.sec}

The 3D-Drift And SHift program (3D-DASH, \citealt{Mowla2022}) is the broadest near-infrared HST survey to date, extending the HST near-infrared imaging to 1.43 square degrees at a median depth of $H = 24.74$ mag. 3D-DASH overlaps with the Cosmic Evolution Survey (COSMOS, \citealt{Scoville2007}) and its mosaic ingests any pre-existing HST/WFC3 imaging in this field. COSMOS2020 (\citealt{Weaver2022}) is the latest release of the COSMOS multi-wavelength catalogue. It collects UV-optical-IR data from various telescopes/surveys (GALEX, CFHT/MegaCam, HST/ACS, Subaru/HSC, VISTA/VIRCAM, Spitzer/IRAC). Together, the cross-matched 3D-DASH+COSMOS2020 catalogue benefits from the high-resolution imaging of 3D-DASH for galaxy shape and size measurements and the consistent multi-band photometry in COSMOS2020 for the SED modelling.  Thanks to its wide area, it sensitively increases the number statistics compared to CANDELS at the bright end.

\section{Physical properties from observations}
\label{physprop.sec}
This Section summarizes the procedures used to obtain derived physical properties of galaxies.  Section\ \ref{SEDmodelling.sec} describes the stellar population modelling of rest-UV to rest-NIR photometry.  Section\ \ref{struc.sec} covers the measurement of structural parameters.  Section\ \ref{Mdust.sec} details the FIR scaling relations adopted to infer dust masses.

\subsection{SED modelling}
\label{SEDmodelling.sec}

To maximize consistency across the different observational surveys, we apply the same template-fitting tools for redshift and stellar population modelling to all samples.  EAZY \citep{Brammer2008} was employed to derive photometric redshifts.  When available, these were replaced by spectroscopic redshifts (17.37\% of SFGs in CANDELS; 14.17\% in 3D-DASH; 4.85\% in KiDS+VIKING+HSC-SSP).  

Stellar population modelling was performed using FAST++ \citep{Schreiber2016}\footnote{\url{https://github.com/cschreib/fastpp}}, a C++ version of the SED fitting code FAST \citep{Kriek2009}.  Specifically, the family of star formation histories considered were delayed tau models ($SFR(t) \propto t \exp(-t/\tau)$), with e-folding times down to $\tau = 300\ \rm{Myr}$ and ages since the onset of star formation varying between 50 Myr and the age of the Universe.  The default setting adopted a fixed Solar metallicity and \citet{Calzetti2000} attenuation law with allowed $V$-band dust attenuation in the range $0 < A_V < 4$.  

Statistical uncertainties on physical properties were inferred via Monte Carlo simulations.  Briefly, fluxes were perturbed according to the photometric uncertainties, and redshift estimation plus SED modelling were repeated on each of the 100 perturbed SEDs for a given object.  Across the considered redshift and mass range, the typical statistical uncertainties propagating from photometric errors (and consistently incorporating redshift errors) amount to 0.02 in $\Delta z / (1+z)$, 0.05 dex in stellar mass, 0.13 dex in SFR and 0.15 mag in $A_V$.  Compared to the width of redshift and mass bins adopted for our analysis ($\Delta z = 0.5$ and 0.5 dex in mass, respectively), these uncertainties are small.  While for a given Monte Carlo realisation some objects near bin boundaries may shift into an adjacent bin or drop out/move into the SFG sample, we verified that such effects have minimal impact on the key relations discussed in this paper.

The above Monte Carlo simulations likely underestimate the full uncertainties on the inferred physical properties.  This is illustrated for example when comparing our results for objects in the COSMOS2020 catalogue with redshifts and stellar population properties derived by \citet{Weaver2022} using the independent code and non-identical settings of LePhare \citep{Arnouts2002, Ilbert2006}. This comparison suggests that characteristic uncertainties may rather be on the order of 0.03 in $\Delta z / (1+z)$, 0.10 dex in stellar mass, 0.23 dex in SFR and 0.3 mag in $A_V$.\footnote{We note that the LePhare run adopted a step-size of 0.1 for the colour excess $E(B-V)$, which translates to a relatively crude grid of visual attenuation levels explored ($\Delta A_V \approx 0.4$).}
We comment on alternative settings of the stellar population modelling in Appendix\ \ref{appendix_stelpop.sec}, and argue that these are unlikely to alter the main conclusions drawn in this paper.

\subsection{Structural measurements}
\label{struc.sec}

Half-light radii ($a$), S\'{e}rsic indices ($n$) and projected axis ratios ($q$) in CANDELS \citep{van2012} and 3D-DASH \citep{Cutler2022} are measured in HST's F160W ($H$) band using GALFIT \citep{Peng2002}, which fits two-dimensional S\'{e}rsic surface brightness profiles convolved with the PSF to galaxy images.
Structural parameters for HSC-SSP as quantified on their deepest and highest quality $i$-band imaging are provided as part of the HSC pipeline products \citep{Aihara2019}.  These again take into account PSF convolution, but follow a slightly different algorithm, described in detail by \citet{Bosch2018}.  Briefly, separate S\'{e}rsic fits with $n=1$ and $n=4$ are executed, after which the linear combination of the two that best fits the image is determined.  As such, the pipeline provides bulge-to-total ratios ($B/T$), which we translate where relevant to corresponding single S\'{e}rsic indices.  We implement this mapping empirically by drawing from the $(B/T)_{\rm HSC}$ -- $n_{\rm CANDELS}$ distribution for galaxies in common to both surveys.  As a sanity check, we further verified the quality of HSC axis ratio and size measurements by comparing those available for sources in the COSMOS field (realised at HSC-Wide depth) with in-hand measurements on HST/ACS F814W imaging.  No evidence for systematic offsets in $q$ were found, and any systematics in size measurements were restricted to the 10\% level for SFGs in the mass and redshift range where we exploit the larger number statistics offered by ground-based samples.  We thus conclude that the heterogeneous resolution from combining high image-quality ground-based and diffraction-limited space-based imaging across different masses/redshifts is not impacting our overall analysis.

We note that in our modelling the different wavelengths (and indeed rest-wavelengths) probed across our full sample will be accounted for self-consistently, in the sense that the observed galaxy sizes will be reproduced by model half-light (rather than model half-mass) radii at the appropriate rest-frame wavelength.  We return to the distinction between half-light and half-mass radii in Section\ \ref{size.sec}.  

\subsection{Dust masses from FIR scaling relations}
\label{Mdust.sec}
Sections\ \ref{SEDmodelling.sec} and\ \ref{struc.sec} outline the means to locate each of the SFGs in our sample in the multidimensional $q-\log(a)-\log(n)-A_V$ space.  When we come to modelling the distribution of SFGs across this space, we will require knowledge of the dust content of main sequence SFGs at different masses and redshifts as inferred from FIR observations, either to compare to as a reference, or to implement as a hard prior.

To this end, we apply scaling relations composed by \citet{Tacconi2020} based on a compilation of dust continuum and CO surveys carried out with the {\it Herschel}, ALMA, NOEMA, and IRAM far-IR/sub-mm telescopes, together spanning $\sim 90\%$ of cosmic history and a similar range in mass as explored in this study.  Because \citet{Tacconi2020} express their scaling relations in terms of the {\it molecular gas} content of galaxies, we follow \citet{Tacconi2018} in applying a metallicity-dependent dust-to-gas ratio to obtain the corresponding {\it dust} mass.  For a given redshift $z$ and stellar mass $M_*$ we hence compute the dust mass of a main sequence SFG using the following set of equations:
\begin{equation}
    \begin{array}{l}
         \log({\rm{sSFR_{MS}}}[{\rm{Gyr^{-1}}}]) \equiv \log({\rm{SFR_{MS}}}/M_*[{\rm{Gyr^{-1}}}])=\\
         \indent (-0.16-0.026\times t_c[{\rm{Gyr}}])\times(\log(M_*/M_\odot)+0.025) \\ 
         \indent - (6.51-0.11\times t_c[{\rm{Gyr}}]) + 9\\ 
         {\rm{with}} 
         ~ \log(t_c[{\rm{Gyr}}]) = \\
         \indent 1.143 - 1.026\times \log(1+z) -0.599\times \log^2(1+z) \\
         \indent +0.528\times \log^3(1+z)
    \end{array}
\end{equation}
yields the SFR of a main sequence galaxy according to \citet{Speagle2014}.
\begin{equation}
    \begin{array}{l}
         \log(t_{\rm{depl}}[{\rm{Gyr}}]) \equiv \log(M_{\rm{gas}}/{\rm{SFR}}[{\rm{Gyr}}]) = 0.21  \\
         \indent - 0.98 \times \log(1+z) + 0.03 \times (\log(M_*/M_\odot)-10.7)
    \end{array}
\end{equation}
yields the corresponding molecular gas mass according to \citet{Tacconi2020}, and
\begin{equation}
    \begin{array}{l}
         \delta_{\rm{dg}} \equiv M_{\rm{dust}}/M_{\rm{molgas}}
    = 10^{-2+0.85\times(12+\log({\rm{O/H}})-8.67)}  \\
         {\rm{with}} ~ 12 + \log({\rm{O/H}}) = 8.74 - 0.087 \times (\log(M_*/M_\odot) \\
         \indent  - 10.4 - 4.46 \times \log(1+z) + 1.78 \times \log^2(1+z))^2 
    \end{array}
    \label{dg.eq}
\end{equation}
yields the corresponding dust mass following \citet{Tacconi2018}.

\section{Methodology}
\label{methodology.sec}

This Section introduces how we model the joint distribution of galaxies' structural and dust properties in $q-\log(a)-\log(n)-A_V$ space. Section \ref{intrinsic_shapes.sec} explains how we can derive the intrinsic shape of galaxies from their distribution in the projected shape -- size plane ($q-\log(a)$). Section\ \ref{SKIRT.sec} introduces the functionality of SKIRT, enabling radiative transfer calculations on model galaxies of arbitrary complexity (or simplicity).  For didactic purposes, we illustrate the behaviour in terms of inclination-dependent attenuation and inclination- and wavelength-dependent half-light radii using a select number of toy model galaxies. Finally, we combine the intrinsic shape and radiative transfer modelling and explain in Section\ \ref{model.sec} how we parameterise population models for galaxy ensembles and constrain them with observations.

\subsection{Intrinsic shapes from projected axial ratios}
\label{intrinsic_shapes.sec}

We model any SFG ensemble (e.g., defined within a bin of mass and redshift) as a family of ellipsoids observed from random viewing angles. The shape of any such ellipsoid can be defined by its principle axis ratios $B/A$ (the intermediate-to-major axis ratio) and $C/A$ (the minor-to-major axis ratio).  Observed from a random angle $(\theta, \phi)$ on a sphere, the corresponding projected axis ratio $q$ on the sky is then given by equations (3) - (6) in \citet{Zhang2022}.  When parameterising an absolute scale of such ellipsoid, we will follow SKIRT convention in defining the intrinsic size $A$ as the semi-major axis length of an ellipse that comprises half of the projected stellar mass distribution in the face-on view.  This 2D definition is tightly linked to $A_{\rm 3D}$, the semi-major axis length of the ellipsoid containing half of the 3D stellar mass distribution ($A_{\rm 3D}/A_{\rm 2D} \approx 1.33$, with only percent-level variations across the range in S\'{e}rsic indices considered).

To acccount for the fact that not all SFGs in an ensemble will be of identical shape, we parameterise the aforementioned family of ellipsoids by the mean and standard deviation of a Gaussian intrinsic ellipticity ($E$) and triaxiality ($T$) distribution, $\langle E \rangle$, $\sigma_E$, $\langle T \rangle$, $\sigma_T$. Here, the ellipticity $E$ relates to the ellipsoid's minor-to-major axis ratio $C/A$ as $E = 1 - C/A$. The triaxiality $T$ is defined as $T=(1-B^2/A^2)/(1-C^2/A^2)$. Additionally, we allow for covariance between the ellipticity and size of the galaxy, parameterised by ${\rm{Cov}(E, \log(A))}$.  The intrinsic shape modelling accounts for measurement errors on the projected axial ratios $q$ (see equation 11 in \citealt{Zhang2022}).  We go over the metric being optimized in our modelling in Section\ \ref{model.sec}.

\subsection{Radiative transfer with SKIRT}
\label{SKIRT.sec}
\subsubsection{SKIRT functionality and settings}
\label{SKIRTsettings.sec}

We use SKIRT (\citealt{Camps2020}) to determine the impact of dust content and galaxy structure on the net visual attenuation $A_V$ (i.e., difference between galaxy-integrated $V$-band magnitudes with and without dust), and on galaxies' observed half-light radii.

SKIRT is a state-to-the-art Monte Carlo code that simulates the radiation transfer (scattering, absorption and emission) in dusty media. Its flexible functionality allows running it on outputs from smoothed particle hydrodynamics (SPH) simulations (see, e.g., \citealt{Camps2016} and \citealt{Popping2022} for applications) as well as on idealized toy model galaxies that can be set up in a modular form within the code itself \citep{Camps2015}.\footnote{\url{https://skirt.ugent.be}}  

Here, we apply the latter functionality.  Cameras are set up to observe toy model galaxies under different viewing angles, and at different wavelengths. We do not model the (computationally more expensive) thermal emission from heated dust, as this information, emerging in the infrared, is also lacking for the galaxies in our observational sample. We make use of the THEMIS dust model (The Heterogeneous dust Evolution Model for Interstellar Solids, \citealt{Jones2017}), which specifies the dust composition and grain size distribution.  For simplicity, we work throughout this study with model galaxies that feature homogeneous stellar populations.  Whereas real galaxies may feature stellar population gradients, recent studies based on observations \citep{Miller2022} and on cosmological galaxy formation simulations \citep{Popping2022} suggest that their impact on colour profiles is significantly sub-dominant compared to dust.  

In accordance with Section\ \ref{intrinsic_shapes.sec}, we work with ellipsoidal geometries of varying thickness $C/A$.  We adopt radial density profiles that project to S\'{e}rsic profiles of index $n$, and execute SKIRT runs for $n=0.5$, 1, 2, 4 and 8.  We generate toy models with a variety of dust contents, and where relevant relate dust surface density and dust mass via $\Sigma_{\rm dust} = 0.5 M_{\rm dust} / (\pi A B)$.  In constructing a SKIRT library of toy model galaxies observed under different viewing angles, we further vary the relative dust and stellar disk size ratio, $R_{\rm{dust}}/R_{\rm{star}}$, and the relative scaleheights of the dust and stellar distributions, $C_{\rm{dust}}/C_{\rm{star}}$.  Finally, additional dimensions to the library capture realisations where a fraction of dust ($f_{\rm clump,\ dust}$) and a fraction of stars ($f_{\rm clump,\ star}$) is placed in dense clumps, the location of which is drawn from the overall density profile.  

Together, this makes for a library of over 10,000 toy model galaxies, each observed under an array of 20 different viewing angles from face-on to edge-on.  From it, we compose lookup tables of the effective dust attenuation in the $V$-band, $A_V$, and measurements at different rest-wavelengths of the ratio of half-light to half-mass radii, $R_{\rm light}/R_{\rm mass}$, for each of the library entries. Rather than rerunning the radiative transfer (RT) calculations for each custom galaxy type and viewing angle in each iteration of the Monte Carlo Markov Chain (MCMC) fitting to the observed galaxy distributions (Section\ \ref{model.sec}), we simply interpolate the relevant $A_V$ and $R_{\rm light}/R_{\rm mass}$ values from the above lookup tables.

Due to computational cost, we run the SKIRT library with axi-symmetric disk galaxies, and approximate the $A_V$ of triaxial galaxies ($T\neq0$) viewed from a random angle $(\theta, \phi)$ by the $A_V$ of an equivalent axi-symmetric galaxy with the same dust mass column $\Sigma_{\rm dust,\ projected} = 0.5 M_{\rm dust}/{\rm Area_{projected}} = 0.5 M_{\rm dust}/(\pi a^2 \times q)$, where $a$ and $q$ are the projected semi-major axis and axial ratio of the half-light ellipse.  We verified using select RT runs on triaxial ellipsoids that the error $\sigma(A_V)$ of this approximation is less than 0.02 mag and thus smaller than the precision of 0.05 mag in our modelling of the observed SEDs.  Our library further lacks a separate dimension for the absolute galaxy size, as it was found that results for a galaxy half the size (and otherwise identical parameters) compared to a galaxy with four times ($2^2 \times$) the surface density were consistent.

\begin{figure}
\centering
\includegraphics[width=\linewidth]{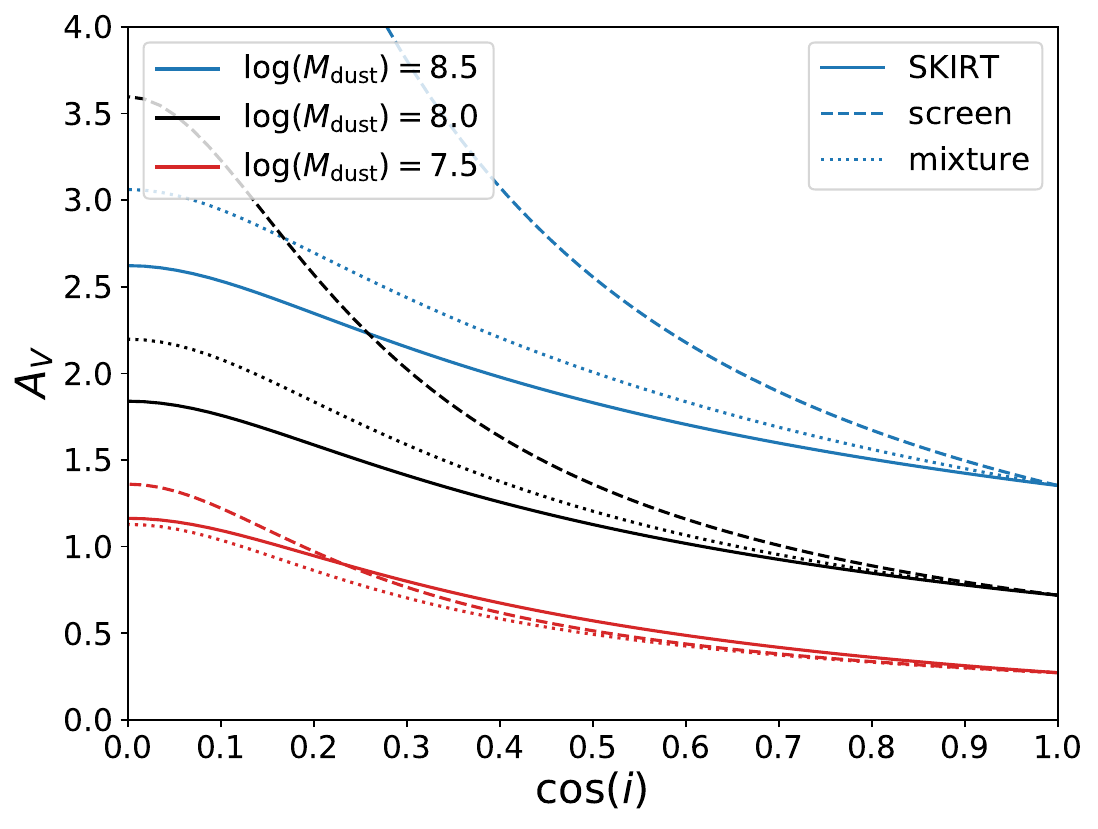}
\caption{Comparison of the inclination-dependent attenuation $A_V$ for toy models with different dust masses. Other galaxy parameters are kept identical to the reference model in black (see main text). Solid lines are the results of SKIRT radiative transfer accounting for scattering and absorption, with identical star and dust geometry as described in the text. Dashed curves represent analytical prescriptions corresponding to a foreground dust configuration, with face-on attenuations matched to the SKIRT results by construction.  Likewise, dotted lines illustrate the analytical case of a homogeneous mixture of stars and dust.
}
\label{fig:mass}
\end{figure}

\subsubsection{Lessons from toy models}
\label{SKIRTlessons.sec}

Before explaining how we will model the observed galaxy populations at different masses and redshifts, we here explore a few toy model galaxies set up in SKIRT (\citealt{Camps2020}) to gain intuition about how different parameters can influence dust attenuation. For example, Figure \ref{fig:mass} compares toy models with different dust masses while keeping other parameters fixed.  In this case, the model galaxies shown are axisymmetric ellipsoids with exponential disk profiles, half-mass radii of 4 kpc, and a thickness $C/A = 0.2$.  The solid curves show the outcome of SKIRT $A_V$ calculations (i.e., the difference between galaxy-integrated $V$-band magnitudes with and without dust) for cameras with different inclinations $i$.  As expected, an increase in dust mass leads to a higher dust attenuation.  Also as anticipated, a galaxy of given dust content is attenuated more when observed closer to edge-on (lower values of $\cos(i)$).  

In detail, the latter scaling with viewing angle differs quantitatively from that of more simplistic analytic prescriptions.  To illustrate this, Figure \ref{fig:mass} shows in dashed and dotted curves the predictions from a foreground dust screen and a homogeneous mixture of dust and stars as outlined by \citet{Calzetti1994}.  In both cases, we anchor the face-on attenuation $A_{V, 0}$ to that computed by SKIRT, and next compute by what factor the attenuation $A_{V, i}$ increases with increasing inclination angle $i$ ($i=0^{\circ}$ being face-on; $i=90^{\circ}$ edge-on).  For the screen prescription, this scaling factor is taken to be equal to the factor by which the length of the sightline through the ellipsoid has increased compared to the face-on view.  In the case of an axisymmetric system, this factor is given by $A_{V, i}/A_{V, 0}=1/\sqrt{(C/A)^2 \sin^2 i + \cos^2 i}$, where $i$ is the inclination angle and $C/A$ is the disk thickness, as applied for example in the analysis by \citet{Zuckerman2021}.  
In the homogeneous mixture configuration, the effective optical depth, $\tau_{\rm eff}$, relates to the full optical depth through the mixture, $\tau$, as $\tau_{\rm eff} = - \ln \left[ (1 - e^{-\tau}) / \tau \right]$.  We thus take the effective face-on attenuation as computed by SKIRT, calculate from it the corresponding full optical depth, scale it up for any arbitrary inclination to account for the increased sightline through the ellipsoid, and convert this back to the effective attenuation for that viewing angle. This yields attenuation levels that increase more slowly with increasing total dust column than the screen case, albeit still not identical to what is computed with SKIRT.  The differences owe to the RT accounting for both absorption and scattering effects, and working on a 3D geometry with radially declining stellar and dust density profiles.  In contrast, the homogeneous mixture considers the effects of absorption on a uniform density slab of mixed stars and dust.  

We conclude from Figure\ \ref{fig:mass} that SKIRT is able to capture inclination-dependent attenuation effects, and that these can differ quantitatively from simpler analytical prescriptions due to more complete physics (absorption + scattering) and the ability to set up more realistic spatial distributions of sources and obscuring material.  For a given geometry, SKIRT can further compute the relation between the input dust mass and the resulting effective attenuation.  Turned around, at fixed dust mass SKIRT can be used to assess how different structural parameters influence attenuation.  This is presented in Appendix\ \ref{appendix_geometry.sec}.
Specifically, Figure\ \ref{fig:toygeometries} considers how the inclination-dependent attenuation changes when varying parameters such as the S\'{e}rsic index $n$, disk thickness $C/A$, dust vs stellar scaleheight or scalelength ratio, or when clumpiness in the dust (and stellar) spatial distribution is introduced.  We note that, for our observational samples, empirical constraints on the S\'{e}rsic index and disk thickness (via the projected axis ratio distribution) are available.  For the remaining structural parameters, observational constraints are either scarce (in the case of $R_{\rm dust}/R_{\rm star}$; e.g., \citealt{Tadaki2017b, Tadaki2020}) or lacking completely ($C_{\rm dust}/C_{\rm star}$ and clumpiness measures). In any case, they are unavailable for the large galaxy ensembles making up our sample.

\begin{figure}
\centering
\includegraphics[width=\linewidth]{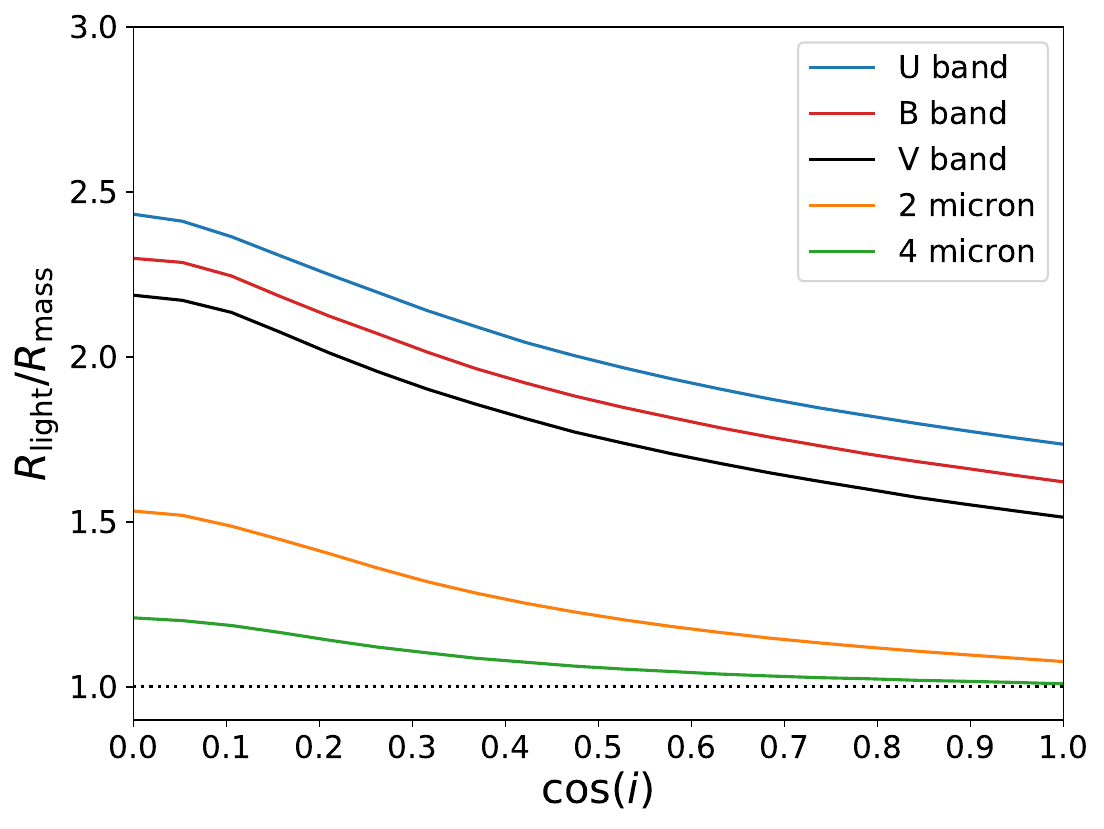}
\caption{Ratio between half-light radius and half-mass radius for the SKIRT reference model galaxy, plotted as a function of inclination, and with colours marking different rest-frame wavelengths.  Centrally enhanced dust columns yield larger half-light radii, especially at shorter rest-wavelengths and for more inclined views.}
\label{fig:R_half}
\end{figure}

One more inference can already be drawn from inspecting the aforementioned reference toy model galaxy.  Since the density profile of both dust and stars declines with increasing distance from the centre, the projected dust column towards inner regions of the galaxy is larger than in its outskirts.  As a result of enhanced central absorption (and scattering), less than half of the light is observed to emerge from within the projected stellar half-mass radius.  Or conversely, the half-light radius of the galaxy exceeds the half-mass radius.  Figure\ \ref{fig:R_half} illustrates how, for our case example, the ratio of half-light to half-mass radius depends on the rest-wavelength of the synthetic observation, and further shows a dependence on inclination, such that relatively speaking the light-weighting effects on galaxy size measurements are more modest for face-on viewing angles ($\cos(i) = 1$) than for edge-on viewing angles ($\cos(i) = 0$).  We emphasize that in this exercise, by construction, all $R_{\rm light}/R_{\rm mass}$ variations and colour gradients implied by it come about due to dust attenuation effects, as no intrinsic stellar population (e.g., age) gradients were introduced.  We further note that, for the galaxy size (4 kpc) and dust content ($10^8\ M_{\odot}$) chosen, rest-optical sizes can exceed the half-mass radii by a factor of $\sim 2$, with such differences only dropping below the $\sim 10\%$ level at wavelengths well into the rest-frame near-infrared.  Of course, the impact of such light-weighting effects on the observed galaxy size could be made arbitrarily high or negligible in a toy model example as this one, by varying the dust content and/or size of the galaxy considered.  We will therefore return to implications on galaxy size measurements informed by the population modelling of our observed samples of SFGs in Section\ \ref{size.sec}.  For now, we remind the reader that for consistency, when fitting to observed galaxy size distributions, we use the SKIRT models' half-light radii quantified at the same rest-wavelengths as probed by the observations.
\\

\subsection{Overall modelling approach and parameterisation}
\label{model.sec}

We develop our model in steps of increasing geometric complexity. Initially, we set up a model of simplified geometry to see how well it matches the observations. Our fitting target is the observed joint distribution of $q-\log(a)-\log(n)-A_V$, where the projected axial ratio $q$, the projected semi-major axis length $a$ and the S\'{e}rsic index $n$ come from the 2D surface brightness fitting, and the visual dust attenuation $A_V$ is obtained from the SED modelling. A given model realisation consists of 10,000 mock galaxies, with their dust and structural parameters (listed below) drawn from Gaussian distributions.  Each of these toy model galaxies is mock-observed under a random viewing angle, making use of the SKIRT lookup tables (see Section\ \ref{SKIRT.sec}) to obtain measures of $A_V$ and $R_{\rm light}/R_{\rm mass}$.

The log-likelihood function is formulated so as to minimize the difference between the distribution in $q-\log(a)-\log(n)-A_V$ space of a given observed galaxy sample and the equivalent distribution of the population of toy model galaxies that is constructed to reproduce their sample properties. In detail, we separate this multi-dimensional space into six parameter planes: $q-\log(a)$, $q-A_V$, $\log(a)-A_V$, $q-\log(n)$, $\log(a)-\log(n)$ and $A_V-\log(n)$. We then calculate the log-likelihood in these parameter planes separately and add them up to obtain a single metric, the total log-likelihood, ${\rm{lnlike}_{tot}}$, which is maximized via the MCMC technique as implemented in {\tt emcee} \citep{Foreman2013}.  Here the log-likelihood function lnlike for an individual 2D parameter plane is calculated as:
\begin{equation}
    \begin{array}{l}
         {\rm{lnlike}} = \sum_i (n_i\times \log(m_i+1)),
    \end{array}
\end{equation}
thus following closely the approach of intrinsic shape modelling in $q -\log(a)$ space as outlined by \citet{Zhang2022}.
For each bin $i$,  $n_i$ is the number of observed galaxies in that bin and $m_i$ is the number predicted by the model with specific parameters.

In our modelling with simplified geometries (Section\ \ref{smoothidenticalmodel.sec}), we allow the following free parameters to define the stellar geometry: $\langle E \rangle$, $\sigma E$, $\langle T \rangle$, $\sigma T$, $\langle \log(A) \rangle$, $\sigma \log(A)$, ${\rm{cov}}(\langle E \rangle, \langle \log(A) \rangle)$, $\langle \log(n) \rangle$, $\sigma \log(n)$ and force the dust geometry to follow an identical spatial distribution as the stars. Additionally, we have two parameters quantifying the Gaussian distribution of dust surface densities among the model galaxy population: $\langle\log(\Sigma_{\rm{dust}})\rangle$ and $\sigma \log(\Sigma_{\rm{dust}})$. At this stage, both dust and stars are assumed to follow smooth and identical distributions.

Once allowing more complex geometries, degeneracies arise between the star-dust geometry and dust mass. From Section\ \ref{altgeometry.sec} onwards, we will therefore impose priors on the dust content inferred from far-IR observations.  To do so, we apply the scaling relations composed by \citet[][see Section\ \ref{Mdust.sec}]{Tacconi2020} to obtain the dust mass of a typical main sequence SFG at the considered mass and redshift.  With the characteristic dust content set according to the above procedure, our modelling no longer features $\langle\log(\Sigma_{\rm{dust}})\rangle$ as a free parameter.  However, we do still leave the freedom to have a distribution in dust properties around the far-IR prior, parameterised by $\sigma \log(\Sigma_{\rm{dust}})$.

At this stage, more freedom will be added to the dust geometry. One model keeps the system smooth (i.e., without clumpiness), but changes the scale-height or radius of the dust disk with respect to the stellar disk using the parameters $C_{\rm{dust}}/C_{\rm{star}}$ and $R_{\rm{dust}}/R_{\rm{star}}$, respectively. An alternative model introduces clumpiness by placing some fraction of dust ($f_{\rm{clump,\ dust}}$) and some fraction of the stellar mass ($f_{\rm{clump,\ star}}$) into clumps.  This alternative model keeps the same height and radius for the dust and star components. We discuss the results obtained with clumpy geometries in depth in Section\ \ref{clumpymodel.sec}.

\begin{figure*}
\centering
\includegraphics[width=\linewidth]{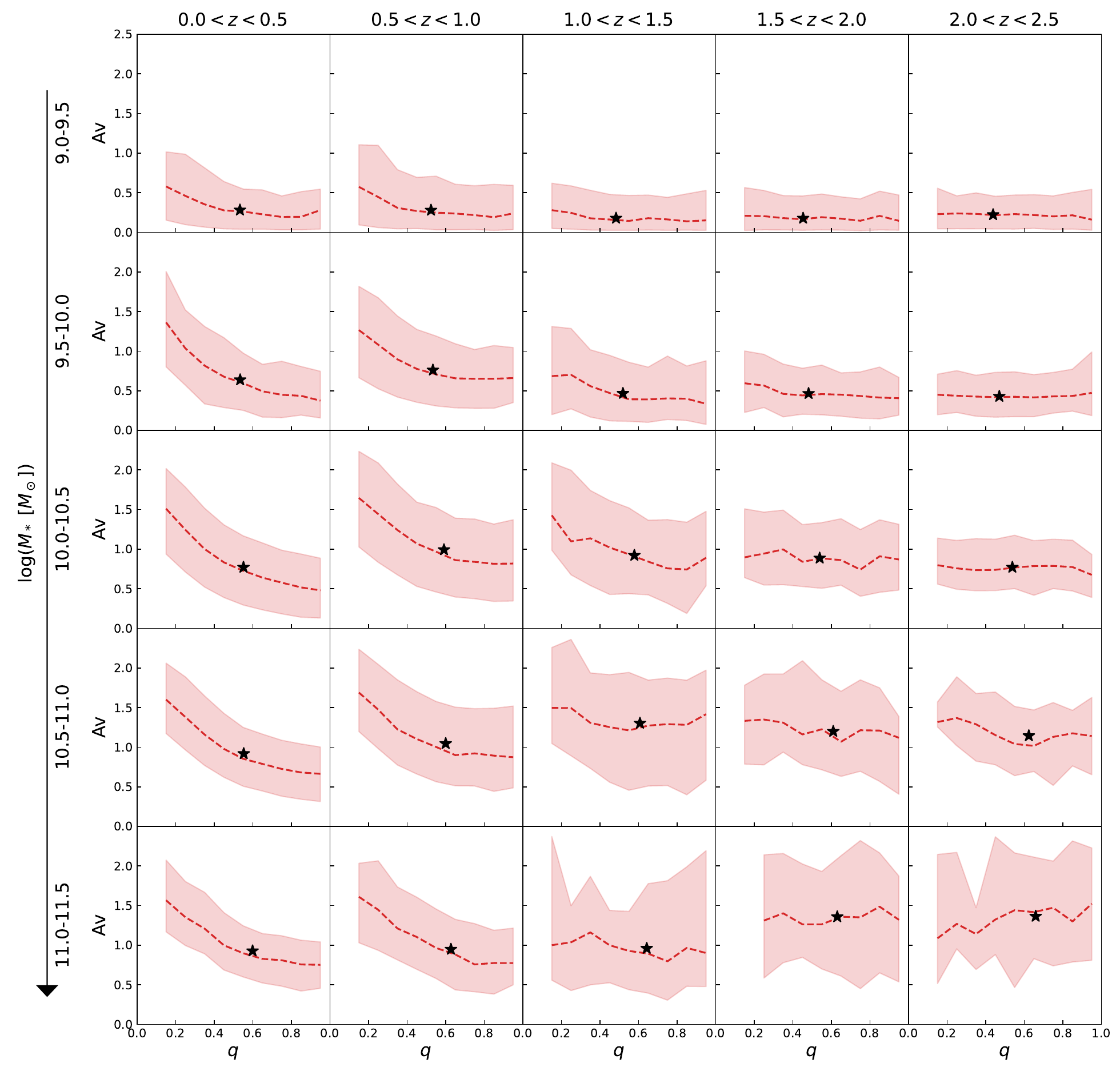}
\caption{Projected axial ratio $q$ versus dust attenuation $A_V$ for the observed SFGs split into different bins of redshift and stellar mass. Dashed lines represent the running median of the $q-A_V$ distribution, while polygons mark the central 68th percentile range. Black stars denote the median of the $(q, A_V)$ values in each $(z, M_*)$ bin. Negative correlations stem from enhanced dust columns and thus increased reddening for more inclined viewing angles.
}
\label{fig:q_Av_SFG_obs}
\end{figure*}

\section{Results}
\label{results.sec}

We now proceed to present the trends of observed attenuation in Section\ \ref{Avobs.sec}, followed by our modelling results in Sections\ \ref{smoothidenticalmodel.sec} to \ref{size.sec}.  Specifically, the results obtained from population modelling with smooth and identical geometries for stars and dust are presented in Section\ \ref{smoothidenticalmodel.sec}.  We then consider two families of alternative star-dust geometries in Section\ \ref{altgeometry.sec}: one where distributions are smooth but distinct in scaleheight or scalelength for stars versus dust (Section\ \ref{diffshapemodel.sec}), and one (our favoured model) in which clumpiness is introduced (Section\ \ref{clumpymodel.sec}).  Finally, Section\ \ref{size.sec} addresses the implications for observed galaxy sizes and their wavelength dependence.  As a visual guide, cartoons of the different considered star-dust geometries are depicted in the top row of Figure\ \ref{fig:summary}. In our summary (Section\ \ref{summary.sec}), we further populate this figure with a schematic overview of our overarching conclusions.

\subsection{Dust attenuation in observations}
\label{Avobs.sec}

In Figure\ \ref{fig:q_Av_SFG_obs}, we consider the relation between the visual attenuation $A_V$ as inferred from SED modelling and the projected axial ratio $q$.  We show the relation separately for samples of SFGs split into five redshift bins spanning the range $0<z<2.5$, and five bins of stellar mass ranging from $10^9\ M_{\odot}$ and $10^{11.5}\ M_{\odot}$.  In several of the panels, a clear and statistically significant negative correlation is observed.  Such an observed trend is reminiscent of the results obtained by \citet{Patel2012} and \citet{Zuckerman2021}, who both noted the enhanced presence of elongated shapes in the region of the rest-frame $UVJ$ colour-colour plane where more dust-obscured SFGs are located \citep[see, e.g.,]{Wuyts2007, Williams2009}.  The interpretation is that those sources with lower $q$ are more elongated in projection on the sky because they are seen under a higher inclination.  Their inclined nature in turn results in a larger projected dust column, which manifests itself in a higher degree of effective dust reddening and attenuation.  Implicitly, this explanation assumes the dust distribution to be relatively diffuse and volume filling, such that dust in the system is not only obscuring the starlight emitted locally, but also that from the background stellar distribution along the line of sight.  From Figure\ \ref{fig:q_Av_SFG_obs} we can further appreciate that at fixed mass the slope of such a negative $q-A_V$ relation is seen to reduce as we consider higher redshifts.  This is despite the fact that SFGs at cosmic noon are known to host a richer ISM than similar-mass counterparts in the local Universe \citep[e.g.,]{Tacconi2020}.

It is also worthwhile to consider how the typical $A_V$ of SFGs varies with redshift and mass (black stars in Figure\ \ref{fig:q_Av_SFG_obs}).  At fixed redshift, we observe the median attenuation to increase with increasing mass.  Indeed, such mass dependence has been previously recovered using a range of obscuration diagnostics for galaxies from the nearby Universe out to $z \sim 4$ \citep{Heinis2014, Pannella2015, McLure2018, Qin2019b}.  In contrast, when considering the longitudinal trend, median $A_V$ values only show a modest variation with redshift.  This is consistent with findings by \citet{Shapley2022}, who juxtaposed a lack of evolution in attenuation tracers to the relatively steep evolution in dust content, at least at the massive end, implied by far-IR observations.  In detail, whereas minor changes in $A_V$ with cosmic time are seen among our sample, they are not necessarily varying in the same direction for different mass bins: if anything, slight increases of $A_V$ with redshift are seen at the massive end, compared to slight declines at lower masses.

Through our population modelling with simplified star-dust geometries (Section\ \ref{smoothidenticalmodel.sec}), we will quantify this apparent tension between attenuation and far-IR results alluded to by \citet{Shapley2022}, after which we will evaluate alternative star-dust geometries in order to reconcile the two (Section\ \ref{altgeometry.sec}).

\subsection{Smooth models with identical star and dust geometries}
\label{smoothidenticalmodel.sec}

\begin{figure}
\centering
\includegraphics[width=\linewidth]{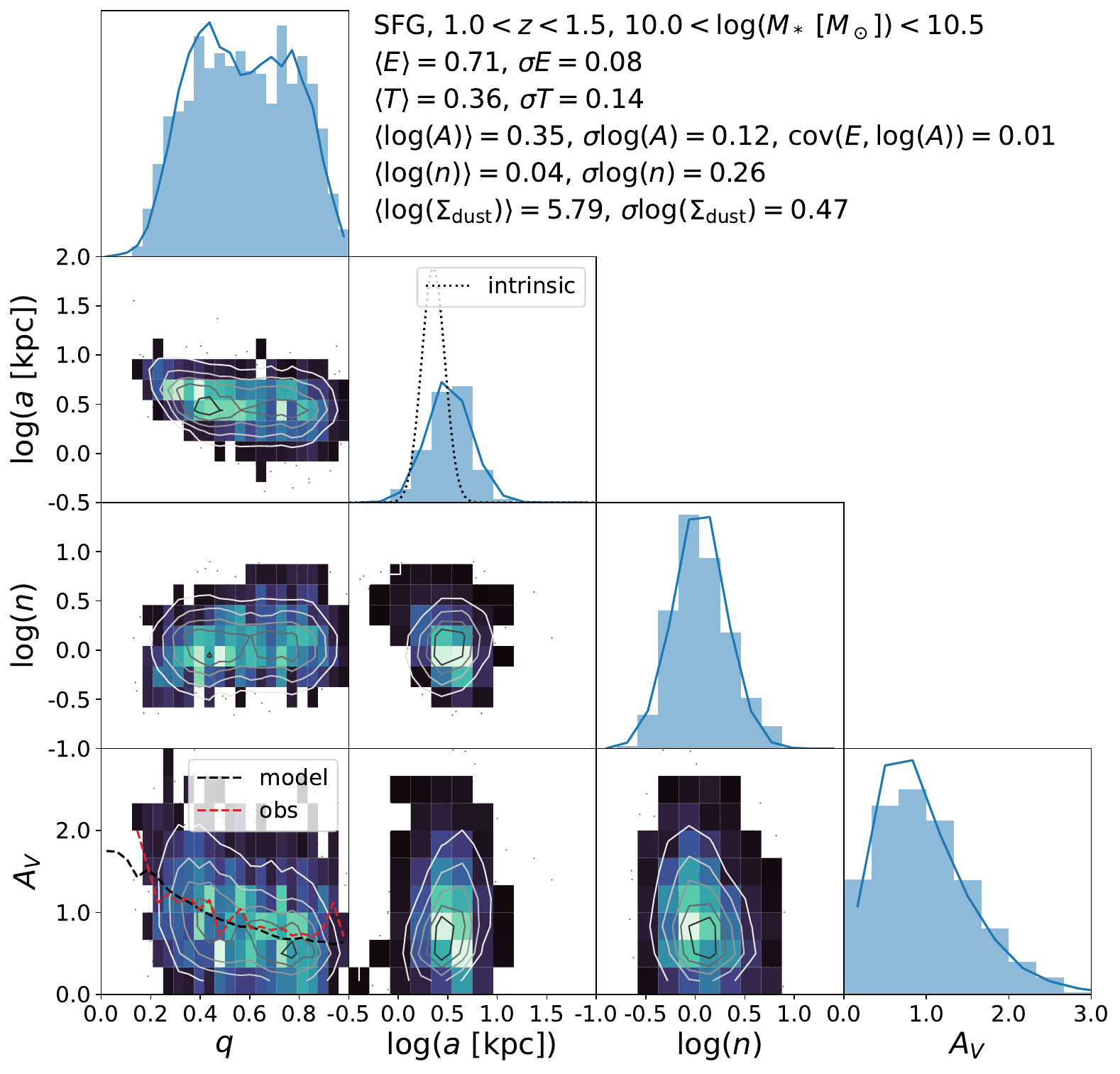}
\caption{$q-\log(a)-\log(n)-A_V$ corner plot of our galaxy sample at $1.0<z<1.5$ with stellar mass $10.0<\log(M_*/M_\odot)<10.5$. Histograms show the observed distribution, and solid blue lines + contours represent the best-fit model distribution.  The model employed here leaves dust content as a free parameter, has identical dust and stellar geometry, and features no clumpiness.  The black dashed line in the $\log(a~[{\rm{kpc}}])$ histogram panel represents the distribution of intrinsic half-mass radii of the galaxy population according to the best-fit model (smaller than the half-light radii due to dust attenuation). The median $A_V$ vs q relation in the model and observations is further illustrated with black and red dashed lines. 
}
\label{fig:case_example_easy}
\end{figure}

Having discussed the relation between structural parameters, viewing angles and the observed levels of dust attenuation in a phenomenological manner, we now turn to modelling of the actual observations.  We do so for the same ensembles of SFGs shown in Figure\ \ref{fig:q_Av_SFG_obs}, defined by mass (bins of 0.5 dex width) and redshift (bins of $\Delta z = 0.5$).  In this Section, we employ the simplest star-dust geometry, namely one in which for each toy model galaxy stars and dust follow the same, smooth spatial distribution.  The dust content is left to vary freely.  A case example of such modelling is presented in Figure\ \ref{fig:case_example_easy}.  It is shown that for the considered bin of redshift ($1.0<z<1.5$) and stellar mass ($10.0<\log(M_*/M_\odot)<10.5$), a model galaxy population can be constructed which reproduces the observed joint distribution of $q-\log(a)-\log(n)-A_V$ in a satisfactory manner.

\begin{figure}
\centering
\includegraphics[width=\linewidth]{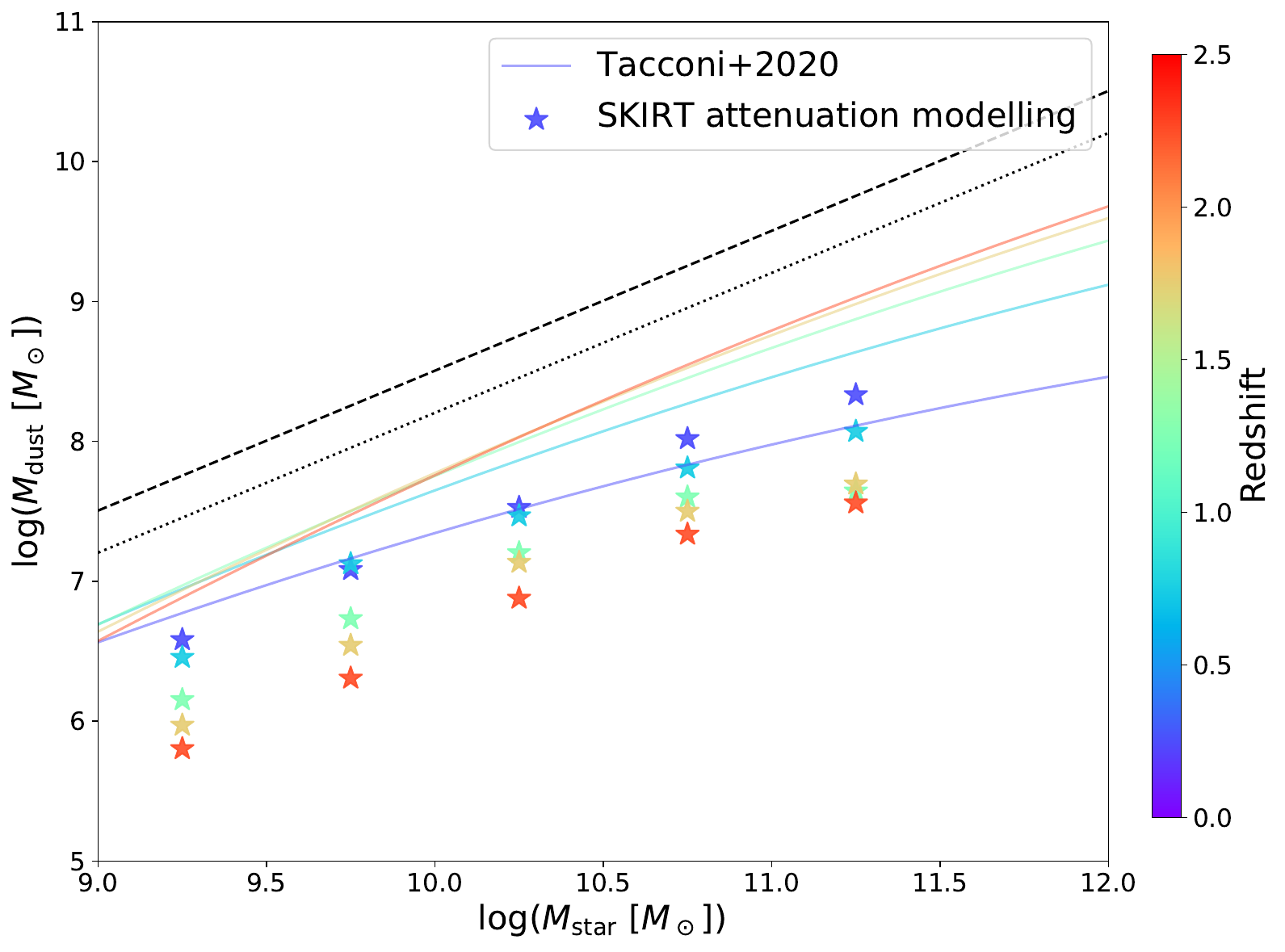}
\includegraphics[width=\linewidth]{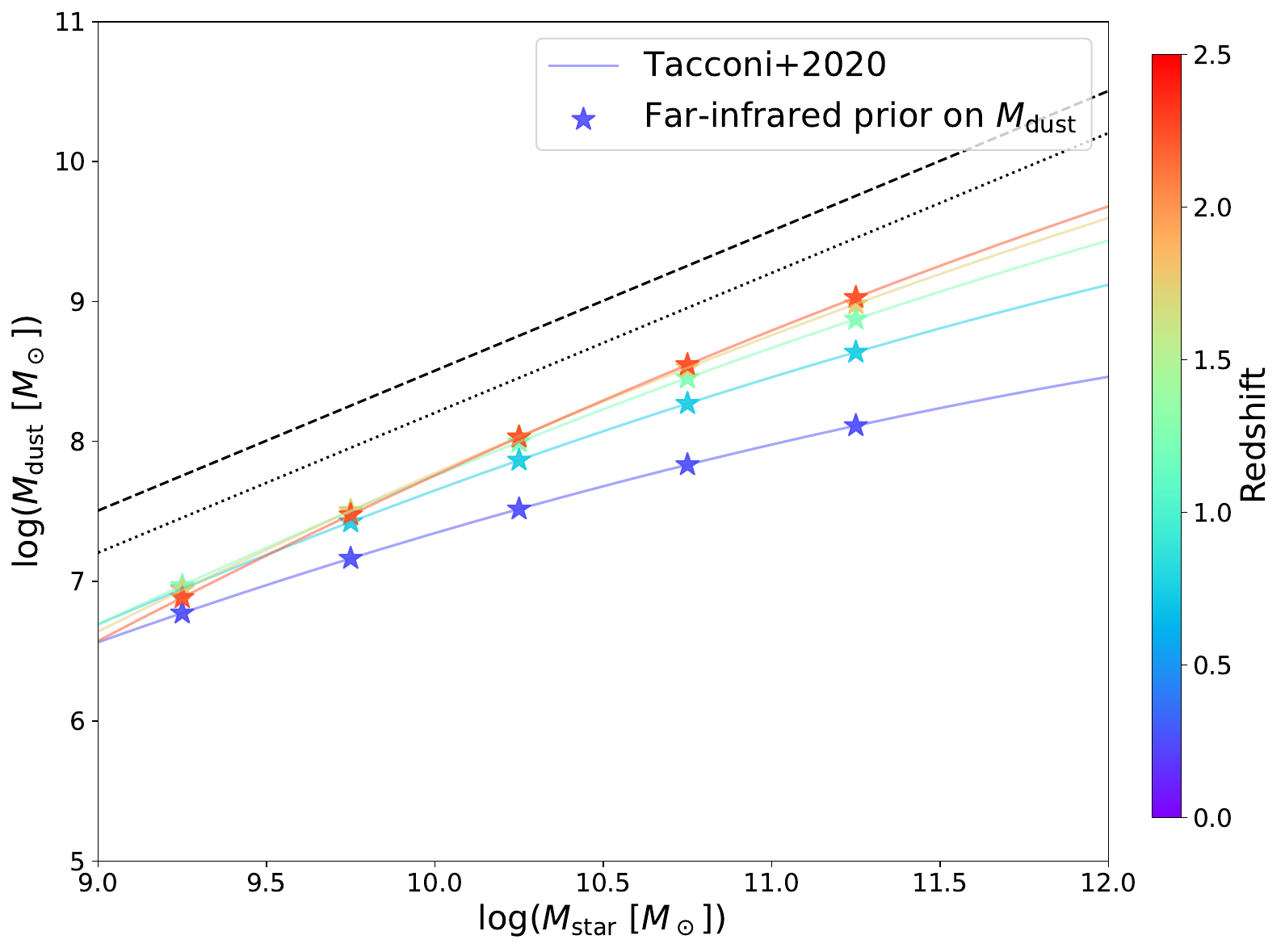}
\caption{{\it Top}: Dust mass vs stellar mass colour-coded by redshift. Star symbols mark the best-fit $M_{\rm dust}$ values obtained via our attenuation modelling under the assumption of identical and smooth spatial distributions for dust and stars. The black dashed and dotted lines correspond to $100\%$ and $50\%$ of the metals returned to the ISM being converted into dust. Solid lines indicate the dust mass of main sequence SFGs from empirical far-IR scaling relations (\citealt{Tacconi2020}; adopting a metallicity-dependent dust-to-gas ratio). {\it Bottom}: Similar to the top panel, but star symbols now correspond to models in which far-IR dust mass priors have been imposed.
}
\label{fig:logMdust_logM_easy}
\end{figure}

\begin{figure}
\centering
\includegraphics[width=\linewidth]{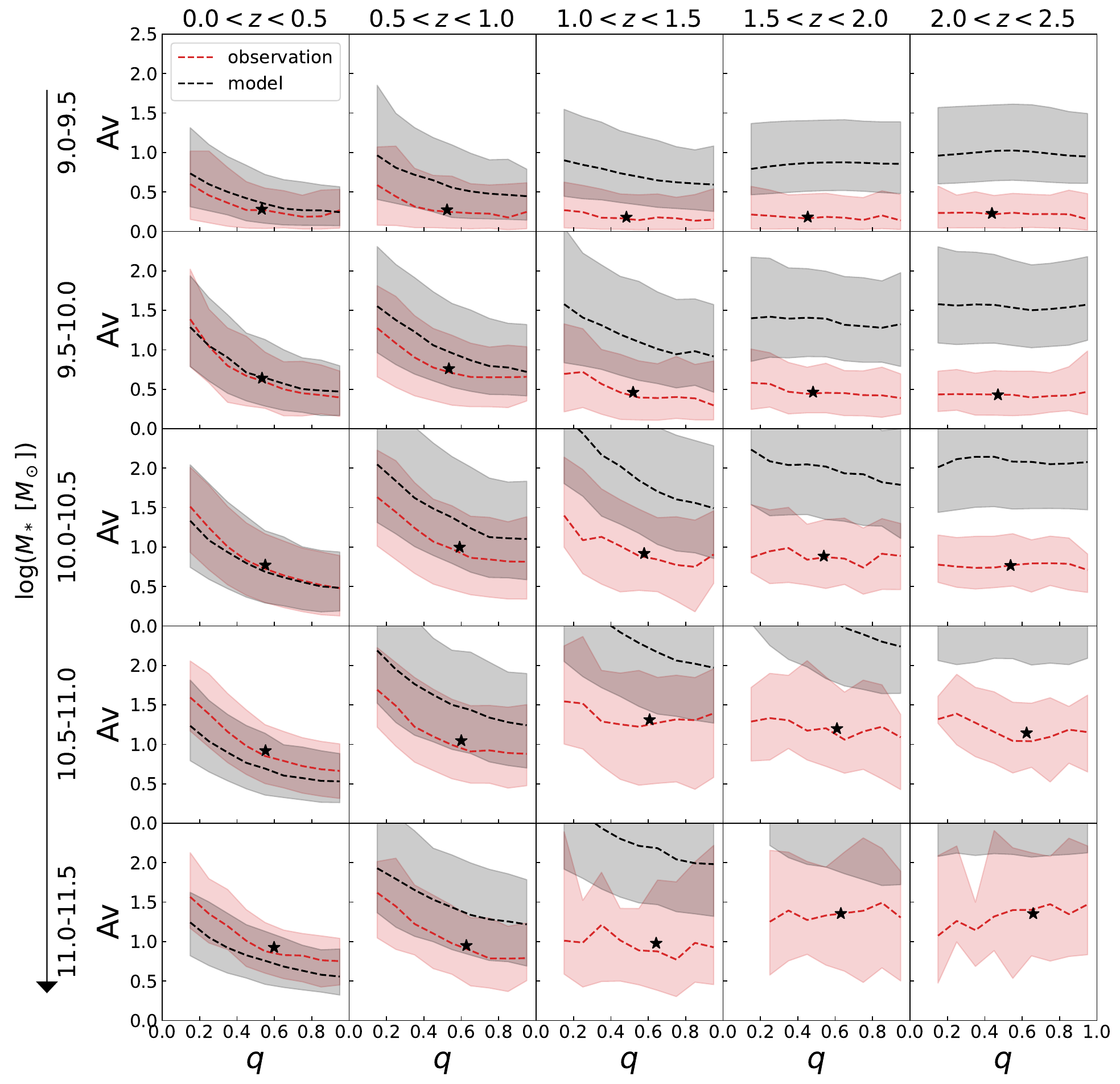}
\caption{Idem to Figure\ \ref{fig:q_Av_SFG_obs}, but alongside the observations (red) showing the $q-A_V$ distribution of a model realisation (black) in which the structural parameters are set to reproduce the observed size, S\'{e}rsic $n$ and $q$ distributions, in which dust masses are fixed according to far-IR scaling relations \citep{Tacconi2020}, and in which the spatial distribution of dust and stars is kept identical and smooth.  This figure illustrates the inability to match the observed dust attenuation while simultaneously imposing far-IR dust masses and structural parameters that are consistent with the observations. More freedom in the adopted star-dust geometry is required to obtain a successful model.
}
\label{fig:q_Av_SFG_easy_T20_}
\end{figure}

Carrying out an equivalent analysis for each of our $(z,M_{\rm star})$ bins, we evaluate the results obtained for the median $M_{\rm dust}$ in Figure\ \ref{fig:logMdust_logM_easy} ({\it star symbols in the top panel}).  For reference, we also illustrate lines of $M_{\rm dust} = y M_{\rm star}$ (dashed black line) and $M_{\rm dust} = 0.5 y M_{\rm star}$ (dotted black line), where we adopted a metal yield of $y = 0.032$, appropriate for a \citet{Chabrier2003} IMF \citep{Madau2014}.  These can be thought of as upper limits on the dust mass that a galaxy of a given stellar mass can reasonably be expected to host.  The dashed (dotted) black lines correspond to scenarios where all (or half) of the metals ever produced and released into the ISM as part of the star formation process in the galaxy, were retained and depleted onto dust grains.  In reality, part of the metals will be ejected by galactic winds, and among those remaining a substantial fraction will reside in the gaseous phase rather than depleting onto dust grains \citep{Remy-Ruyer2014,DeVis2019,Li2019}.  The dust masses inferred from SKIRT attenuation modelling reassuringly lie well below these (very generous) upper limits.  

Figure\ \ref{fig:logMdust_logM_easy} (top panel) also shows that higher dust masses are found for SFGs of higher stellar mass, as could have been anticipated from the $A_V - M_{\rm star}$ relation described in Section\ \ref{Avobs.sec}.  However, the shortcomings from this model realisation come to light when contrasting the dust masses inferred using SKIRT to observational constraints on the dust content of SFGs from far-IR scaling relations \citep[][coloured curves corresponding to different redshifts]{Tacconi2020}.  The latter typically imply higher dust masses, especially at higher redshifts.  In fact, the dependence of the dust-to-stellar mass ratio on redshift is qualitatively different: SFGs are increasingly ISM and dust rich at higher lookback times according to the far-IR scaling relations, whereas at face value the opposite is implied by our attenuation modelling with simplistic star-dust geometries.  This tension could have been anticipated from the modest evolution in $A_V$ paired with the known size evolution of SFGs \citep[e.g.,]{van2014b}, and has been the topic of recent work by \citet{Shapley2022}, who analysed Balmer decrements (H$\alpha$/H$\beta$) and rest-UV slopes and came to similar conclusions.  Along a similar vein, \citet{Whitaker2017} noted a lack of redshift evolution in the obscured fraction of star formation with stellar mass, despite the larger ISM content and more compact sizes of high-redshift galaxies.

For completeness, we point out that the low-mass, high-redshift data ingested by \citet{Tacconi2020} into their scaling relations are predominantly coming from CO rather than dust continuum observations.  Had we adopted a constant rather than metallicity-dependent dust-to-gas ratio to convert cold gas masses to dust masses, it would have yielded high-z scaling relation curves that run more parallel to the $z \sim 0.25$ one in the $M_{\rm dust} - M_{\rm star}$ plane (i.e., the red curve would shift upward by nearly 0.6 dex at $\log(M_*/M_\odot) = 9.25$, without appreciable change at the massive end). This would exacerbate the discrepancy with the results inferred from attenuation modelling for the low-mass, high-z bins.  However, we consider the metallicity-dependent dust-to-gas ratio (Eq.\ \ref{dg.eq}) adopted as default in our analysis more realistic.  We note that other potential sources of systematic uncertainty in the measurement of dust and/or cold gas masses based on far-IR/sub-mm diagnostics are smaller in size and unable to account for the differences seen in the top panel of Figure\ \ref{fig:logMdust_logM_easy} \citep[see, e.g.,][for a more exhaustive discussion]{Berta2016, Scoville2016, Tacconi2020}.

An alternative way of illustrating the shortcomings of the considered simplistic star-dust geometry is displayed in Figure\ \ref{fig:q_Av_SFG_easy_T20_}.  Here, we keep for each SFG ensemble the best-fit model parameters as obtained in the fitting with free dust content, and merely adjust the mean dust content so as to match the typical $M_{\rm dust}$ of SFGs at the respective mass and redshift as given by the FIR scaling relations (bottom panel of Figure\ \ref{fig:logMdust_logM_easy}).  The result is a significant overestimate of $A_V$ values at $z > 0.5$, which becomes progressively worse towards higher redshifts.  If instead the model parameters outlined in Section\ \ref{model.sec} are re-fit while imposing the FIR priors on $M_{\rm dust}$, the fit is driven towards larger sizes for the toy model galaxies.  This happens in an attempt to reduce the dust column and hence the predicted attenuation, but comes at the expense of a tension with the observed galaxy sizes.  Consequently, the discrepancy with respect to the observations is revealed in a combination of poorly reproduced $A_V$ and $\log(a)$ distributions.  In other words, under the assumed star-dust geometry no toy model galaxy population can simultaneously reproduce the $M_{\rm dust}$, $A_V$ and galaxy size constraints.

In summary, a more complex star-dust geometry is required to reconcile attenuation and far-IR measurements.

\subsection{Alternative star-dust geometries}
\label{altgeometry.sec}

Although our modelling with identical, smooth star and dust geometries was able to reproduce fairly well the observed $q-\log(a)-\log(n)-A_V$ distributions (Section\ \ref{smoothidenticalmodel.sec}), this was achieved at the expense of a severe mismatch in terms of inferred dust masses with respect to constraints from far-IR/sub-mm observations \citep{Tacconi2020}.

This motivates us to turn the exercise around, fix the median dust mass according to the FIR scaling relations (Figure\ \ref{fig:logMdust_logM_easy}, bottom panel), and evaluate the implied adjustments needed to the star-dust geometry to maintain, or improve, the match to the $q-\log(a)-\log(n)-A_V$ observations.  In order to avoid introducing too much degeneracy, we explore variations of one aspect of the geometry at a time, and comment on their physical plausibility.

\subsubsection{Separate, smooth distributions for stars and dust}
\label{diffshapemodel.sec}

{\it Variations in $C_{\rm{dust}}/C_{\rm{star}}$:} At first, we attempt leaving the relative scaleheight of the dust vs stellar distribution ($C_{\rm{dust}}/C_{\rm{star}}$) free.  Because dust disks that are relatively thin compared to the stellar component yield lower $A_V$ and shallower $A_V - q$ relations (Figure\ \ref{fig:toygeometries}, middle-left panel), the fits with realistic dust masses are driven in this direction in order to optimally reproduce the observed attenuation levels.  Specifically, while best-fit $C_{\rm{dust}}/C_{\rm{star}}$ values for the lowest redshift bin are near unity for massive ($\log(M_*/M_\odot) > 10$) SFGs and around 0.5 for those at lower mass, they progressively decrease down to 0.1 - 0.3 for our samples above $z > 1$.  Even with $C_{\rm{dust}}/C_{\rm{star}}$ as low as 0.1, the best-fit models to the lowest mass high-z SFG ensembles do not predict $A_V$ values quite as low as those observed.  We regard this solution as physically implausible given that, if anything, ISM and stellar disks at high redshift (and especially low mass) can be expected to be less differentiated.  SFGs at cosmic noon are characterised by dynamically turbulent, thick structures \citep[e.g.,]{Wisnioski2015,Wisnioski2019,Simons2017,Ubler2019,Tiley2021}, and their young stellar populations are anticipated to be more closely associated with the ISM disk out of which they formed.  In contrast, the settled disks of nearby galaxies are able to host a dynamically colder and thus thinner ISM structure, whereas their stellar content has had ample time to be puffed up by dynamical disturbances such as (minor) mergers, or indeed may feature thick stellar disk components that formed as such during an earlier, more turbulent past \citep[e.g.,]{Shapiro2010,Mackereth2019,Poci2019}.

{\it Variations in $R_{\rm{dust}}/R_{\rm{star}}$:} A more interesting scenario may be one in which the relative scalelength of dust and stellar disks ($R_{\rm{dust}}/R_{\rm{star}}$) is allowed to vary instead.  Here, it is important to note that, while reducing the scalelength of dust {\it and} stars simultaneously has the net effect of increasing dust columns and thus attenuation levels, the same is not true if only the dust scalelength is reduced (Figure\ \ref{fig:toygeometries}, middle right panel).  This can be understood naturally by the fact that, while central dust columns do increase, these enshroud a progressively smaller portion of the stars overall.  If instead the dust distribution is made more extended relative to that of the stars, the anticipated attenuation would depend more sensitively on viewing angle, which is not supported by the observations.  The net result of this behaviour is that in our population modelling with free $R_{\rm{dust}}/R_{\rm{star}}$, values of $R_{\rm{dust}}/R_{\rm{star}} \lesssim 0.4$ are formally preferred for galaxies above $z > 1$, with dust disks that are typically $\sim 3$ times more compact than their stellar counterparts for massive galaxies ($\log(M_*/M_\odot) > 10$) and as much as $\sim 7$ times more compact for low-mass galaxies ($\log(M_*/M_\odot) < 10$).  As we will detail below, we regard especially the latter size difference as uncomfortably large. The match to the observed $q-\log(a)-\log(n)-A_V$ joint distributions are overall decent when adopting this modelling with free $R_{\rm{dust}}/R_{\rm{star}}$, with a minor caveat that the $q - A_V$ relations for the best-fit model are modestly shallower than those observed for massive ($\log(M_*/M_\odot) > 10$) SFGs at low redshift ($z < 1$).

In terms of physical plausibility, resolved observations of the dust reservoirs within the most massive ($\log(M_*/M_\odot) > 11$) high-z SFGs have indeed revealed evidence for compact dust sizes compared to those observed in the rest-frame optical, and indeed even compared to those inferred from stellar mass maps reconstructed from multi-band HST imaging \citep{Hodge2016,Tadaki2017b,Tadaki2020,Puglisi2019}.  Assembling ALMA observations for the largest such sample at $z \sim 2$ to date, \citet{Tadaki2020} find the 870 $\mu$m dust continuum sizes of their 62 SFGs with $\log(M_*/M_\odot) > 11$ to be on average 2.3 (1.9) times more compact than those quantified from rest-optical (stellar mass) maps, with considerable object-to-object variation.  Below $10^{11}\ M_{\odot}$, resolved dust continuum observations of $z \sim 2$ SFGs become rapidly more scarce.  \citet{Nelson2019} study an intermediate mass SFG at $z=1.25$, finding the rest-500 $\mu$m dust emission to be 1.4/1.8/1.4 times more compact than the H$\alpha$/rest-UV/rest-optical light, but if anything somewhat less compact than the stellar mass distribution ($R_{\rm e,M_*}/R_{\rm e,dust} = 0.9$).  Exploiting slightly lower resolution ($0\farcs7$) dust continuum and CO line observations of a more extensive sample of $1.1<z<1.7$ SFGs, \citet{Puglisi2019} recovers a significant spread in FIR-to-rest-optical size ratios, with a larger fraction of compact ALMA sources at the high-mass end.  Most recently, \citet{Magnelli2023} used JWST/MIRI to probe the resolved line emission from Polycyclic Aromatic Hydrocarbons (PAHs).  They find the rest-optical to rest-mid-infrared size ratio of main sequence galaxies to increase with mass, from $\sim 1.1$ at $10^{9.8}\ M_{\odot}$ to $\sim 1.6$ at $10^{11}\ M_{\odot}$ (see also \citealt{Shen2023}).  Any differences between the extent of the stellar mass distribution and the obscured star formation distribution are presumed to be smaller than that.

Indeed, it appears plausible that -if anything- a stronger differentiation between dust and stellar sizes would come about for more massive SFGs, which are likely to have matured further along their inside-out growth path.  We thus conclude that, while differences between dust and stellar radii may contribute to reconciling dust mass and dust attenuation constraints, the required quantitative size differences appear large, especially for less massive galaxies.

\begin{figure*}
\centering
\includegraphics[width=\linewidth]{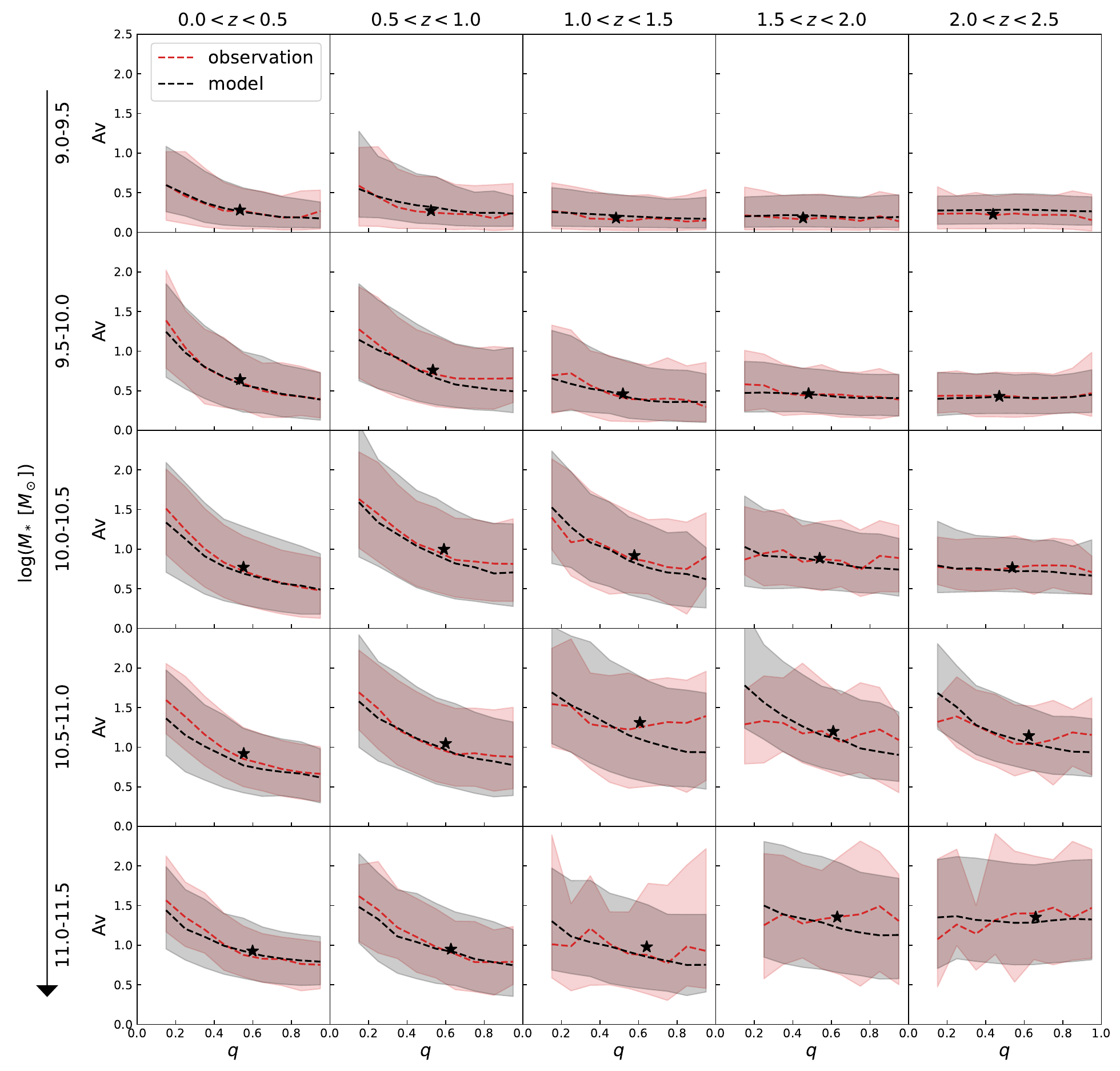}
\caption{Projected axial ratio $q$ vs dust attenuation $A_V$ distribution of the observations (red) and the best-fit models (black) in redshift and stellar mass bins. The model employed here adopts dust masses from far-IR scaling relations \citep{Tacconi2020} as input, and allows for clumpiness while keeping the global geometry of stars and dust identical. Dashed curves denote the running median of the $q-A_V$ distribution, and polygons show the upper 84th and lower 16th percentiles. Black stars mark the median of the observed $(q, A_V)$ values.
}
\label{fig:q_Av_SFG_T20_clumpy}
\end{figure*}

\begin{figure*}
\centering
\includegraphics[width=0.495\linewidth]{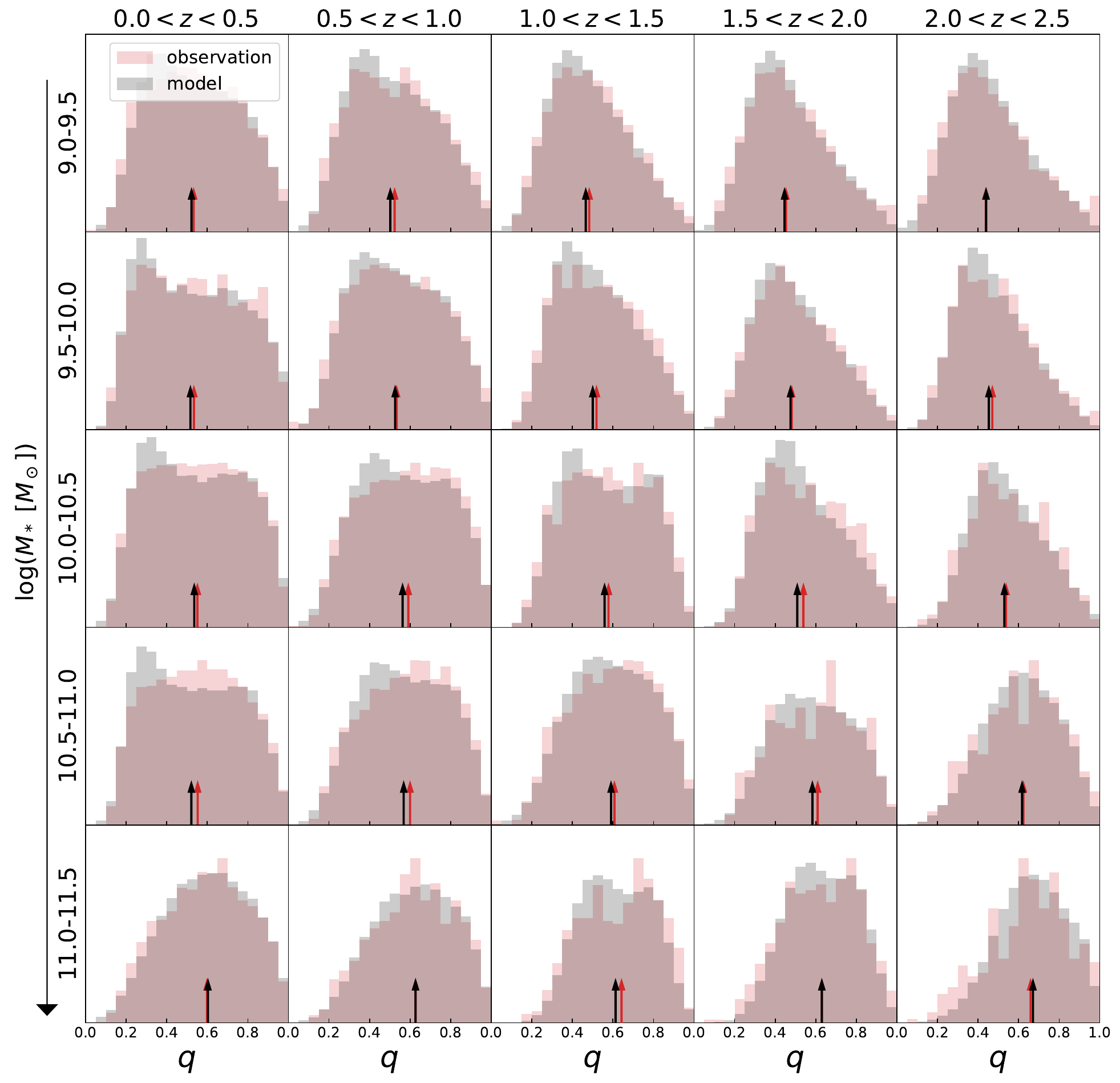}
\includegraphics[width=0.495\linewidth]{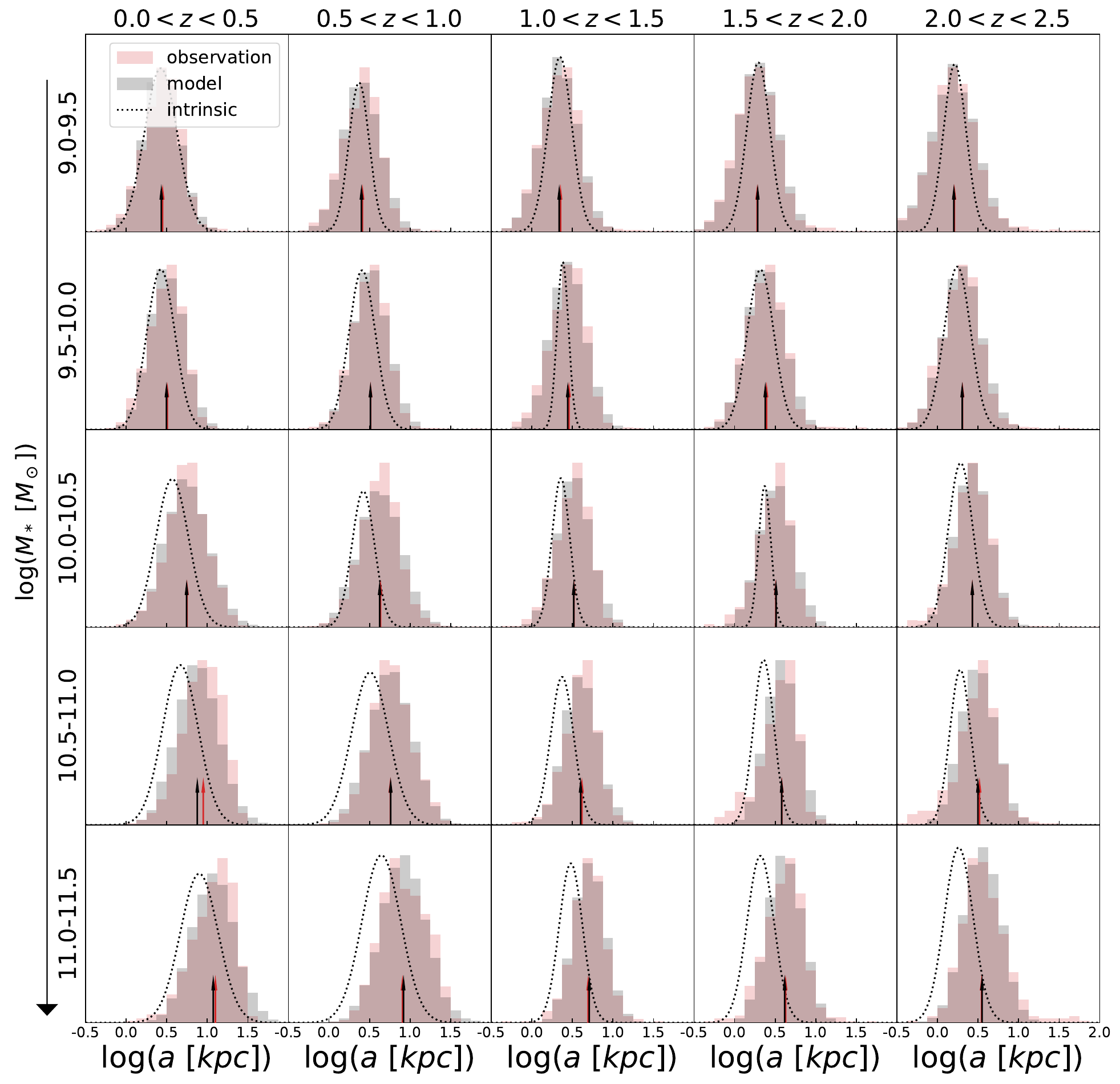}
\includegraphics[width=0.495\linewidth]{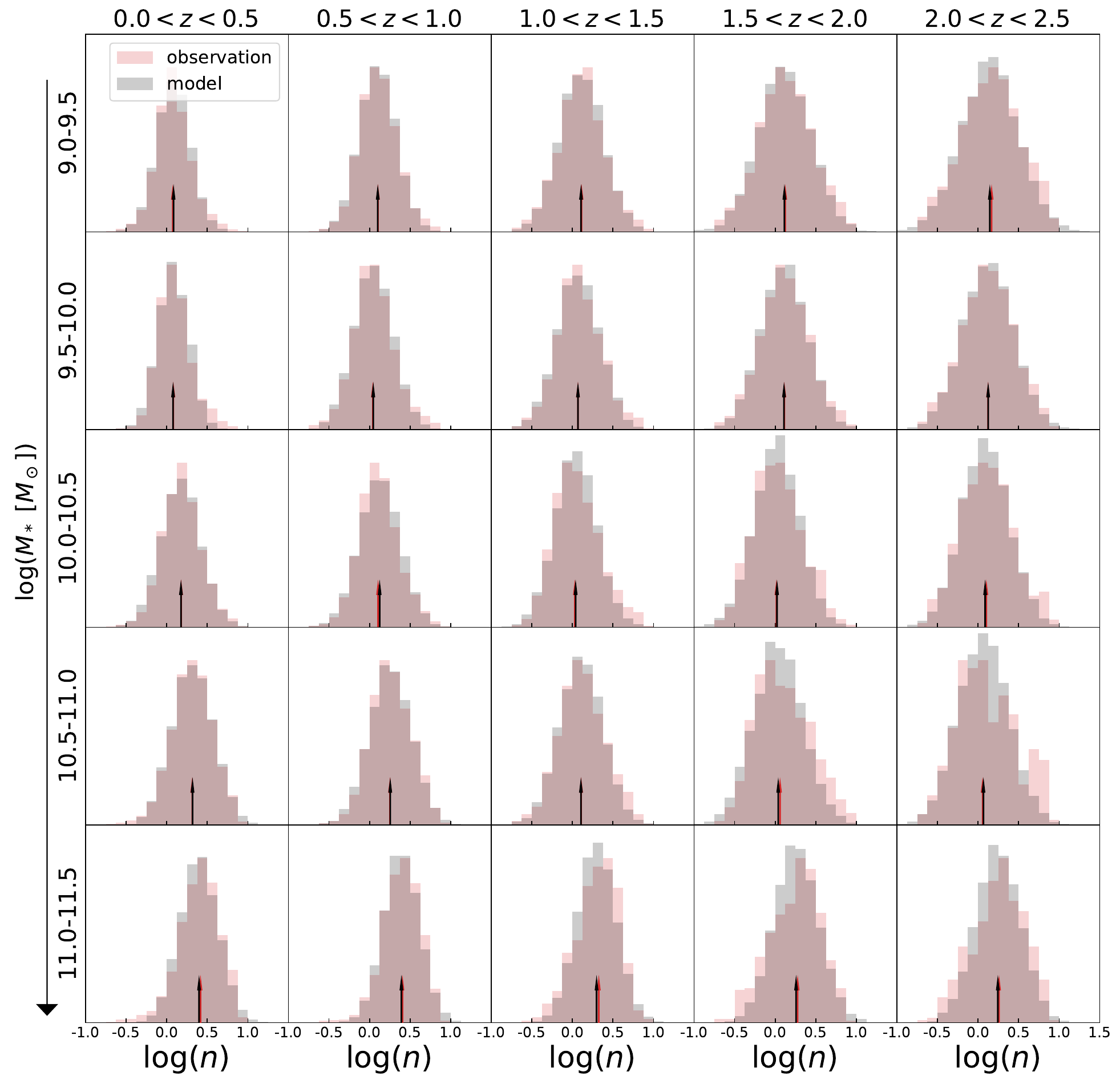}
\includegraphics[width=0.495\linewidth]{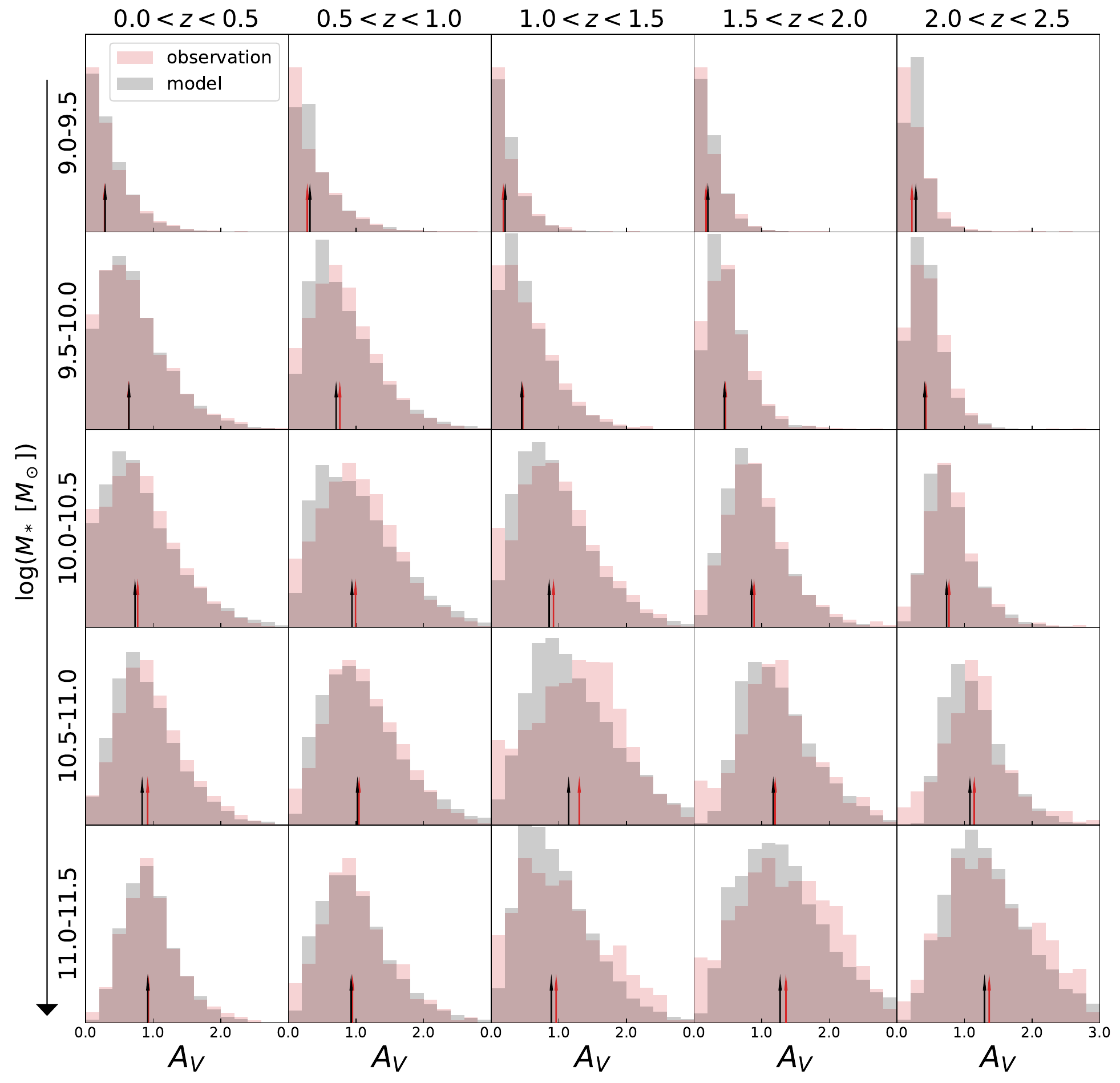}
\caption{Comparison of the distribution of SFG properties, split by redshift and stellar mass.  Observational distributions are shown in red; the corresponding best-fit model in black. Arrows at the bottom of each histogram indicate the median value of the distribution. The model displayed here adopts dust masses from far-IR scaling relations \citep{Tacconi2020} as input, allows for clumpiness, while keeping the global geometry of stars and dust identical.  From left to right and top to bottom, histograms represent the distributions of axial ratios, $q$, the projected semi-major axis half-light radii, $\log(a)$, the S\'{e}rsic index, $\log(n)$, and visual attenuation, $A_V$.  In the size panels, the intrinsic size of the stellar (rather than light) distribution of the model galaxies is also illustrated by the dashed black curve.  Across the mass and redshift space explored, families of model galaxies can be constructed that, when observed from random viewing angles, reproduce simultaneously the observed structural and attenuation properties.  Doing so while imposing far-IR constraints on dust masses was achieved by invoking clumpier dust geometries for higher-redshift SFGs.
}
\label{fig:hist_T20_clumpy}
\end{figure*}

\begin{figure}
\centering
\includegraphics[width=\linewidth]{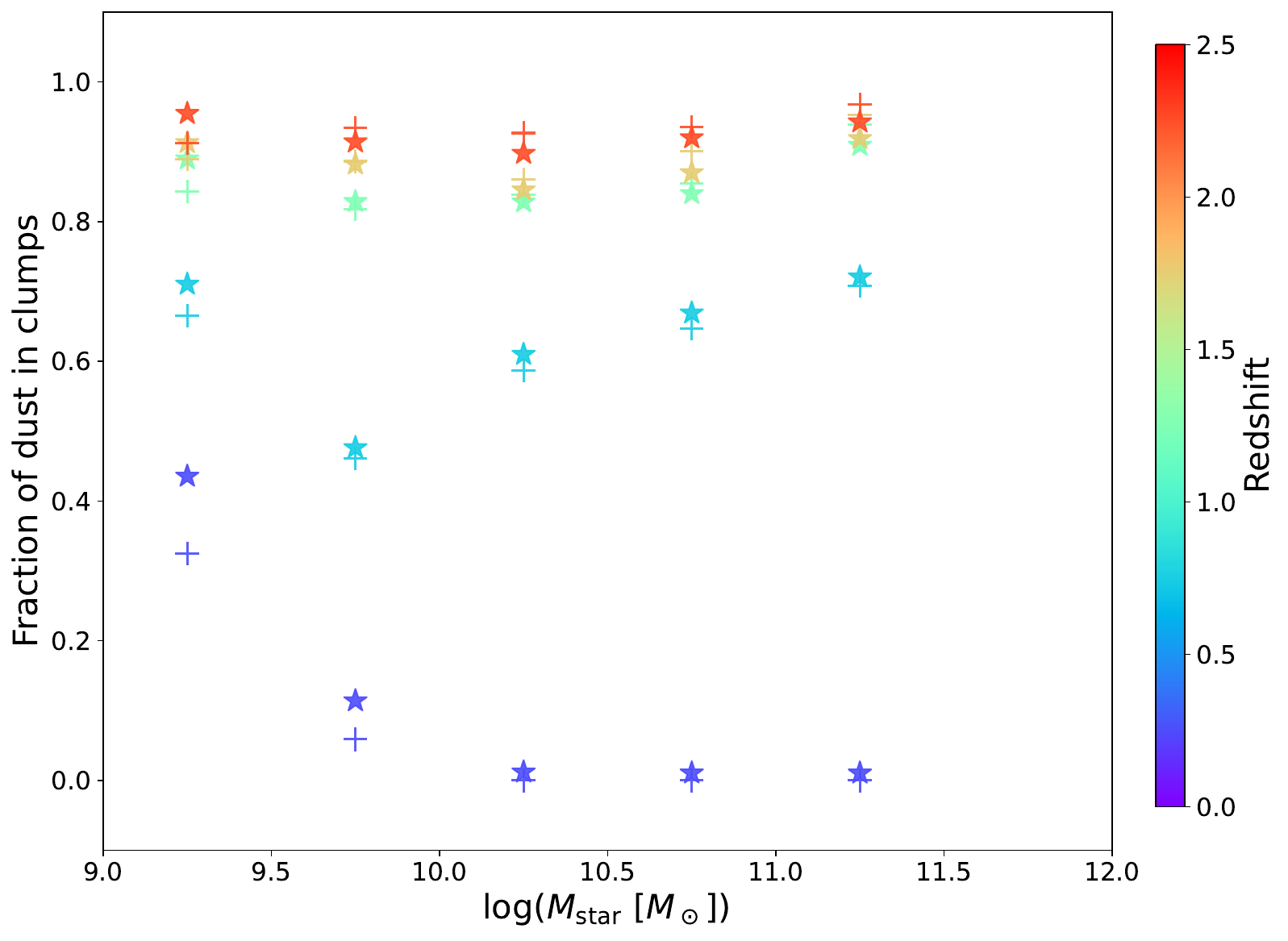}
\caption{Fraction of dust mass in clumps vs stellar mass, colour-coded by redshift. Star symbols show the results of modelling with a diffuse + clumpy geometry, and cross symbols indicate the predicted fraction of dust in clumps from the difference in dust mass between the smooth geometry models and the far-IR scaling relations (\citealt{Tacconi2020}).
}
\label{fig:fdust}
\end{figure}

\subsubsection{Modelling with clumpy ellipsoids}
\label{clumpymodel.sec}

Thus far, we have imposed a smooth distribution of stars and dust.  Our final and favoured scenario relaxes this constraint by introducing clumpiness.  We consider a parameterisation in which the fraction of stars residing in clumps ($f_{\rm clump,\ star}$) and the fraction of dust locked up in clumps ($f_{\rm clump,\ dust}$) is allowed to each vary separately.  In a physical sense, there is no a priori reason why they should be the same, and pragmatically, tying $f_{\rm clump,\ star} = f_{\rm clump,\ dust}$ would have the undesired effect of not altering the effective attenuation as much for a fixed dust mass.

The clump locations are drawn randomly from the underlying smooth density distribution, the dust and stellar clumps are by construction co-located, and since all clumps are set up with identical properties, only their number differs when varying the total mass of dust in clumps. Specifically, they represent dense ISM clumps as found in ultraluminous infrared galaxies (ULIRGs, \citealt{Krumholz2005}), with $10^6\ M_{\odot}$ of dust comprised in structures of 30 pc half-mass radius.  Milky Way type cloud densities (containing $\sim 5$ times less dust in the same volume) were found to provide poorer fits to the high-z galaxy observations as their larger number would render the distribution closer to volume filling, hence mimicking the previously explored smooth models.  For simplicity, we keep the overall density profile of the smooth stellar and dust components identical throughout this exploration.

The results from our modelling with clumpy ellipsoids are presented in Figures\ \ref{fig:q_Av_SFG_T20_clumpy} and\ \ref{fig:hist_T20_clumpy}.  Figure\ \ref{fig:q_Av_SFG_T20_clumpy} shows the median $q - A_V$ relation and central 68th percentiles for observations (in red, reproduced from Figure\ \ref{fig:q_Av_SFG_obs}) and our family of toy model galaxies observed from random viewing angles (in black).  The two show an encouraging agreement, with only few $(z, M_*)$ bins where modest deviations in slope are notable.  Figure\ \ref{fig:hist_T20_clumpy} demonstrates that this agreement, obtained while imposing dust masses from FIR scaling relations (bottom panel of Figure\ \ref{fig:logMdust_logM_easy}), is achieved while simultaneously reproducing the SFGs' axial ratio distributions, observed half-light radii, S\'{e}rsic indices, and attenuation distributions.

The variation in axial ratio distributions across mass and redshift space echoes the previous findings by \citet{van2014b} and \citet{Zhang2019} that low-mass SFGs at cosmic noon have not yet settled in axisymmetric disk configurations.  This is indicated by a $q$ distribution that is skewed to low values.  In comparison, intermediate mass SFGs, especially at lower redshifts, exhibit the broader $q$ distributions characteristic for disks.  Finally, our improved number statistics at the high-mass end suggest that the most massive SFGs tend to be somewhat rounder in shape.  We note that dedicated fits of just the 1D $q$ distribution can in some cases improve the match to observations of this particular structural property, but a family of galaxies with the corresponding intrinsic 3D shapes would perform more poorly in reproducing the distribution of attenuation levels and the $q - A_V$ relation.

In terms of galaxy size, the dashed black curves in Figure\ \ref{fig:hist_T20_clumpy} represent the distribution of projected semi-major axis lengths of the stars.  It is clear that, whereas at low mass these coincide with the corresponding half-light radii (black histograms), they are significantly smaller than the half-light radii for massive SFGs.  We address this in more depth in Section\ \ref{size.sec}.

Whereas SFGs across a wide range in redshift and mass feature a distribution of S\'{e}rsic indices that is centered around $n=1$ (exponential disks) and relatively symmetric in $\log(n)$, a notable shift towards $n>1$ structures is observed among the most massive SFGs that still managed to escape quenching despite their formidable mass.  This trend was already noted in \citet{Wuyts2011}, and is naturally accounted for in our analysis by incorporating the empirical S\'{e}rsic index constraints.

Finally, the $A_V$ distributions in Figure\ \ref{fig:hist_T20_clumpy} illustrate that the best-fit models echo the observational findings already touched on in Section\ \ref{Avobs.sec}.  Namely, that attenuation levels shift to higher values with increasing stellar mass, and comparatively vary less with redshift at fixed mass.

Turning to the fraction of dust in clumps required to achieve the satisfactory agreement with the observed $q-\log(a)-\log(n)-A_V$ joint distributions, these are illustrated in Figure\ \ref{fig:fdust}.  A clear increase with redshift is notable, where for SFGs at cosmic noon of order $\sim 90\%$ of the dust is assigned to dense clumps.  For the lower redshift ($z < 1$) SFGs, where FIR-based and attenuation-inferred dust masses were already in better agreement with simplified, smooth geometries (top panel of Figure\ \ref{fig:logMdust_logM_easy}), a comparatively lower fraction of the dust is allocated to clumps.  In contrast to the significant dust clump fractions, the best-fit fractions of stars assigned to clumps remain low across the range of masses and redshifts explored ($\lesssim 0.2$, and often negligible).

In fact, in the limit where clumps are star-less and sufficiently compact compared to the global dimensions of the galaxy, they merely serve as pockets to hide dust mass without contributing to the net attenuation.  As such, the results presented in this Section could have been anticipated from those obtained in Section\ \ref{smoothidenticalmodel.sec}.  Interpreting the freely fitted dust masses obtained there under the assumption of smooth geometries as only representing the mass contained in a diffuse dust component, we can combine them with the overall $M_{\rm dust, FIR}$ from the FIR scaling relations to obtain
\begin{equation}
f_{\rm clump,\ dust,\ expected} = 1 - (M_{\rm dust, diffuse} / M_{\rm dust, FIR})
\end{equation}
These estimates are denoted with plus symbols in Figure\ \ref{fig:fdust}, and they compare favourably with the RT results based on clumpy geometries.

Considering that SFGs at cosmic noon are characterised by specific star formation rates that are an order of magnitude higher than those of similar-mass SFGs in the nearby Universe \citep[see, e.g.,][and references therein]{Popesso2023}, it may not be surprising that a correspondingly higher fraction of their dust content is comprised within birth clouds.  In due time, these birth clouds are likely to be dispersed, with (part of) their dust content being mixed into a diffuse component.  In detail, however, the evolution of the dust budget and distribution is governed by a complex interplay of processes, with multiple channels for dust production (supernovae, asymptotic giant branch stars and grain growth in the ISM) as well as destruction and redistribution (e.g., supernova shocks, outflows), which we do not claim to disentangle in this study.  We defer the reader to \citet{Popping2017} for a comprehensive overview of these processes in a galaxy formation context.  

As a cautionary note, the high fractions of dust mass locked up in clumps (i.e., not part of a volume-filling diffuse component) shown in Figure\ \ref{fig:fdust} do not necessarily imply that these galaxies would appear as highly substructured in a typical ALMA observation.  As mentioned before, the fraction of stars assigned to the clumps is low in our modelling, and much of the clumps' dust emission would appear blended into a smooth distribution due to the practical resolution limitations of most observing programmes.  This is also reflected in preliminary mock observations we generated with SKIRT, and on which we will report in a forthcoming study dedicated to expanding our modelling to a panchromatic UV-to-submm approach.  Observationally, the jury is still out on how much of high-z galaxies' dust emission can be attributed to substructure, with some reporting clumpy morphologies or (often subdominant) substructure \citep[e.g.]{Swinbank2015,Iono2016,Hodge2019,Dessauges2019,Dessauges2023,Rujopakarn2023} and others noting a lack thereof \citep[e.g.]{Hodge2016,Rujopakarn2019,Ivison2020}.  Of relevance to interpreting these somewhat heterogeneous results is their large range in spatial resolution, from kpc scales down to an exquisite 30 pc in the best gravitational lens cases, and the significant spread in sensitivities achieved (affecting the ability to retrieve real and avoid spurious substructure).  Furthermore, many (but not all) of the aforementioned studies focus on luminous sub-mm galaxies, whose dust morphologies are not necessarily representative of that of the overall SFG population.

We conclude that our results are consistent with clumpier dust geometries in higher-redshift SFGs.  For those bins of massive, high-z SFGs where Figure\ \ref{fig:q_Av_SFG_T20_clumpy} shows a modestly steeper model $q - A_V$ relation than seen in the observations (e.g., $\log(M_*/M_\odot)=10.5-11.0$ at $1.0<z<1.5$ and $1.5<z<2.0$), we note that our modelling with more compact dust disks (Section\ \ref{diffshapemodel.sec}) yielded shallower $q - A_V$ slopes (see also middle-right panel of Figure\ \ref{fig:toygeometries}) and thus a better agreement as far as this projection of the multi-dimensional space of observations is concerned.  For other $(z,M_*)$ bins, on the other hand, the modelling from Section\ \ref{diffshapemodel.sec} did not necessarily outperform the modelling with clumpy geometries assessed here.  We can therefore not rule out that both clumpier dust geometries and more compact dust versus stellar sizes are at play in explaining the dust properties of massive high-z SFGs.

\begin{figure*}
    \centering
    \begin{subfigure}[b]{.4\textwidth}
    \includegraphics[width=\linewidth]{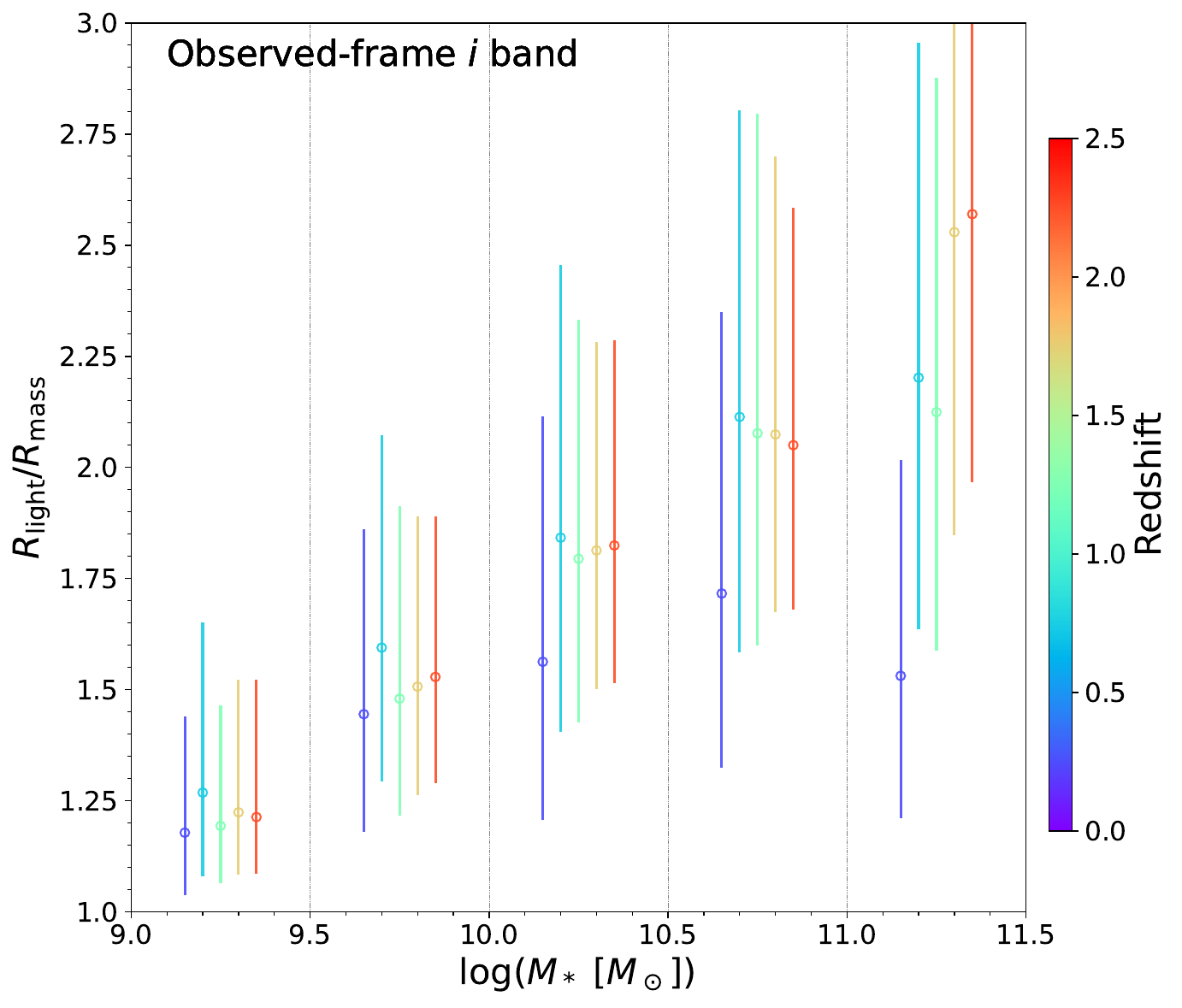}
    \caption{}
    \end{subfigure}
    \begin{subfigure}[b]{.4\textwidth}
    \includegraphics[width=\linewidth]{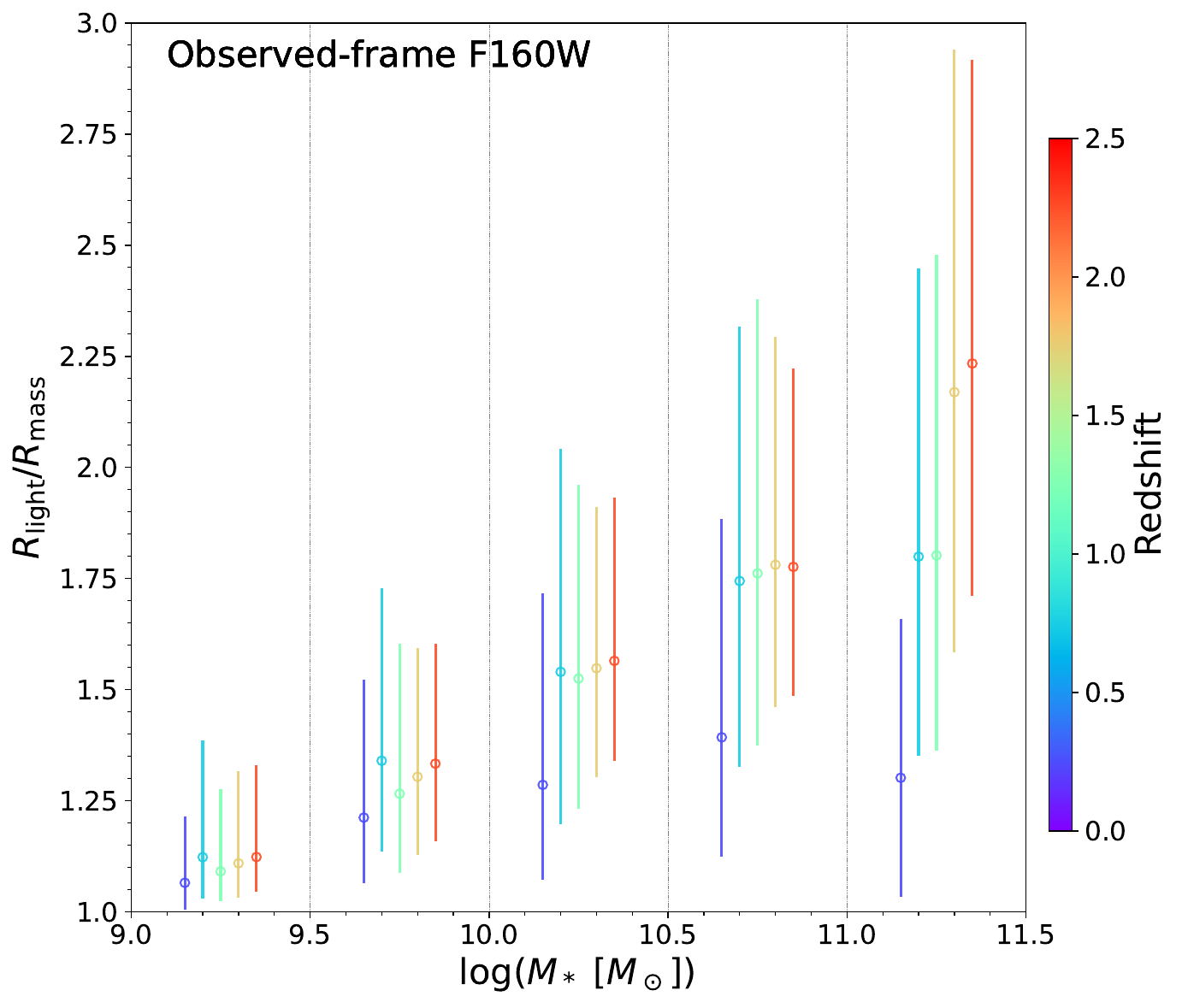}
    \caption{}
    \end{subfigure}
    \centering
    \begin{subfigure}[b]{.4\textwidth}
    \includegraphics[width=\linewidth]{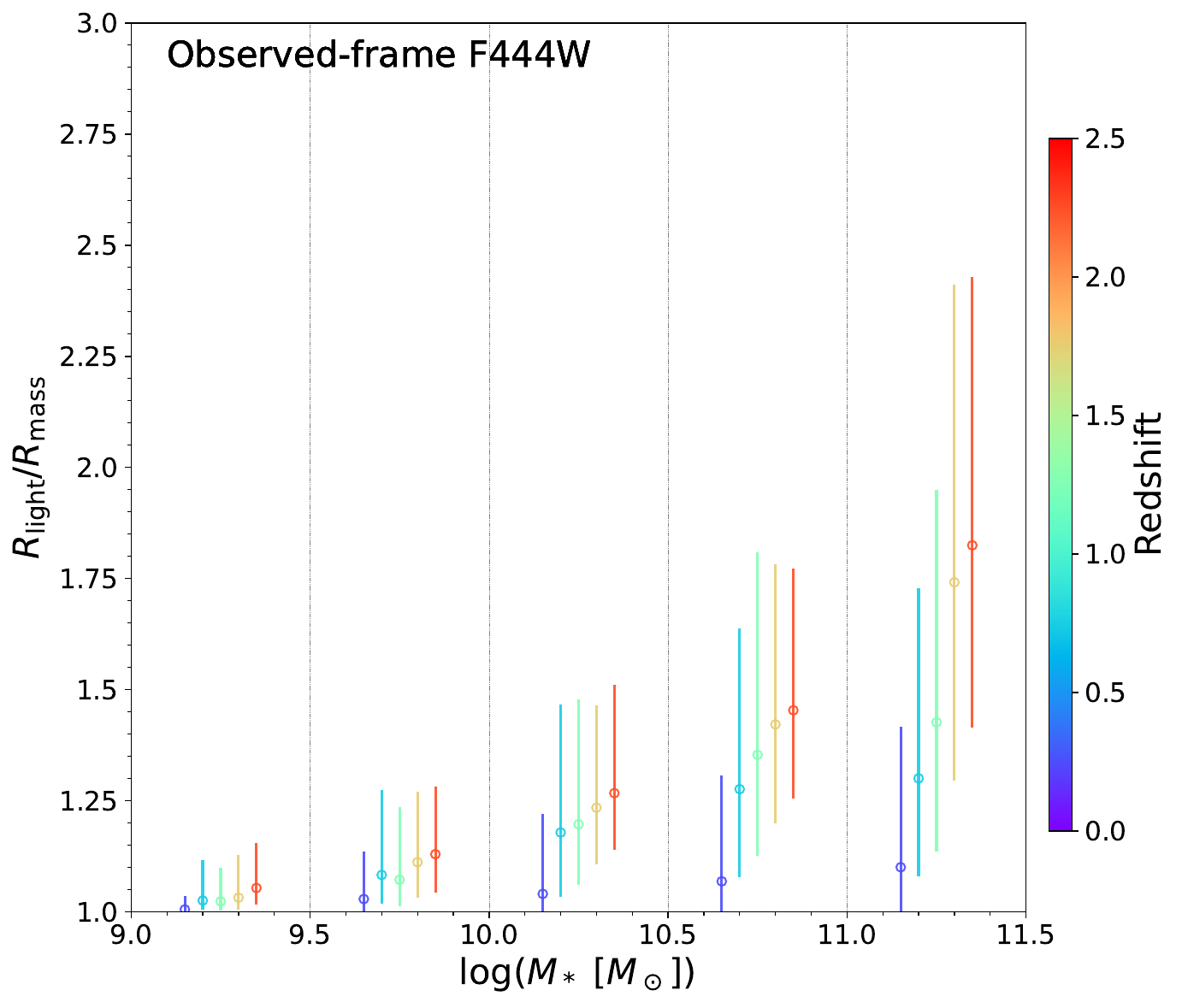}
    \caption{}
    \end{subfigure}
    \begin{subfigure}[b]{.4\textwidth}
    \includegraphics[width=\linewidth]{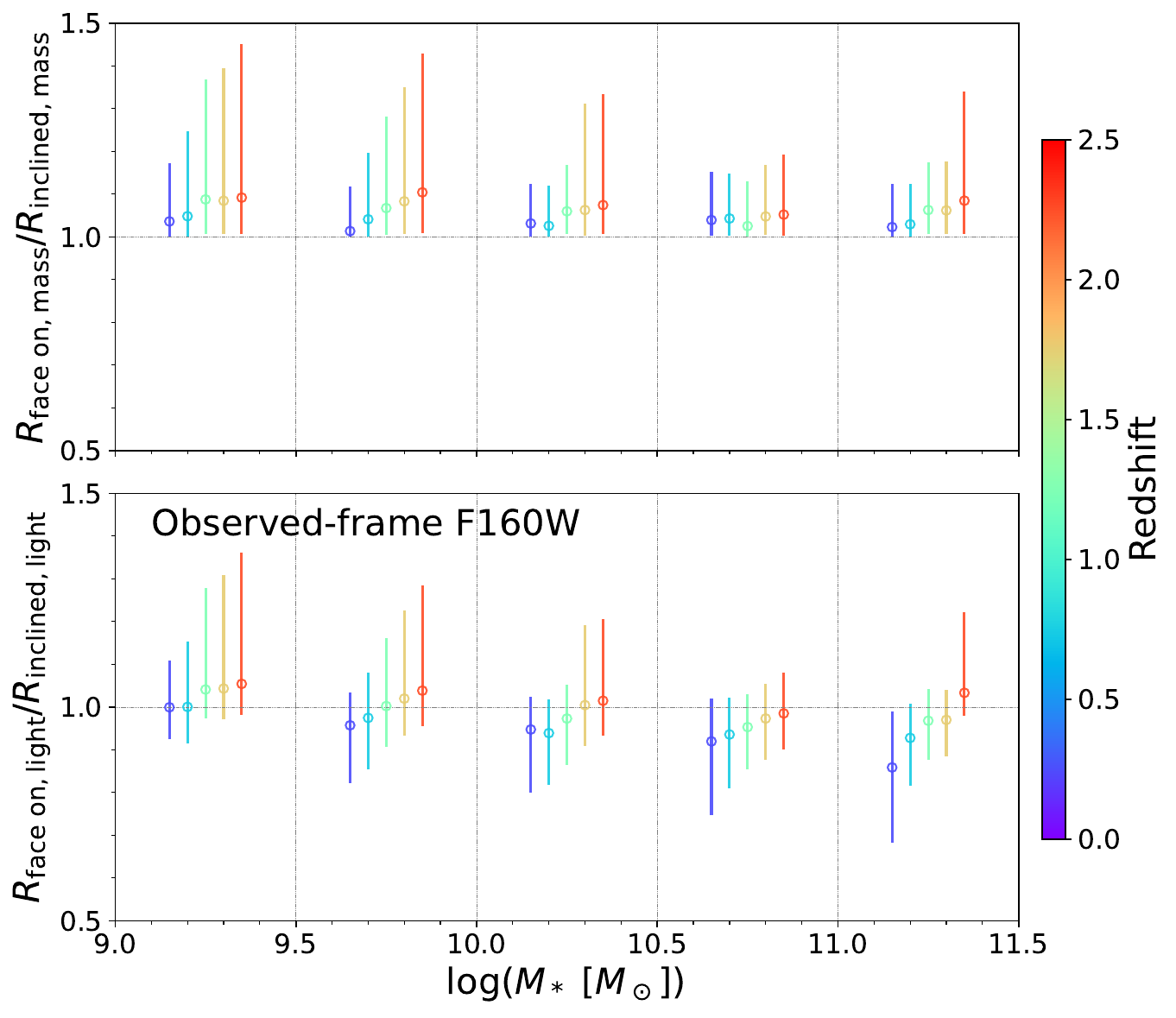} 
    \caption{}
    \end{subfigure}
    \caption{Size comparisons based on the best-fitting population of toy model galaxies using our favoured diffuse + clumpy dust geometry (Section\ \ref{clumpymodel.sec}). Panels (a)-(c): ratio of measured half-light to stellar half-mass radii, $R_{\rm{light}}/R_{\rm{mass}}$, as a function of galaxy mass.  Different panels quantify the half-light radii in different observed-frame wavebands ($i$, F160W and F444W, respectively).  Colour coding denotes the redshift bin.  Circles and bars mark the median and central 68th percentile of the distribution, respectively. (d): Ratio of the observed semi-major axis size seen by a face-on observer versus by observers positioned at random viewing angles.  The top (bottom) panel quantifies the results in the absence (presence) of dust. Overall, size measurements are substantially more influenced by dust attenuation than by orientation.  The impact of dust is largest for massive galaxies, especially when probing shorter rest-wavelengths.}
    \label{fig:RlightRmass}
\end{figure*}

\subsection{Implications for galaxy sizes}
\label{size.sec}

Because the size of a galaxy is a fundamental parameter in characterising its growth stage, we here consider the implications of our analysis for size measurements of the stellar component.

That half-light and half-mass radii in principle can differ was already illustrated by the case example toy model galaxy from Figure\ \ref{fig:R_half}.  However, that specific toy model galaxy was set up with arbitrary size, shape and dust content.  Now, we can evaluate the half-light to half-mass size ratios for the same mock galaxy populations that, when observed from random viewing angles, reproduce the joint $q-\log(a)-\log(n)-A_V$ distribution of SFGs in a certain $(z, M_*)$ bin.  This is done in Figure\ \ref{fig:RlightRmass} (panels a - c), where we show as a function of galaxy stellar mass the median $R_{\rm light}/R_{\rm mass}$ and central 68th percentile of its distribution for a given SFG ensemble.  Different colours mark the different redshift bins, and separate panels illustrate the size ratio as quantified in increasing observed-frame wavelengths ($i$, F160W, F444W).  The model results shown are those obtained with the diffuse + clumpy geometry discussed in Section\ \ref{clumpymodel.sec}.

A few takeaways are notable.  First, the inferred half-light to half-mass size ratios depend steeply on mass, and for massive SFGs can be at the factor $\sim 2$ level.  The amplitude of such size biases decreases with increasing wavelength of observation, and for a given observed-frame waveband with decreasing redshift.  The latter trend is influenced by both evolution in the intrinsic properties of the SFGs (e.g., size, dust content and clumpiness), but importantly also by the fact that depending on redshift different rest-wavelengths are probed.  It is noteworthy that, among massive ($\log(M_*/M_\odot) > 10.5$) SFGs at $z \sim 2$, even imaging in JWST/NIRCam's F444W filter may not be directly probing the stellar mass distribution.  In these dust-rich, massive SFGs the rest-frame $\sim 1.4\ \mu$m emission probed features non-negligible central obscuration leading to $R_{\rm light}/R_{\rm mass}$ correction factors of $\sim 40\%$ and up.

Very similar trends were found when models with varying dust versus stellar scalelengths (Section\ \ref{diffshapemodel.sec}) were explored instead of models that varied the amount of dust confined to dense clumps.  Quantitatively, the $R_{\rm light}/R_{\rm mass}$ ratios are then slightly higher, but only by $\sim 3\%$, 8\% and 11\% at observed-frame $i$, F160W and F444W wavelengths, respectively.  This emphasizes the robustness of the trends, but does mean that empirical measurements of $R_{\rm light}/R_{\rm mass}$ may by themselves only be of limited use in discriminating between these scenarios.

Similar conclusions to those presented in Figure\ \ref{fig:RlightRmass} were drawn by \citet[][see the left-most panel of their Figure 8]{Popping2022}, who applied SKIRT radiative transfer to high-redshift galaxies from the cosmological hydrodynamical simulation TNG50.  Their analysis suggests that, for the same observed waveband, size biases become yet larger (exceeding a factor 4 for massive galaxies observed at 1.6 $\mu$m) when extending the considered redshift range out to $z \sim 5$.  By construction, their approach differs in two key aspects from the one using idealized, toy model galaxies we adopt here.  First, the TNG model galaxy population was not fine-tuned to reproduce at each redshift and mass the observationally constrained distribution of galaxy sizes, shapes, dust content and attenuation.  Second, the TNG galaxies grew organically according to the physics implemented in the simulation.  The structure of TNG galaxies thus emerged self-consistently, and so did the spatial distribution of their stellar populations and ISM.\footnote{The realism of these spatial distributions is subject to resolution effects and dependent on subgrid recipes, which, by lack of explicit treatment of dust, include assumptions regarding the dust-to-metal ratio imposed in post-processing.}  Given that for our toy model galaxies we explicitly assumed homogeneous stellar populations, it is worth noting that in their analysis, \citet{Popping2022} found stellar population induced size variations to be strongly subdominant to those induced by dust effects.

Encouragingly, the implied impact of dust on observed galaxy sizes appears also in line with observational trends, with wavelength-dependent size measurements and biases with respect to stellar half-mass radii being evident when comparing and interpreting rest-UV to rest-optical imaging in HST surveys such as CANDELS \citep[see, e.g.,]{Wuyts2012, Suess2019}.  More recently, the advent of JWST with Cycle 1 lookback surveys such as CEERS \citep{Bagley2022} has enabled a confirmation of the progressively smaller sizes of galaxies with increasing wavelength into the rest-frame near-infrared.  Without discriminating between star-forming and quiescent galaxies, \citet{Suess2022} report sizes of CEERS galaxies at cosmic noon that are 30\% smaller at 4.4 $\mu$m compared to 1.6 $\mu$m at $\sim 10^{11}\ M_{\odot}$ and negligibly smaller at $10^9\ M_{\odot}$.

CANDELS-based analyses of SFGs at $z \sim 1-2$ were influential in pinpointing dust as the principle physical origin of their observed colour gradients \citep{Liu2016, Liu2017}.  These studies paired radially resolved profiles of a single rest-frame colour with global estimates of the galaxy-integrated attenuation.  The longer wavelength baseline introduced by JWST has since enabled the use of colour-colour diagnostics to break age-dust degeneracies at a spatially resolved level, leading to consistent inferences of dust being the dominant cause of the observed colour profiles in SFGs \citep{Miller2022}.

For completeness, we demonstrate in panel (d) of Figure\ \ref{fig:RlightRmass} that other effects, related to the observer's viewing angle, introduce comparatively negligible biases or variations in the measured size.  We address this aspect in two steps, first evaluating the differences in observed semi-major axis sizes in the absence of dust\footnote{For a more in depth analytical and numerical treatment of projection effects and their impact on galaxy size in the absence of dust, applied to galaxies of different S\'{e}rsic index and intrinsic shape, we refer the reader to \citet{vdVen2021}.}, then incorporating the effects of spatially varying attenuation.  In the first case, deviations from axisymmetry are the sole reason why the semi-major axis size does not remain equal across all viewing angles.  By definition, the ratio of stellar semi-major axis size for the face-on and inclined view, $R_{\rm face\ on,\ mass}/R_{\rm inclined,\ mass}$, has a lower bound of unity.  The median deviation above unity, averaged over all viewing angles, remains restricted to $\sim 5\%$ or less across the full range of redshifts and galaxy masses explored.  The distribution of $R_{\rm face\ on,\ mass}/R_{\rm inclined,\ mass}$ extends to the highest values among low-mass, high-redshift galaxies.  These are the systems for which the observed axial ratio distribution hints at significantly prolate, elongated shapes.  They are also the systems for which effective attenuation levels are low, and thus represent the only regime where viewing angle and dust effects are of similar amplitude (and in both cases very minor).

The case example toy model galaxy shown in Figure\ \ref{fig:R_half} illustrates that, even for a given galaxy, the degree to which the half-light radius exceeds the half-mass radius depends on viewing angle.  The sense of this trend is such that the half-light to half-mass size ratio is larger for more inclined systems, given their enhanced projected dust columns.  This effect therefore compensates to some degree the above viewing angle dependence due to deviations from axisymmetry.  Once folding in the impact of dust, the $R_{\rm face\ on,\ light}/R_{\rm inclined,\ light}$ ratios are pulled down relative to $R_{\rm face\ on,\ mass}/R_{\rm inclined,\ mass}$ (bottom-right panel of Figure\ \ref{fig:RlightRmass}), and are no longer exclusively above unity.  The observed semi-major axis size does show some variations with viewing angle, but the observed size relative to that seen by a face-on observer can be both larger or smaller.  Most importantly, viewing angle effects for a typical galaxy remain at the $\sim 5\%$ level, much smaller than the $R_{\rm light}/R_{\rm mass}$ ratios seen in the rest-optical.  Even when probing rest-frame near-infrared wavelengths of (especially massive) SFGs at cosmic noon, the inferred dust-induced $R_{\rm light}/R_{\rm mass}$ ratios dominate over any variations of the observed size with viewing angle.

\section{Summary}
\label{summary.sec}

\begin{figure}
\centering
\includegraphics[width=\linewidth]{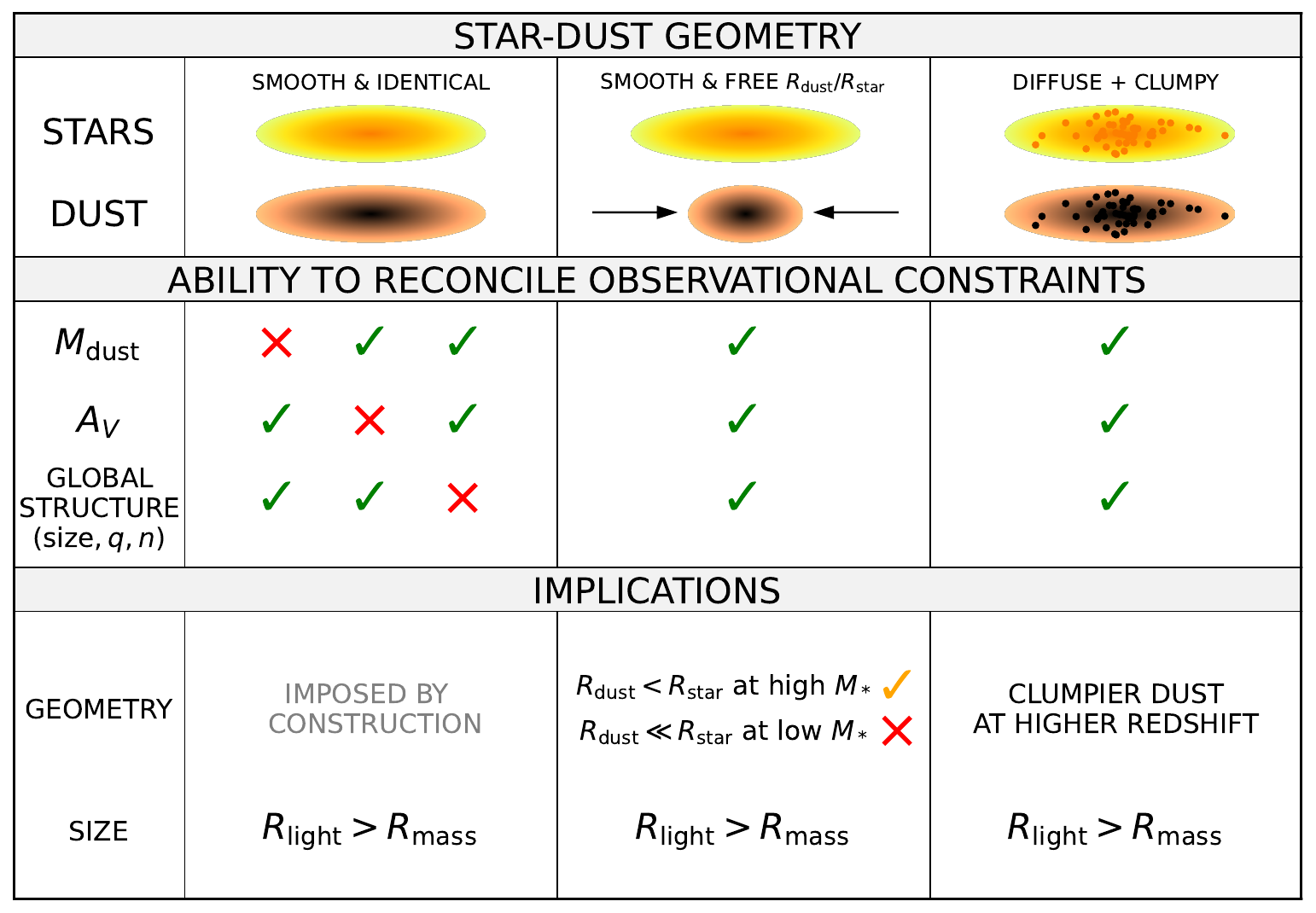}
\caption{Schematic overview of three star-dust geometry families explored as part of our analysis.  {\it Left column:} When adopting the same, smooth spatial distributions for stars and dust, no population of toy model galaxies can simultaneously match both dust mass constraints from far-infrared/sub-mm observations, attenuation constraints from UV-to-near-infrared SED modelling and structural parameters quantified from galaxies' 2D surface brightness profiles.  Models with realistic global structure that reproduce the attenuation feature implausibly small dust masses, especially at higher redshifts.  Conversely, models that impose FIR-based constraints on $M_{\rm dust}$ severely overpredict the attenuation and/or sizes of high-redshift SFGs.  {\it Middle column:} Models in which the dust scalelength is allowed to vary with respect to the stellar one fare better in reconciling the observational constraints.  They do so by invoking compact dust configurations, qualitatively akin to what is observed among high-mass SFGs at cosmic noon.  However, quantitatively the required size differences between dust and stellar distributions are large, uncomfortably so among lower mass SFGs.  {\it Right column:} Models in which a fraction of dust (and a modest fraction of stars) is placed in dense clumps are successful in simultaneously satisfying empirical constraints on $M_{\rm dust}$, $A_V$ and the global structure of SFGs.  They imply that the fraction of dust locked in dense clumps increases with increasing redshift.  All star-dust geometries explored lead to half-light radii that exceed the stellar half-mass radii due to dust-induced light weighting effects.}
\label{fig:summary}
\end{figure}

We investigate the dust properties (mass, geometry, and effective attenuation) of star-forming galaxies from cosmic noon to the present day.  To do so, we compile a large sample of $0<z<2.5$ SFGs from high image quality surveys of tiered area and depth: HSC-SSP, 3D-DASH, and CANDELS.  We model the observed joint distribution of axial ratios, sizes, S\'{e}rsic indices and attenuation using families of toy model galaxies treated with radiative transfer to shed light on the typical star-dust geometry within SFGs across cosmic time.  A synopsis of the models explored, their relative success in reconciling different observational constraints, and their implications regarding galaxies' star-dust geometry and size, are depicted in Figure\ \ref{fig:summary}.  Our main conclusions are:\\

$\bullet$ The effective dust attenuation depends on a galaxy's orientation with respect to the observer. Edge-on galaxies are more opaque than their face-on counterparts. However, the slope of the $q - A_V$ relation becomes shallower with increasing redshift.

$\bullet$ Whereas attenuation levels are observed to rise with galaxy mass at a given epoch, there are at fixed mass only modest variations in dust attenuation with redshift.  Paired with the size evolution of SFGs (more compact at earlier times) and scaling relations based on far-infrared/sub-mm observations (more ISM-rich SFGs at cosmic noon), this poses an apparent tension when interpreting the observed (inclination-dependent) attenuation with the simplest possible star-dust geometries.  Mock galaxy populations in which the star and dust geometries are imposed to be identical and smooth lead to inferred dust masses that fall significantly below those anticipated on the basis of far-IR observations of higher-redshift SFGs.

$\bullet$ Imposing instead dust masses from far-IR scaling relations, we are able to reproduce the observed $q-\log(a)-\log(n)-A_V$ distributions, provided additional flexibility to the assumed star-dust geometry is introduced.  Specifically, a scenario in which SFGs at higher redshifts comprise a larger fraction of their dust content in dense clumps is adequate in alleviating the aforementioned tension.  We further explore the impact of geometries with different scalelengths for stars and dust, and find that additional contributions from more compact dust (relative to stellar) distributions cannot be excluded.

$\bullet$ Our results have direct implications for the size measurements of SFGs.  Dust-induced light weighting effects will render half-light radii large compared to the corresponding stellar half-mass radii.  This effect is most prominent among massive SFGs when probed at shorter rest-wavelengths (at the factor $\sim 2$ level when probing rest-wavelengths $\lesssim 0.5\ \mu$m).  Any variations in observed semi-major axis size with orientation are comparatively minor (typically at the level of $\sim 5\%$).  While less affected by dust than shorter wavebands, even imaging in JWST/NIRCam's F444W band may not be taken as directly probing the stellar mass distribution in the most massive SFGs at cosmic noon.

This work constitutes a first step in analysing and interpreting multi-wavelength diagnostics of attenuation and galaxy structure jointly.  Further investigations will be able to build on this work by folding in new observational constraints.  JWST imaging will continue to more robustly document the resolved spectral energy distributions within galaxies, while in its various spectroscopic modes JWST is adding insight regarding the spatial distribution of nebular attenuation.  Finally, the spatial extent of dust emission will be pinned down across a growing range of galaxy properties, by probing line emission from PAH molecules with JWST/MIRI in the mid-IR, and by probing the Rayleigh-Jeans tail of the dust continuum emission in the (sub)mm regime with NOEMA and ALMA.

\section{Acknowledgements}
We thank the authors of SKIRT, Maarten Baes and Peter Camps, for making their radiative transfer code publicly available.  J.Z. gratefully acknowledges support from the China Scholarship Council (grant no. 201904910703).  S.W. acknowledges support from the Chinese Academy of Sciences President's International Fellowship Initiative (grant no. 2022VMB0004).  Support from NASA STScI grants HST-GO-16259, and HST-GO-16443 is gratefully acknowledged.\\
This paper is based on observations made with the NASA/ESA HST, obtained at the Space Telescope Science Institute, which is operated by the Association of Universities for Research in Astronomy, Inc., under NASA contract NAS 5-26555.\\
The KiDS production team acknowledges support from: Deutsche Forschungsgemeinschaft, ERC, NOVA and NWO-M grants; Target; the University of Padova, and the University Federico II (Naples).\\
The Hyper Suprime-Cam (HSC) collaboration includes the astronomical communities of Japan and Taiwan, and Princeton University. The HSC instrumentation and software were developed by the National Astronomical Observatory of Japan (NAOJ), the Kavli Institute for the Physics and Mathematics of the Universe (Kavli IPMU), the University of Tokyo, the High Energy Accelerator Research Organization (KEK), the Academia Sinica Institute for Astronomy and Astrophysics in Taiwan (ASIAA), and Princeton University. Funding was contributed by the FIRST program from the Japanese Cabinet Office, the Ministry of Education, Culture, Sports, Science and Technology (MEXT), the Japan Society for the Promotion of Science (JSPS), Japan Science and Technology Agency (JST), the Toray Science Foundation, NAOJ, Kavli IPMU, KEK, ASIAA, and Princeton University. \\
This paper makes use of software developed for the Large Synoptic Survey Telescope. We thank the LSST Project for making their code available as free software at  http://dm.lsst.org\\
This paper is based in part on data collected at the Subaru Telescope and retrieved from the HSC data archive system, which is operated by the Subaru Telescope and Astronomy Data Center (ADC) at National Astronomical Observatory of Japan. Data analysis was in part carried out with the cooperation of Center for Computational Astrophysics (CfCA), National Astronomical Observatory of Japan. The Subaru Telescope is honored and grateful for the opportunity of observing the Universe from Maunakea, which has the cultural, historical and natural significance in Hawaii. \\

\section{DATA AVAILABILITY}

The data underlying this article can be accessed from the HSC-SSP (\url{https://hsc-release.mtk.nao.ac.jp}) and MAST (\url{https://archive.stsci.edu/hlsp/3d-dash}; \url{https://archive.stsci.edu/hlsp/candels}) data archives.  The derived data generated in this research will be shared on reasonable request to the corresponding author.

\bibliographystyle{mnras}
\bibliography{library}

\appendix

\section{Variations in stellar population modelling}
\label{appendix_stelpop.sec}

Throughout this paper, we have used on the one hand estimates of the effective attenuation derived from stellar population modelling of the observed SEDs, and on the other hand the actual net visual attenuation (difference between $V$-band magnitude with or without dust) as computed on toy model galaxies using SKIRT.  Implicitly, comparison of the two assumes that they are measuring the same quantity.  We here raise a few caveats regarding this assumption, highlighting scope for future investigations, but argue that the uncertainties associated with $A_V$ estimation are unlikely to alter the overarching conclusions drawn from our analysis.

The assumptions underlying our SED modelling procedure outlined in Section\ \ref{SEDmodelling.sec} are simplistic.  This is to some extent motivated by pragmatic considerations, such as computation time when applied to large samples and consistency with a vast body of literature.  The modest constraining power of broad-band SEDs further implies that introducing more freedom in the SED fit would also open up more degeneracies.  In detail, however, each of the settings, ranging from the adopted Solar metallicity and universal \citet{Calzetti2000} attenuation law to the chosen family of star formation histories, can be questioned and indeed may lead to scatter and/or systematics.  Tests of SED modelling on mock-observed simulated galaxies illustrating these effects have been presented by, e.g., \citet{Wuyts2009}, \citet{Mitchell2013} and \citet{Hayward2015}.

From an observational perspective too, non-universality of the attenuation law in SFGs has been proposed with typically steeper slopes for galaxies of higher specific star formation rates \citep{Kriek2013,Reddy2018,Reddy2023}, and spectroscopic studies have begun to chart the mass and redshift dependence of stellar metallicities in SFGs out to cosmic noon and beyond \citep{Cullen2019, Kashino2022}.  Inspired by these works, we subjected galaxies from the CANDELS survey to a set of alternative SED modelling procedures, systematically varying the stellar metallicity ($\log(Z/Z_{\odot}) = -0.7, -0.4, 0.0, 0.3$) and exploring a few different attenuation laws using the parameterisation by \citet{Noll2009}: $(\delta,E_b) = (-0.4, 1.61), (-0.2, 1.23), (0.0, 0.85)$.  The latter effectively sample the range in attenuation law slope, $\delta$, and 2175$\angstrom$ bump strength, $E_b$, sampled by the composite galaxy SEDs from \citet{Kriek2013}.

As anticipated, lowering the metallicity (at fixed attenuation law slope) yields higher $A_V$, because the intrinsic stellar population is bluer and therefore more dust is required to reproduce the observed galaxy colour.  On the other hand, adopting a steeper attenuation law (at fixed metallicity) yields lower $A_V$, as less dimming of the $V$-band light is required to produce the same redness.  Considering the SFG population at cosmic noon, for which the aforementioned studies argued for both steeper attenuation curves and lower metallicities, especially at low masses, these two effects will work in opposite directions, and may thus be expected to compensate to some degree.

\begin{figure}
\centering
\includegraphics[width=\linewidth]{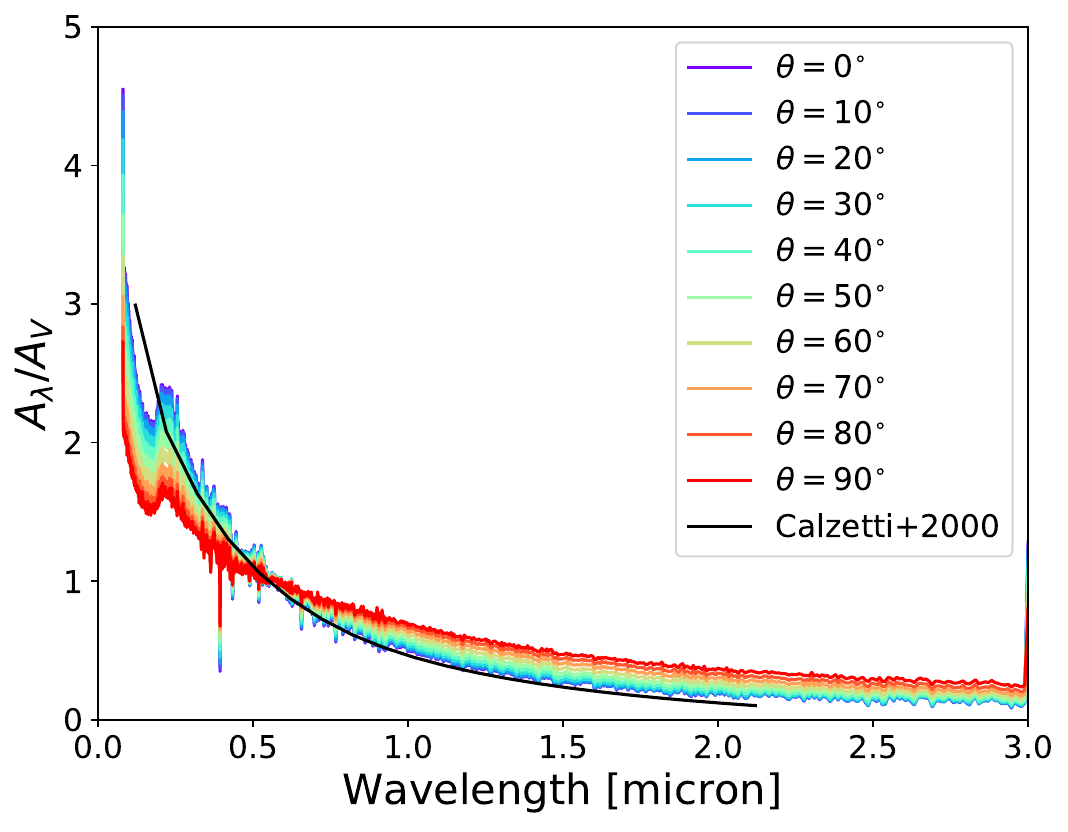}
\caption{Dust attenuation law for the same model galaxy at different viewing angles, as computed with SKIRT radiative transfer (see text for details).}
\label{fig:Calzetti_law}
\end{figure}

We experimented with two approaches to quantify the net effect.  In a first effort, we selected for each galaxy the $A_V$ of whichever cell in the stellar metallicity -- attenuation law grid yielded the fit with the lowest $\chi^2$, irrespective of the significance by which the fit was improved.  In the second approach, we adopted a smoothly varying $Z_*(z,M_*)$, defined as being 0.2 dex lower than the gas-phase metallicity given by Equation\ \ref{dg.eq} \citep[see, e.g.,]{Peng2014}, and similarly let the attenuation law slope $\delta$ vary from 0 at low-redshift/high-mass to -0.4 at high-redshift/low-mass.  Taking our default modelling results as a reference, we found median $\Delta A_V = A_{V,{\rm free}} - A_{V,{\rm reference}}$ of 0.0 (16th and 84th percentiles of -0.2 and 0.3, respectively) for approach 1.  Interpolating $A_V$ values from the $(Z_*, \delta)$ grid according to the smoothly varying metallicity and attenuation law scaling (approach 2), we obtained a median $\Delta A_V = A_{V,{\rm smooth\ (Z_*, \delta)\ scaling}} - A_{V,{\rm reference}}$ of 0.07 (16th and 84th percentiles of -0.13 and 0.26, respectively).  For individual $(z,M_*)$ bins, systematic offsets in $A_V$ were typically restricted to the $\pm 0.2$ range for both approaches, with the exception of low-mass ($\log(M_*/M_\odot)<10$) SFGs at $z<0.5$, where $\Delta(A_V)\approx 0.3 - 0.5$ were seen.  While admittedly ad hoc in their implementation, our two tests suggest that the overall attenuation trends are recovered relatively robustly.

Systematics in $A_V$ estimates as a consequence of the adopted star formation history were found to be of similar amplitude, albeit not necessarily in the same direction.  This was assessed by contrasting our reference SED modelling results to results obtained with Bagpipes \citep{Carnall2018}, where we left metallicity as a free parameter and explored both log-normal star formation histories and the piece-wise constant (also known as non-parametric) star formation history introduced by \citet{Leja2019}.

While a full exploration of attenuation curves in the context of our SKIRT calculations is outside the scope of this work, we do illustrate in Figure\ \ref{fig:Calzetti_law} that in detail, attenuation curves may not only vary from galaxy to galaxy, but also for a given galaxy with orientation towards the observer.  For the model parameters that best describe the SFG ensemble at $z = 1.0 - 1.5$ and $\log(M_*/M_\odot) = 10 - 10.5$, we show with different colours the effective attenuation law obtained with SKIRT cameras set up under different inclinations.  While the \citet{Calzetti2000} law is bracketed by the range of attenuation curves shown, it is evident that even for a single galaxy the attenuation curve is not universal.  Instead, a trend towards progressively greyer (i.e., shallower) slopes is seen with increasing inclination.  Indeed, observational evidence for greyer attenuation law slopes among more inclined systems has previously been reported for nearby and intermediate redshift galaxies \citep{Wild2011, Salim2018, Barisic2020}.  For a more in-depth investigation of such viewing angle and variable attenuation curve effects, we refer the reader to \citet{Trayford2020}, who apply SKIRT radiative transfer to simulated galaxies from the EAGLE cosmological simulation.

\section{Inclination-dependent attenuation for different geometries}
\label{appendix_geometry.sec}

\begin{figure*}
\centering
\includegraphics[width=0.44\textwidth]{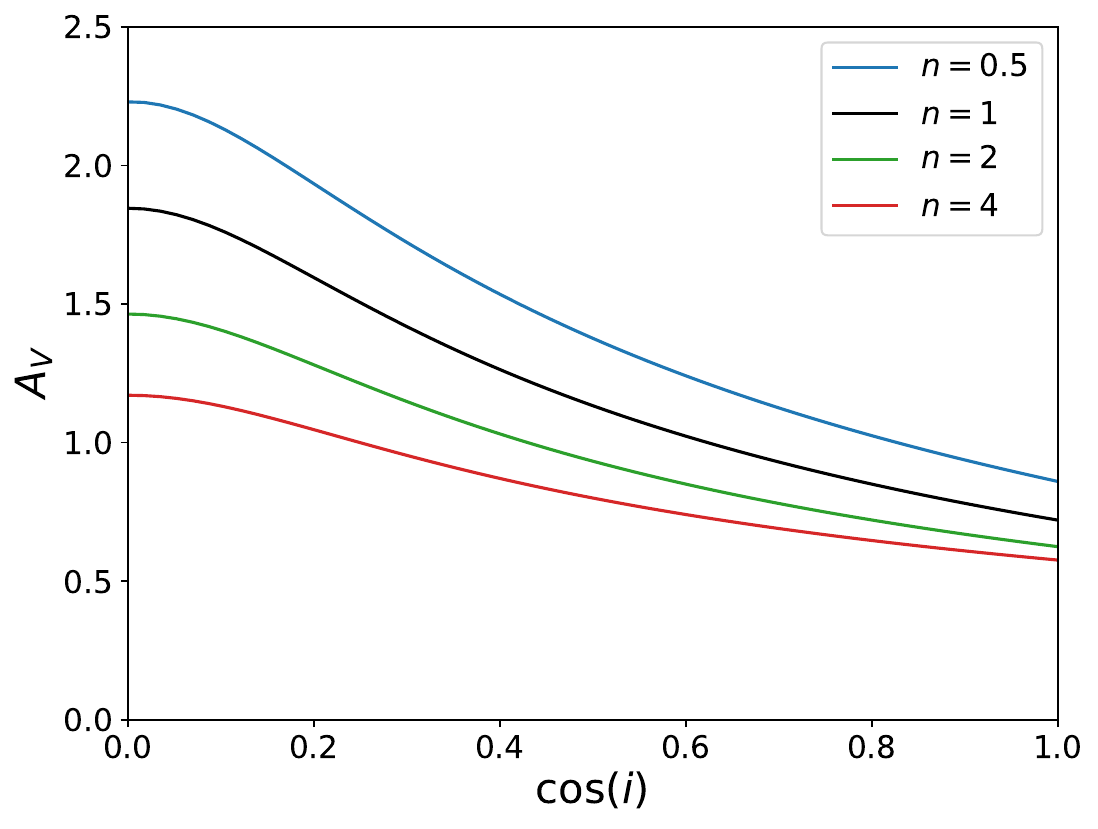}
\includegraphics[width=0.44\textwidth]{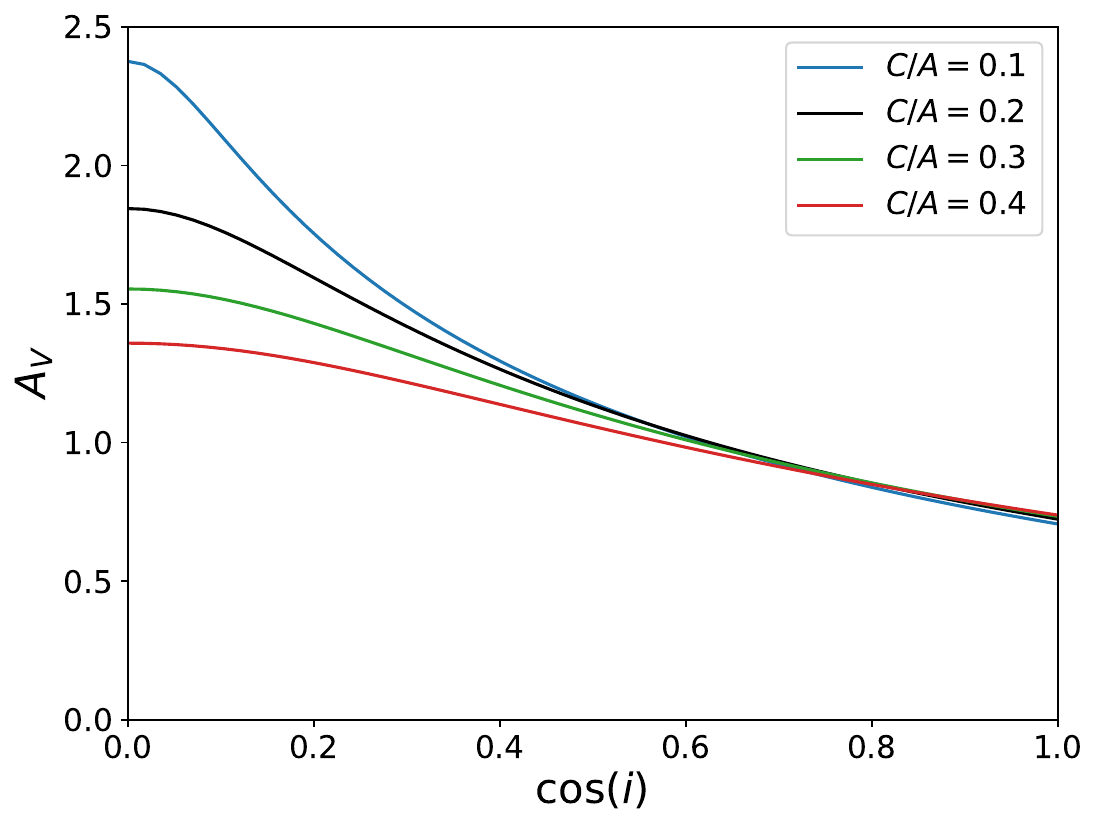}
\includegraphics[width=0.44\textwidth]{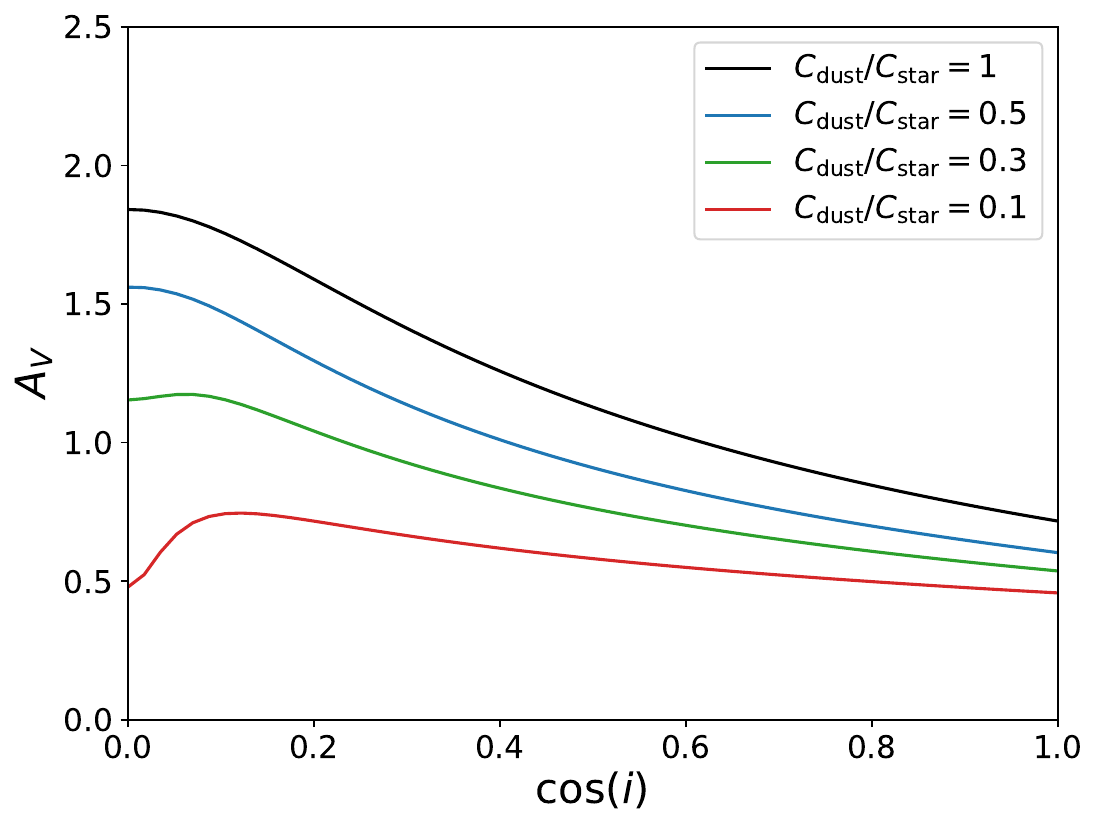}
\includegraphics[width=0.44\textwidth]{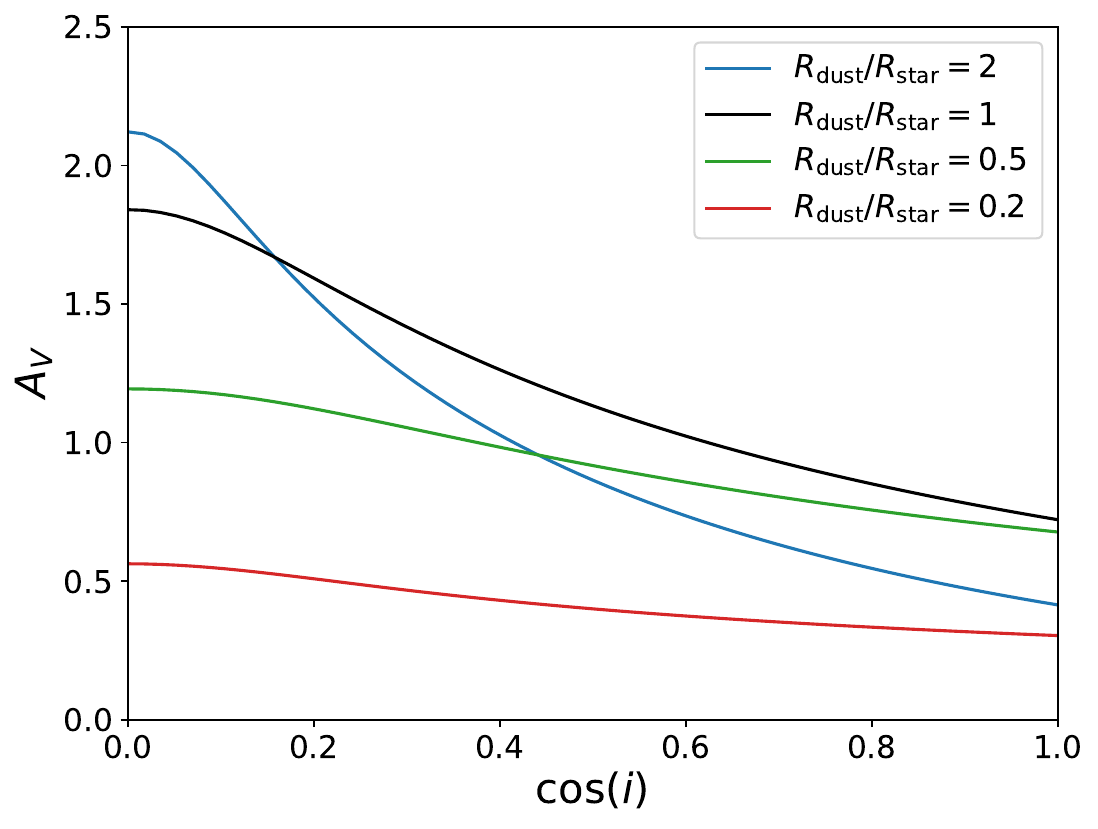}
\includegraphics[width=0.44\textwidth]{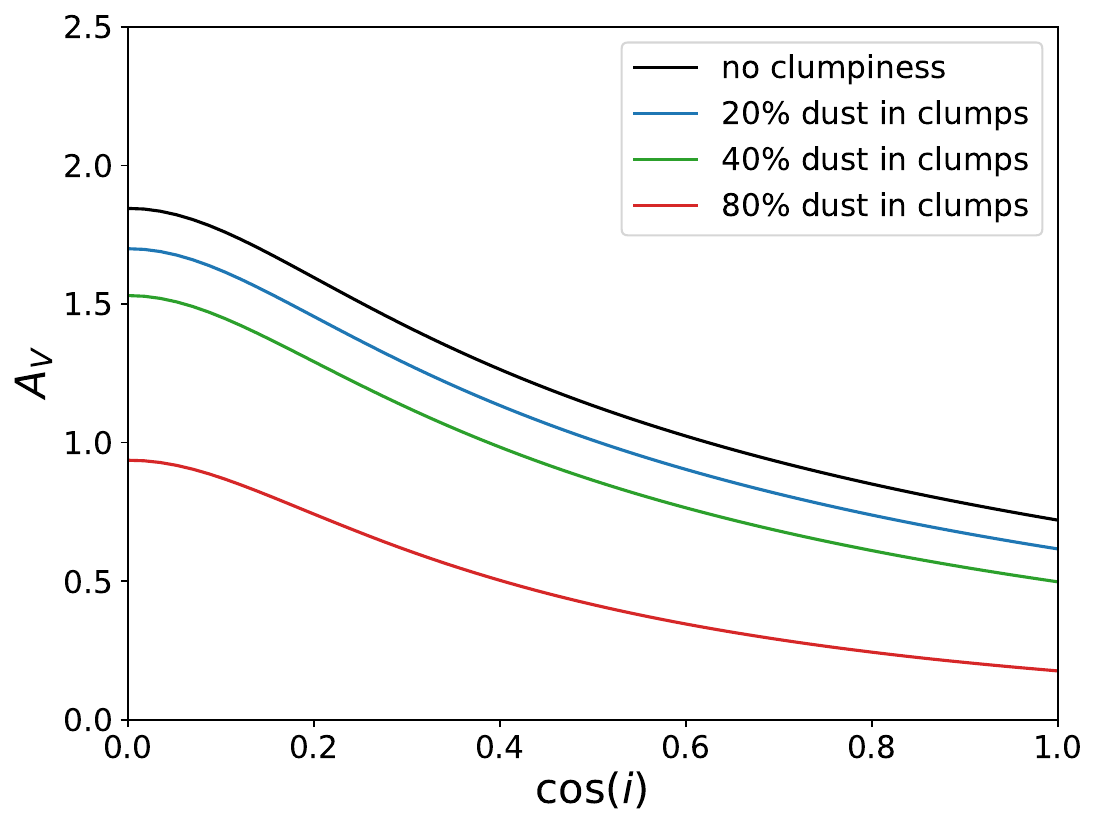}
\includegraphics[width=0.44\textwidth]{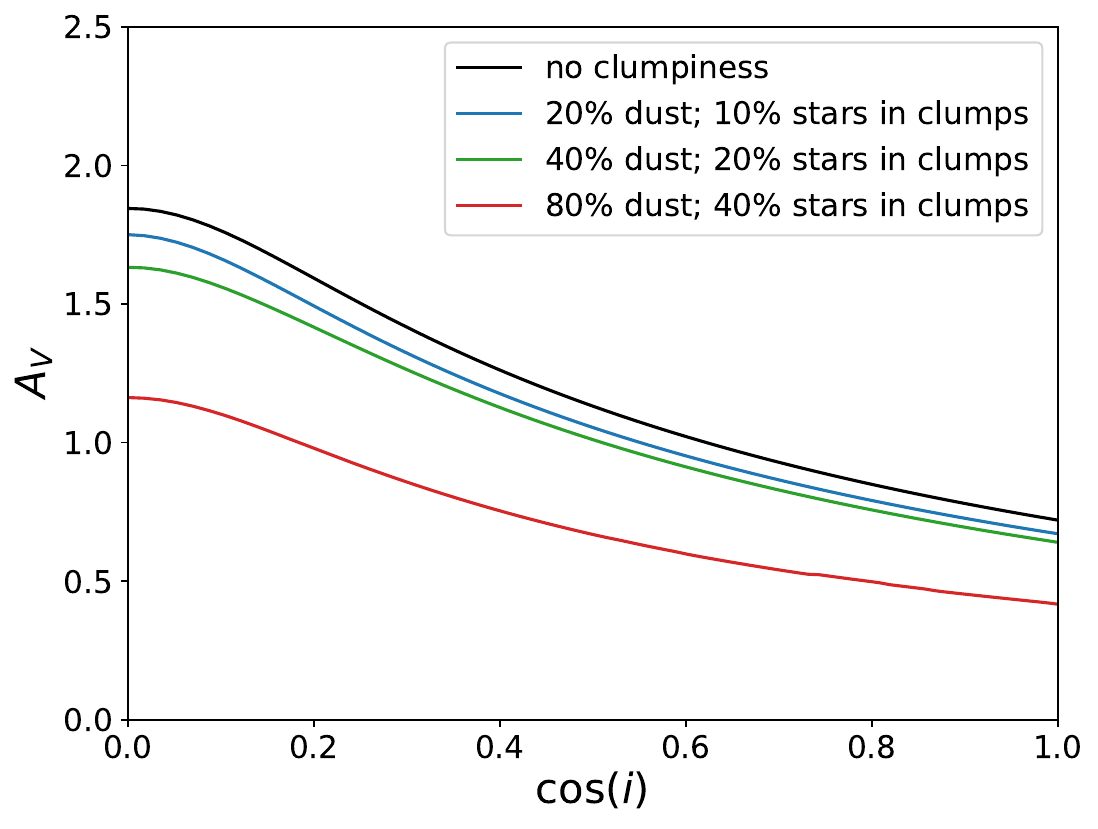}
\caption{{\it From left to right and top to bottom:} Comparison of inclination-dependent attenuation for toy models with different S\'{e}rsic indices $n$, thickness $C/A$, relative scaleheight of dust and stars $C_{\rm dust}/C_{\rm star}$, relative scalelength of dust and stars $R_{\rm dust}/R_{\rm star}$, different fractions of dust in clumps, and different fractions of dust and stars in clumps.  In each panel, all parameters other than the one listed in the legend, are kept fixed to that of the reference model shown in black.
}
\label{fig:toygeometries}
\end{figure*}

Whereas Figure\ \ref{fig:mass} presented the variation in effective attenuation for toy model galaxies of identical geometry with different dust content, we here explore the opposite.  Figure\ \ref{fig:toygeometries} shows in each panel with a black curve the attenuation for a reference toy model galaxy observed under different inclinations ($i = 0\deg$ is face-on; $i = 90\deg$ is edge-on).  The reference corresponds to an axisymmetric ellipsoidal galaxy with smooth stellar and dust distribution, $n=1$ S\'{e}rsic profile, 4 kpc size, and thickness $C/A = 0.2$, which contains $10^8\ M_{\odot}$ of dust.  In each panel, the reference model is juxtaposed to the results from separate SKIRT runs on galaxies with identical dust mass, but distinct structural properties.  The structural parameter that is varied with respect to the reference model is indicated in the legend of each panel.  For instance, the top-left panel shows that shallower radial density profiles (lower $n$) lead to overall higher attenuation and a steeper inclination dependence compared to cuspier profiles (higher $n$).  Adjusting the disk thickness $C/A$ has little effect on the face-on attenuation, but impacts the edge-on attenuation due to enhanced dust columns when the disk is thinner ({\it top-right panel}).  In both these experiments, the density profile of stars and dust was kept tied.  

When setting up smaller scaleheights for the dust disk compared to the stellar distribution ({\it middle-left panel}), this again reduces the inclination dependence and overall attenuation.    This is best pictured for a razor-thin dust disk, which, when seen edge-on, would barely affect the starlight emitted above and below the midplane.  Variations in the relative scalelength of the dust and stellar distribution ({\it middle-right panel}) have a perhaps less trivial impact.  More compact dust configurations, while of higher surface density, only obscure the inner part of the galaxy effectively, leading to a net reduction in attenuation, especially for edge-on viewing angles.  On the other hand, model galaxies with more extended dust distributions feature a reduced dust column and thus lower attenuation of starlight when seen face-on, while their dust content acts more closely like a foreground screen and thus produces higher $A_V$ when seen edge-on.

Finally, the bottom panels of Figure\ \ref{fig:toygeometries} evaluate the variation from the reference model when part of the dust ({\it bottom-left panel}), or part of the dust and a proportional fraction of the stars ({\it bottom-right panel}), is placed in dense clumps.  The location of these clumps is drawn randomly from the galaxy's overall density profile.  In the former case, the dense clumps contain dust yet no stars. Given that they are far from volume filling, more dust in clumps then effectively reduces the diffuse component responsible for attenuating the starlight.  In the latter case, as now also a portion of the stars are assigned to the same dusty clumps, these dense dust components will attenuate the starlight emitted from within the clouds, hence increasing the galaxies' net attenuation.

\end{document}